\documentclass[a4paper,12pt,oneside,openright]{revtex4-1}
\usepackage{graphicx}
\usepackage{amsmath,mathtools,amsfonts}
\usepackage{hyperref}
\usepackage{microtype}
\usepackage{subfig}
\usepackage{multirow}
\usepackage{tikz}
\usepackage{float}
\usepackage[normalem]{ulem}
\usepackage{multirow}

\graphicspath{{./figs/}}

\newcommand\york{Department of Physics and Astronomy, York University, Toronto, Ontario, M3J 1P3, Canada}

\newcommand\mainz{PRISMA$^+$ Cluster of Excellence \& Institut f\"ur Kernphysik,
Johannes Gutenberg-Universit\"at Mainz}

\newcommand\cern{Theoretical  Physics  Department,  CERN,  CH-1211  Geneva  23,  Switzerland}

\newcommand\kmyork{Department of Mathematics and
Statistics, York University, Toronto, Ontario
M3J 1P3, Canada and CSSM, University of Adelaide,
Adelaide, SA, 5005, Australia}
\begin{document}

\title{A lattice investigation of exotic tetraquark channels}

\author{R.J. Hudspith$^1$}
\author{B. Colquhoun$^2$}
\author{A. Francis$^3$}
\author{R. Lewis$^2$}
\author{K. Maltman$^4$}

\affiliation{$^1$\mainz}
\affiliation{$^2$\york}
\affiliation{$^3$\cern}
\affiliation{$^4$\kmyork}

\date{\today}

\begin{abstract}  
We perform an $n_f=2+1$ lattice 
study of a number of channels where past 
claims exist in the literature for
the existence of strong-interaction-stable 
light-heavy tetraquarks.
We find no evidence for any such deeply-bound 
states, beyond the $J^P=1^+$, $I=0$ $ud\bar{b}\bar{b}$ and $I=1/2$ 
$ls\bar{b}\bar{b}$ states already 
identified in earlier lattice studies. We
also describe a number of systematic
improvements to our previous lattice 
studies, including working with larger 
$m_\pi L$ to better suppress possible 
finite volume effects, employing extended 
sinks to better control excited-state 
contamination, and expanding the number 
of operators used in the GEVP analyses.
Our results also allow us to rule out
several phenomenological models which
predict significant tetraquark binding 
in channels where no such binding is found.
\end{abstract}

\maketitle

\section{Introduction}

Results from multiple recent 
lattice studies~\cite{Bicudo:2015kna,Bicudo:2016ooe,Francis:2016hui,Junnarkar:2018twb,Leskovec:2019ioa}  
now rather firmly establish the existence of an exotic, doubly bottom, $I=0$, $J^P=1^+$
$ud\bar{b}\bar{b}$ tetraquark state, bound with respect to $BB^*$, and hence 
strong-interaction stable. Though results for the binding energy vary, all correspond to 
masses below $BB$ 
threshold, making the state stable with respect to, not only strong-interaction, but 
also electromagnetic, decays. The results of Refs.~\cite{Francis:2016hui,Junnarkar:2018twb} also predict a strong-interaction-stable flavour $\bar{3}_F$, $I=1/2$, $J^P=1^+$ strange partner.
Refs.~\cite{Eichten:2017ffp,Braaten:2020nwp}, using 
heavy quark symmetry arguments, supplemented by
either phenomenological input~\cite{Eichten:2017ffp}
or phenomenological plus lattice 
input~\cite{Braaten:2020nwp} for leading finite 
heavy mass corrections, concur on the
strong-interaction stability of both the $I=0$ and $I=1/2$ doubly bottom 
states {\footnote{Other arguments favoring the possibility of bound doubly bottom tetraquarks, may be found in Refs.~\cite{Manohar:1992nd} and \cite{Czarnecki:2017vco}.}}. The binding in
these channels appears to be driven by 
a combination of the attractive short-distance 
colour Coulomb interaction between two  $\bar{b}$ antiquarks in a colour $3_c$
and the spin-dependent interaction of light quarks in the $\bar{3}_F$, spin $0$, colour
$\bar{3}_c$ ``good light diquark'' 
configuration, whose attractive character 
is known phenomenologically from 
observed heavy baryon splittings~\cite{Francis:2016hui,Francis:2018jyb}. 
Neither of these interactions is 
accessible in a state consisting of two well-separated heavy mesons.

While the existence of this $\bar{3}_F$ of
$J^P=1^+$ strong-interaction-stable tetraquark 
states seems well established theoretically, 
detecting these states, and confirming this 
prediction, remains experimentally challenging. 
Production rates are likely to be very low, given
that two $b\bar{b}$ pairs must first be produced. While the doubly charmed $\Xi_{cc}$ baryon has 
now been observed by LHCb~\cite{Aaij:2017ueg,Aaij:2018gfl},
and double-$\Upsilon$ production has been reported by CMS~\cite{Khachatryan:2016ydm}, to date, not even bottom-charm, let alone doubly
bottom baryon states have been detected. The recently proposed inclusive search strategy, 
in which a $B_c$ meson originating from a displaced vertex would signal the presence of doubly bottom hadron production at the LHC~\cite{Gershon:2018gda}, though clearly of 
interest, would still leave open the question of whether such a signal contained a doubly bottom tetraquark component, or was due entirely to doubly bottom baryon production. 

In this paper we use lattice QCD calculations 
to investigate whether strong-interaction-stable 
(hereafter ``bound'') tetraquarks exist in other channels likely to be more amenable to 
experimental detection. A number of such channels have been investigated in the 
literature using various approaches,
including QCD-inspired models~\cite{Ader:1981db,Heller:1986bt,Carlson:1987hh,Zouzou:1986qh,Lipkin:1986dw,SilvestreBrac:1993ss,Semay:1994ht,Pepin:1996id,Brink:1998as,Vijande:2003ki,Janc:2004qn,Vijande:2006jf,Vijande:2007rf,Ebert:2007rn,Zhang:2007mu,Vijande:2009kj,Yang:2009zzp,Carames:2011zz,Silbar:2013dda,Karliner:2017qjm,Caramees:2018oue,Chen:2018hts,Deng:2018kly,Park:2018wjk,Yang:2019itm,Huang:2019otd,Hernandez:2019eox,Tan:2020ldi,Lu:2020rog}
and QCD sum rules~\cite{Navarra:2007yw,Du:2012wp,Chen:2013aba,Wang:2017uld,Chen:2017rhl,Agaev:2018khe,Agaev:2019lwh,Agaev:2019kkz,Agaev:2019wkk,Sundu:2019feu,Agaev:2020zag,Wang:2020jgb,Tang:2019nwv}, and we will 
study a range of channels where claims for the
existence of such bound tetraquark states have
been made. Our investigations will also serve to test those aspects of the models not 
constrained by fits to the ordinary meson and baryon spectrum as well as approximations
employed in implementing the QCD sum rule framework in studies of tetraquark channels.

In the case of model studies, the parameters of the models (most typically constituent
quark models) are fixed from earlier fits to the ordinary meson and baryon spectrum. 
Interactions of a constituent quark pair in a colour $\bar{3}_c$ or a constituent quark and
antiquark in a colour $1_c$ configuration are,
thus, phenomenologically constrained. In 
tetraquark (and other multi-quark) channels, 
however, additional colour configurations,
which have totally unconstrained constituent 
interactions, are also present. In those 
channels already identified by
the lattice as supporting deeply bound, doubly bottom tetraquark states, where the physics of
the binding is believed to be dominated by a combination of the colour Coulomb attraction
in the anti-diquark colour $3_c$ and 
the diquark attraction in the
good-light-diquark $\bar{3}_c$ configuration, 
these dominant 
interactions {\it are} phenomenologically constrained, and one would, therefore, expect 
the models to successfully predict tetraquark binding in these channels. This is, indeed, typically
the case. However, even in these channels, differences in the assumed form of the model 
interactions, which, by construction, produce no differences in the ordinary hadron spectrum
(since the model parameters have been fit to ensure the spectrum is reproduced) can produce significant differences in an exotic channel. An example is provided by the comparison of results for  predicted binding in the $I=0$, $J^P=1^+$ $ud\bar{b}\bar{b}$ channel produced by a range of dynamical chiral~\cite{Pepin:1996id,Vijande:2003ki,Vijande:2006jf,Vijande:2007rf,Vijande:2009kj,Yang:2009zzp,Deng:2018kly,Tan:2020ldi} and non-chiral\cite{Carlson:1987hh,Zouzou:1986qh,SilvestreBrac:1993ss,Semay:1994ht,Vijande:2007rf,Brink:1998as,Janc:2004qn,Ebert:2007rn,Vijande:2009kj,Yang:2009zzp,Carames:2011zz,Silbar:2013dda,Park:2018wjk,Hernandez:2019eox,Lu:2020rog}
quark models. The latter typically employ a purely one-gluon-exchange (OGE) form for the spin-spin
interaction, while spin-spin interactions for the 
former are produced by a combination of effective 
Goldstone-boson exchange, acting between the light constituents only, and nominal OGE acting between 
all constituents. Even if the tetraquark state is 
entirely dominated by the $3_c$ anti-diquark, $\bar{3}_c$ good-light-diquark configuration, 
this configuration includes both $1_c$
and $8_c$ $\bar{b}$-light quark pairs. The fits of the models to the ordinary meson and
baryon spectrum place no constraints on these 
residual $8_c$ heavy-light interactions. In
addition, the colour dependence of the OGE interaction, and the 
absence of Goldstone-boson-exchange contributions to the spin-dependent 
heavy-light interactions in the chiral quark model framework mean one must expect 
significant differences in the residual 
heavy-light interactions, and hence in the predictions for the tetraquark binding, in the two classes of models. This expectation is 
also borne out in the literature, where the dynamical non-chiral models of Refs.~\cite{Carlson:1987hh,Zouzou:1986qh,SilvestreBrac:1993ss,Semay:1994ht,Brink:1998as,Janc:2004qn,Ebert:2007rn,Vijande:2009kj,Yang:2009zzp,Carames:2011zz,Park:2018wjk,Hernandez:2019eox,Lu:2020rog} produce binding 
energies between $54$ and $160$ MeV, in 
contrast to the results of the dynamical chiral models of Refs.~\cite{Pepin:1996id,Vijande:2003ki,Vijande:2009kj,Yang:2009zzp,Deng:2018kly,Tan:2020ldi}, which lie between $214$ and $497$ MeV.

Different systematic questions arise for  predictions generated using QCD sum rules,
where the underlying dispersive 
representations are rigorously valid but 
practical applications require approximations 
on both the spectral and OPE sides. All tetraquark
studies we are aware of work with 
Borel-transformed sum rules and the SVZ spectral
ansatz, which consists of a single low-lying 
``pole'' (characterized by its mass, $M$, and 
coupling, $f$, to the relevant interpolating
current) and a continuum (approximated using 
the operator product expansion (OPE)) for 
$s$ above a ``continuum threshold''
$s_0$~\cite{Shifman:1978bx,Shifman:1978by}.
The OPE side is taken as input, and the sum
rules used to fix $M$, $f$, and $s_0$.
The OPE, however, involves not just known 
quantities, like quark masses, $\alpha_s$ 
and $\langle \bar{q}q\rangle$, but also
unknown higher dimension condensates. 
Contributions up to dimensions $D=8$ 
or $10$ are retained in recent tetraquark 
analyses. As stressed in the earliest of 
the sum rule studies of doubly heavy 
tetraquarks~\cite{Navarra:2007yw}, 
estimating such higher $D$ contributions 
(typically in terms of products of known
lower-dimension condensates) requires use of 
the factorization approximation. The accuracy
of this approximation is not known in general,
but for the $I=1$ vector and axial-vector 
current-current two-point functions,
extractions of the two relevant $D=6$ 
four-quark condensates from finite-energy sum
rule analyses of OPAL~\cite{Ackerstaff:1998yj}
and ALEPH~\cite{Davier:2013sfa} hadronic 
$\tau$ decay data found it to be off by as 
much as a factor of
$\sim 6$~\cite{Boito:2012cr,Boito:2014sta}. 
On the spectral side of the sum rules, the 
single-narrow-resonance-plus-continuum 
SVZ ansatz may not be
suitable for all channels. It would, 
for example, represent a poor choice for 
the $I=0$, $J^P=0^-$ channel, which has 
not one, but three narrow, low-lying $\eta$ 
resonances. Its use in exotic
channels, where limited prior knowledge of 
the qualitative features of the spectral
distribution is likely, thus has the 
potential to produce difficult-to-assess 
systematic uncertainties. 
One way to investigate this issue is to 
consider sum rules for different
interpolating operators with the same
quantum numbers. The associated spectral
functions will all receive contributions
from all states with these quantum numbers,
with only the contribution strengths
operator dependent. Finding 
compatible pole masses and similar 
continuum thresholds from all the sum 
rules would increase confidence in the 
reliability of the SVZ ansatz. 
Finding incompatible pole masses
would, in contrast, signal either that the 
approximations used on the OPE sides 
were unreliable, or that the
channel has more than one narrow state,
raising questions about the suitability of
the use of the SVZ ansatze.

The lattice approach is ideally suited to investigating channels with a deeply bound
tetraquark ground state, and to testing model and sum rule predictions for the 
existence of such states. Provided one employs interpolating operators with reasonable ground
state overlap, the sizeable gap to the lowest 
meson-meson threshold increases the likelihood the ground state will dominate the
corresponding two-point function at moderate Euclidean times, 
before the signal is lost in noise. Judicious source and sink choices (discussed in more 
detail in Sec.~\ref{sec:walls_boxsink} below)
are also relevant to improving the ground state signal. 
Weakly bound states represent more of a challenge because finite volume (FV) effects have the potential to produce a FV analogue of
what will become a meson-meson scattering state in the continuum that is shifted below
the continuum meson-meson threshold. FV
effects are expected to be small for bound states, but not 
necessarily negligible for continuum scattering states. They are typically handled
either by extrapolations using 
simulations at multiple volumes, or by 
working with ensembles where the product $m_\pi L$, with $L$ the lattice length, is 
large enough to suppress the dominant 
``round-the-world'' FV effects. A general rule of thumb is that this requires $m_\pi L>4$. The fact that the good-light-diquark configuration is believed to play an important role in binding for the tetraquarks so far identified and that, 
phenomenologically, the associated contributions to binding are expected to grow with {\it decreasing} light quark mass, also puts a 
premium on working with $m_\pi$ as close to the physical point as possible, subject 
to maintaining a detectable signal and keeping FV effects under control. 

For the $32^3\times 64$ ensembles used in our previous studies of the $J^P=1^+$, $\bar{3}_F$ $ud\bar{b}\bar{b}$ and $\ell s\bar{b}\bar{b}$
($\ell =u,d$) channels~\cite{Francis:2016hui}
and $J^P=1^+$, $I=0$ $ud\bar{c}\bar{b}$ channel~\cite{Francis:2018jyb}
the $m_\pi L$ values were $6.1$ for $m_\pi\simeq 415$ MeV and $4.4$ for $m_\pi\simeq 299$ MeV, but only $2.4$ for $m_\pi\simeq 164$ MeV. For 
the current study, which focuses on searching for channels with bound, non-molecular (i.e.,
non-weakly-bound) tetraquark ground states, we have thus generated a new ensemble with the
same lattice spacing but a larger
volume ($48^3\times 64$) and a pion mass, $m_\pi\simeq 192$ MeV, still close to physical,
but with $m_\pi L=4.2>4$. With only this single $m_\pi$, we will be unable to identify channels
in which a shallow bound state might exist at physical $m_\pi$, but not at the slightly higher
$m_\pi$ of our simulation. Channels identified as having a moderate-to-deeply bound
tetraquark ground state for $m_\pi\simeq 192$ MeV, will, however, certainly also have such a ground state at slightly lower, physical $m_\pi$.

We now turn to the channels to be investigated in this paper. The likelihood of binding is
increased by restricting our attention to those
channels where no relative spatial excitations
are required, and where light quark pairs have
access to the favourable $\bar{3}_F$, spin 
$0$, colour $\bar{3}_c$, good-light-diquark 
configuration. The antiquark pair must 
then have access to the $3_c$ configuration
and, with no relative spatial excitation, have
spin $1$ if the two antiquarks are the same. 
Both spin $0$ and $1$ are possible if the antiquarks are different. These considerations lead us to investigate channels with either 
$I=0$ or $1/2$, $J^P=1^+$ and two 
identical antiquarks, or $I=0$ 
or $1/2$, $J^P=0^+$ or $1^+$ and two
non-identical antiquarks.

In what follows, we will first briefly revisit
our previous study~\cite{Francis:2016hui} of 
the doubly bottom, $J^P=1^+$, $\bar{3}_F$ 
channels already well established to support 
bound tetraquark states. Our goal here is not to 
re-establish this fact, but rather to illustrate
an important improvement to the methods employed 
in Ref.~\cite{Francis:2016hui}. In that analysis
(as well as in our analysis of the $I=0$, 
$J^P=1^+$ $ud\bar{c}\bar{b}$ 
channel~\cite{Francis:2018jyb}) local sinks
were combined with gauge-fixed wall sources.
With this setup, the resulting ground state 
effective mass plateaus were short 
and reached at later Euclidean times.
With Wall-Local effective masses typically
plateauing from below, this leaves open the 
possibility that the true plateaus had not yet 
been reached and the 
resulting binding energies overestimated. 
Improving the ground state signal is thus
desirable, and we have succeeded in 
accomplishing this goal using the extended
``Box-Sink'' construction, described in 
Sec.~\ref{sec:walls_boxsink} below. The results 
for the bound doubly
bottom $\bar{3}_F$ ground states, at 
the $m_\pi$ of this study, are thus included 
primarily to illustrate this improvement, 
and to motivate the use of the Box-Sink 
construction in the other channels considered
here, which are identified and discussed in more
detail in Sec.~\ref{channels_review}. 
For the doubly bottom channels, where binding is 
already clearly established, the ultimate goal
is to carry out simulations at not just 
the larger volume and single $m_\pi$ of the
current study, but also, 
at this same volume, for multiple $m_\pi$, 
in order to provide updated results for the 
binding, extrapolated to physical $m_\pi$. 
The results of this extended update, which is
in progress, will be reported elsewhere. 

The rest of this paper is organized as
follows. In Sec.~\ref{channels_review},
we list the additional channels 
to be studied and review existing 
predictions for tetraquark binding in each.
Sec.~\ref{sec:operators} provides a list
of the operators used in these studies and 
Sec.~\ref{sec:walls_boxsink} a discussion of
our gauge-fixed wall sources, and the 
extended, Box-Sink construction, introduced 
to improve our ground state signals. 
Sec.~\ref{sec:lattice} provides details 
of our lattice setup and calculations,
and Sec.~\ref{sec:results} the results 
of these calculations. Our final conclusions,
together with a discussion of the implications
of our results, are given in 
Sec.~\ref{sec:conclusions}.

\section{Channels of interest and existing tetraquark predictions therein}\label{channels_review}

In this section, we specify the additional
channels to be considered in this study, 
providing a review of existing tetraquark binding
predictions for, and the reasons for an interest 
in, each such choice. The predictions are to be tested against results from the lattice in Sec.~\ref{sec:conclusions}.

The existence of sizeable $B_c$ data sets 
already establishes the existence of significant
simultaneous production of $b\bar{b}$ and 
$c\bar{c}$ pairs. A bound tetraquark in one 
of the two mixed-heavy charm-bottom channels
is thus expected to be much more amenable to 
experimental detection than either of the 
$J^P=1^+$, $\bar{3}_F$ doubly bottom 
analogues discussed above. 
Replacing one of the two $\bar{b}$ antiquarks by 
a $\bar{c}$, however, is expected to
both reduce the Coulomb attraction of the 
$3_c$ heavy antiquark pair and increase 
the residual spin-dependent heavy-light
interactions, which heavy baryon spectrum
splittings suggest is likely to further
reduce binding. The bottom-charm states
(if any) are thus qualitatively expected
to be more weakly bound than their doubly bottom counterparts~\cite{Francis:2018jyb}. 

The mixed-heavy $I=0$, $J^P=1^+$ $ud\bar{c}\bar{b}$ channel was investigated
previously on the lattice~\cite{Francis:2018jyb}. Evidence of a
possible bound ground state with binding between $\sim 15$ and $61$ MeV was found, though with 
rather short, late-time plateaus. The heavy 
quark symmetry arguments of 
Refs.~\cite{Eichten:2017ffp,Braaten:2020nwp}, which
produce binding compatible with that found
from the lattice for the doubly bottom $I=0$, $J^P=1^+$ ground state, in contrast, 
predict no binding. A number of non-chiral models~\cite{Zouzou:1986qh,SilvestreBrac:1993ss,Semay:1994ht,Caramees:2018oue,Park:2018wjk,Lu:2020rog} 
also find an either unbound or only weakly bound ground state, with binding energies 
between $-1$ MeV (unbound) and $20$ MeV (depending on the details of the potential used) and $23$ MeV reported in 
Refs.~\cite{Semay:1994ht} and \cite{Caramees:2018oue}, respectively. 
Significantly larger binding is found in most recent chiral model studies, where 
Refs.~\cite{Deng:2018kly,Yang:2019itm} and 
\cite{Tan:2020ldi}, for example, report ground
states bound by $171\pm 12$ MeV, 
$217$ MeV and $199$ MeV, respectively. QCD sum
rules, in contrast, produce inconclusive results, 
with binding of $60\pm 100$ MeV reported in
Ref.~\cite{Chen:2013aba}.

The $J^P=0^+$, $I=0$, $ud\bar{c}\bar{b}$ analogue has not yet been investigated
on the lattice. The heavy quark symmetry arguments of Refs.~\cite{Eichten:2017ffp,Braaten:2020nwp} 
again predict no binding in this channel. Non-chiral models~\cite{Lipkin:1986dw,SilvestreBrac:1993ss,Semay:1994ht,Karliner:2017qjm,Caramees:2018oue,Park:2018wjk,Lu:2020rog} predict an either 
unbound or only weakly bound ground state, with Refs.\cite{Semay:1994ht,Karliner:2017qjm} and \cite{Caramees:2018oue} quoting binding
energies of between $-11$ MeV (unbound) and $13$ MeV (depending on the details of the
potential used), $11\pm 13$ MeV, and $23$ MeV, respectively. Significantly larger binding is found
in most recent chiral model studies, with
Refs.~\cite{Deng:2018kly,Yang:2019itm,Tan:2020ldi},
for example, reporting bindings of $136\pm 12$ MeV,
$196$ MeV and $178$ MeV, respectively. Two
recent QCD sum rule studies differ in their
pole mass results, with Ref.~\cite{Chen:2013aba}, which includes
OPE contributions up to $D=8$, finding 
$7.14\pm 0.10$ GeV, while Ref.~\cite{Agaev:2018khe}, which includes
contributions up to $D=10$, finds
$6660\pm 150$ MeV. The latter corresponds
to a very deep, $485\pm 150$ MeV, binding relative to $DB$ threshold. 

The $J^P=0^+$ and $1^+$ $\ell s\bar{c}\bar{b}$ 
channels have yet to be
studied on the lattice. Analogous $J^P=1^+$,
$\ell s\bar{b}^\prime\bar{b}$ states, with variable
$\bar{b}^\prime$ mass as low as $\sim 0.6\, m_b$,
have, however, been considered, for $m_\pi = 299$ MeV 
only, in Ref.~\cite{Francis:2018jyb}. With both
$\bar{b}$ and $\bar{b}^\prime$ treated using
NRQCD, this study could not be extended to
$\bar{b}^\prime$ masses as low as $m_c$; the observed variable $\bar{b}^\prime$ mass
dependence, however, was argued to make
a bound $\ell s\bar{c}\bar{b}$ state extremely 
unlikely in the $J^P=1^+$ channel at this pion 
mass. Both the heavy quark symmetry arguments of Refs.~\cite{Eichten:2017ffp,Braaten:2020nwp} and
non-chiral model of Ref.~\cite{Lu:2020rog}
predict no tetraquark bound state in either this channel or the $J^P=0^+$ analogue. In 
contrast, the QCD sum rule study of Ref.~\cite{Wang:2020jgb}, which includes
contributions up to $D=8$,
finds binding energies of $200\pm 130$ and $180\pm 110$ MeV for the $J^P=0^+$ and $1^+$
ground states, respectively. Even deeper 
$J^P=0^+$ bindings, exceeding $400$ MeV,
are reported in the QCD sum rule studies of 
Refs.~\cite{Agaev:2019lwh} and
~\cite{Agaev:2020zag}, which include
OPE contributions up to $D=10$.

The reported observation by the D0 
collaboration~\cite{D0:2016mwd,Abazov:2017poh}
of a narrow state, $X(5568)$, decaying to $B_s\pi^\pm$, and hence having four distinct
flavours, prompted speculation that the 
related singly-heavy $ud\bar{s}\bar{b}$ and $ud\bar{s}\bar{c}$ channels might support
bound tetraquark states. The initial argument was based on the expectation that the putative
$X(5568)$ should have a $U$-spin-interchanged $SU(3)$ $ud\bar{s}\bar{b}$ partner with 
similar mass, coupled with the observation that the threshold, $5773$ MeV, for the lowest-lying, 
two-meson $ud\bar{s}\bar{b}$ state, $BK$, lies well above the reported $X(5568)$ mass. While 
this initial motivation is weakened by the fact 
that searches by the LHCb~\cite{Aaij:2016iev}, CMS~\cite{Sirunyan:2017ofq}, 
CDF~\cite{Aaltonen:2017voc} and ATLAS~\cite{Aaboud:2018hgx} collaborations 
failed to confirm the D0 result, a number of
model and QCD sum rule studies exist predicting bound tetraquark ground states 
in the $I=0$, $J^P=0^+$ and $1^+$,
$ud\bar{s}\bar{b}$ and/or $ud\bar{s}\bar{c}$ channels. The quark colour delocalization
screening model, for example, predicts $J^P=0^+$ and $1^+$ $ud\bar{s}\bar{b}$ 
ground states bound by $74$ and $58$ MeV relative to the respective two-meson, $BK$ 
and $B^*K$, thresholds~\cite{Huang:2019otd}.
Another $SU(3)$ chiral quark model study, that of
Ref.~\cite{Chen:2018hts}, obtains similar
bindings, $70$ and $68$ MeV, for the $J^P=0^+$ and $1^+$ bottom-strange states. Somewhat
smaller bindings, $19$ and $16$ MeV,
respectively, are found in the alternate chiral quark model study of Ref.~\cite{Tan:2020ldi}. 
A bound tetraquark with even lower mass, 
$5380\pm 170$ MeV (corresponding to a binding
of $394\pm 170$ MeV relative to $BK$ threshold), 
is also found for the $J^P=0^+$
bottom-strange channel in the QCD sum rule study 
Ref.~\cite{Agaev:2019wkk}. In fact, the only
bottom-strange sector investigation we are aware 
of which does not produce a bound tetraquark state
is the non-chiral-model, $I=0$, $J^P=1^+$ channel 
study of Ref.~\cite{Zouzou:1986qh}. This
study also found no bound tetraquark in the $I=0$,
$J^P=1^+$ $ud\bar{s}\bar{c}$ channel. An absence
of binding was also found for the related
$I=0$, $J^P=0^+$ $ud\bar{s}\bar{c}$ channel in 
the sum rule study of Ref.~\cite{Agaev:2019wkk}. 
To our knowledge, the only other 
$ud\bar{s}\bar{c}$ channel study is that of
Ref.~\cite{Tan:2020ldi}, which finds $I=0$,
$J^P=0^+$ and $1^+$ states bound by $15$ and $9$ 
MeV, respectively. Any bound state
found in these channels would be of
considerable phenomenological interest,
since it would involve light degrees of 
freedom in the novel strangeness $+1$, colour
$3_c$ configuration about which no phenomenological information is currently known.

While it might appear natural to include the $I=0$, $J^P=1^+$ 
$ud\bar{c}\bar{c}$ and $I=1/2$, $J^P=1^+$ $\ell s\bar{c}\bar{c}$  
channels in our study, we have not done so,
since evidence already exists that bound doubly charmed tetraquark states, even if they
exist in these channels, will be at most weakly bound. This evidence comes from both the lattice and 
the heavy-quark-symmetry arguments of 
Refs.~\cite{Eichten:2017ffp,Braaten:2020nwp},
whose results are compatible with those of
lattice studies for the doubly bottom bound
states. The latter predicts no binding in
either of the doubly charmed $J^P=1^+$
channels. On the lattice, in the $I=0$, $J^P=1^+$ $ud\bar{c}\bar{c}$ channel, (i) the 
results of Ref.~\cite{Cheung:2017tnt} clearly establish the absence of binding at 
$m_\pi =391$ MeV, 
(ii) Ref.~\cite{Junnarkar:2018twb} (which 
reaches a lower, $257$ MeV, pion mass for the
coarsest of the three lattices studied) finds
an extrapolated, physical-$m_\pi$ binding of
$23\pm 11$ MeV, sufficiently small for the
authors to comment that further FV studies 
are required to reach a firm conclusion
concerning binding, and (iii) results for 
the $b^\prime$ mass dependence of the 
binding energies for $I=0$, $J^P=1^+$ 
$ud\bar{b}^\prime\bar{b}^\prime$ states at 
$m_\pi = 299$ MeV, reported in 
Ref.~\cite{Francis:2018jyb}, strongly
disfavour the possibility that binding
survives to physical $m_\pi$ and $b^\prime$
masses as low as $m_c$. Only one lattice
result, whose binding, $8\pm 8$ MeV, is 
compatible with a two-meson-threshold
ground state, exists for the $I=1/2$, $J^P=1^+$ $\ell s\bar{c}\bar{c}$ channel~\cite{Junnarkar:2018twb}. 
QCD sum rule studies also find no evidence 
for binding in either of the
non-strange~\cite{Navarra:2007yw,Du:2012wp,Wang:2017uld,Tang:2019nwv}, or strange~\cite{Du:2012wp,Wang:2017uld,Tang:2019nwv}
doubly charmed $J^P=1^+$ channels.
Non-chiral models whose
$I=0$, $J^P=1^+$ $ud\bar{b}\bar{b}$ 
bindings are compatible with those found
on the lattice, similarly, predict either 
unbound~\cite{Zouzou:1986qh,Brink:1998as,Ebert:2007rn,Vijande:2009kj,Karliner:2017qjm,Park:2018wjk,Deng:2018kly}, or only weakly bound~\cite{Janc:2004qn,Vijande:2007rf,Yang:2009zzp}, 
doubly charmed $J^P=1^+$ ground states. 
While many chiral quark model studies
predict significant binding in the
$I=0$, $J^P=1^+$ $ud\bar{c}\bar{c}$ 
channel~\cite{Pepin:1996id,Vijande:2003ki,Vijande:2007rf,Vijande:2009kj,Yang:2009zzp,Yang:2019itm,Tan:2020ldi}, 
the models underlying these predictions typically also over-bind the $I=0$, $J^P=1^+$, $ud\bar{b}\bar{b}$ ground state relative
to results known from the lattice.
We conclude that the absence of
deeply bound tetraquark states in the
two doubly charmed channels is already
established, and hence have not included
these channels in the current study. Additional, larger volume ensembles would be
required to reliably investigate such channels,
where at-best-weak binding makes FV
studies mandatory. In future, we plan to generate
additional ensembles with similarly low
$m_\pi$, but larger volume than considered here,
and will revisit the doubly charmed $J^P=1^+$ 
channels when these become available.

Two final channels are studied in this paper.
These are the triply-heavy $J^P=1^+$ 
$uc\bar{b}\bar{b}$ and $sc\bar{b}\bar{b}$ 
channels, considered previously only in the
lattice study of Ref.~\cite{Junnarkar:2018twb}, 
which found ground state masses compatible 
with the corresponding lowest two-meson
thresholds. These channels are, of course, 
not among those where a bound tetraquark state, 
if it existed, would be more amenable to 
current experimental detection than the
theoretically established doubly bottom
states. They are included here only to
provide an additional, independent check of the 
results of Ref.~\cite{Junnarkar:2018twb}.

We close this section by stressing
that deeply bound tetraquark ground states are predicted by some models (especially chiral quark
models) and/or some QCD sum rule analyses in 
all of the $I=0$, $J^P=0^+$ and $1^+$ 
$ud\bar{c}\bar{b}$, $I=1/2$, $J^P=0^+$ 
$\ell s\bar{c}\bar{b}$, and
$I=0$, $J^P=0^+$ and $1^+$ $ud\bar{s}\bar{b}$
channels. Examples where predicted binding 
exceeds $100$ MeV can be found in (i) the recent sum rule studies of the $I=0$, 
$J^P=0^+$ $ud\bar{c}\bar{b}$, $I=1/2$, 
$J^P=0^+$ $\ell s\bar{c}\bar{b}$, and $I=0$, 
$J^P=0^+$ $ud\bar{s}\bar{b}$ channels, which 
produce ground states 
$485\pm 150$ MeV~\cite{Agaev:2018khe},
$407\pm 160$ MeV~\cite{Agaev:2019lwh} 
or $467\pm 150$ MeV~\cite{Agaev:2020zag},
and $397\pm 170$ MeV~\cite{Sundu:2019feu},
respectively, below the corresponding lowest
two-meson thresholds and (ii) recent chiral 
quark model studies of the $ud\bar{c}\bar{b}$
sector, which produce an $I=0$, $J^P=0^+$ 
ground state $136\pm 12$  
MeV~\cite{Deng:2018kly}, $196$ 
MeV~\cite{Yang:2019itm}, or $178$ 
MeV~\cite{Tan:2020ldi} below $BD$ threshold
and an $I=0$, $J^P=1^+$ ground state 
$171\pm 12$ MeV~\cite{Deng:2018kly}, $217$ 
MeV~\cite{Yang:2019itm} or $199$ 
MeV~\cite{Tan:2020ldi} below $B^*D$ threshold.
States bound this deeply should be readily
amenable to detection in the multi-operator 
lattice analyses detailed below. 

It is also worth emphasizing that chiral and
non-chiral quark models, all of which admit
parametrizations which successfully reproduce the ordinary meson and baryon spectra, make 
qualitatively different predictions for
the tetraquark channels discussed above.
The non-chiral models, once one moves beyond
the doubly bottom sector, predict only a few,
usually at-most-weakly-bound tetraquark
candidates. The majority of chiral model 
studies, in contrast, predict a larger number
of much more deeply bound states. The lattice 
studies below should thus allow us to rule 
out at least one of these classes of 
models as providing an acceptable 
phenomenological representation of QCD 
in the low-energy regime.


\section{Operators}\label{sec:operators}

In this paper we work with the following set of 
operators having couplings to states 
with two quarks and two antiquarks with quark 
flavours $\psi,\phi,\theta,$ and $\omega$, and 
having either ``meson-meson'' or 
``diquark-antidiquark'' spin-colour-flavour 
structure:  
\begin{equation}
\begin{gathered}
D(\Gamma_1,\Gamma_2) = (\psi_a^T C\Gamma_1 \phi_b)(\bar{\theta}_a C\Gamma_2 \bar{\omega}^T_b ),\\
E(\Gamma_1,\Gamma_2) = (\psi_a^T C\Gamma_1 \phi_b)(\bar{\theta}_a C\Gamma_2 \bar{\omega}^T_b - \bar{\theta}_b C\Gamma_2 \bar{\omega}^T_a),\\
M(\Gamma_1,\Gamma_2) = (\bar{\theta}\Gamma_1 \psi)(\bar{\omega}\Gamma_2 \phi),\qquad N(\Gamma_1,\Gamma_2) = (\bar{\theta}\Gamma_1 \phi)(\bar{\omega}\Gamma_2 \psi),\\
O(\Gamma_1,\Gamma_2) = (\bar{\omega}\Gamma_1 \psi)(\bar{\theta}\Gamma_2 \phi),\qquad
P(\Gamma_1,\Gamma_2) = (\bar{\omega}\Gamma_1 \phi)(\bar{\theta}\Gamma_2 \psi).\\
\end{gathered}
\end{equation}
The Dirac structures $\Gamma_1$ and $\Gamma_2$
relevant to the channels we consider are
specified below. The short-hand terminologies
``meson-meson'' and ``diquark-antidiquark'' 
are meant only to
compactly characterize the spin-colour-flavour structures, and should not be over-interpreted; 
there is nothing, for example, to prevent one 
of the ``meson-meson'' 
operators from coupling to a state with a
spatially extended diquark-antidiquark structure 
or one of the ``diquark-antidiquark'' operators 
from coupling to a meson-meson scattering state. 
An illustration is provided by the deeply
bound ground state in the doubly bottom, $I=0$, $J^P=1^+$ channel, which was found to couple
strongly to both the ``meson-meson'' and
``diquark-antidiquark'' operators considered in
Ref.~\cite{Francis:2016hui}.

\begin{table}[ht]
\begin{tabular}{c|ccc}
\toprule
Type $(\psi \phi \theta \omega)$ & $I(J)^{P}$ & Diquark-Antidiquark & Dimeson \\
\hline
\multirow{5}{*}{$udcb/udsb/udsc$} & \multirow{5}{*}{$0(1)^+$}
& \multirow{3}{*}{ $D(\gamma_5,\gamma_i),D(\gamma_t\gamma_5,\gamma_i\gamma_t)$ } &
$M(\gamma_5,\gamma_i)-N(\gamma_5,\gamma_i)$\\
& &
& $M(I,\gamma_i\gamma_5)-N(I,\gamma_i\gamma_5)$\\
& & \multirow{3}{*}{$E(\gamma_5,\gamma_i),E(\gamma_t\gamma_5,\gamma_i\gamma_t)$}  & $O(\gamma_5,\gamma_i)-P(\gamma_5,\gamma_i)$\\
& & & $O(I,\gamma_i\gamma_5)-P(I,\gamma_i\gamma_5)$\\
& & & $\epsilon_{ijk}M(\gamma_j,\gamma_k)$\\
\hline
\multirow{2}{*}{$udbb$} & \multirow{2}{*}{$0(1)^+$}
& \multirow{2}{*}{$D(\gamma_5,\gamma_i),\;D(\gamma_t\gamma_5,\gamma_i\gamma_t)$} &
$M(\gamma_5,\gamma_i)-N(\gamma_5,\gamma_i)$\\
& & & $M(I,\gamma_i\gamma_5)-N(I,\gamma_i\gamma_5)$\\
\hline
\multirow{3}{*}{$lsbb/ucbb/scbb$} & \multirow{3}{*}{$\frac{1}{2}(1)^+$}
& \multirow{3}{*}{$D(\gamma_5,\gamma_i),\;D(\gamma_t\gamma_5,\gamma_i\gamma_t)$} &
$M(\gamma_5,\gamma_i),\;M(I,\gamma_i\gamma_5)$\\
& & & $N(\gamma_5,\gamma_i),\;N(I,\gamma_i\gamma_5)$\\
& & & $\epsilon_{ijk}M(\gamma_j,\gamma_k)$\\
\hline
\multirow{4}{*}{$uscb$} & \multirow{4}{*}{$\frac{1}{2}(1)^+$}
& \multirow{2}{*}{$D(\gamma_5,\gamma_i),\;D(\gamma_t\gamma_5,\gamma_i\gamma_t)$} &
$M(\gamma_5,\gamma_i),\;M(I,\gamma_i\gamma_5)$\\
& & 
& $N(\gamma_5,\gamma_i),\;N(I,\gamma_i\gamma_5)$\\
& & \multirow{2}{*}{$E(\gamma_5,\gamma_i),\;E(\gamma_t\gamma_5,\gamma_i\gamma_t)$} & $O(\gamma_5,\gamma_i),\;O(I,\gamma_i\gamma_5)$\\
& & & $\epsilon_{ijk}M(\gamma_j,\gamma_k)$\\
\botrule
\end{tabular}
\caption{$J^P=1^+$ operators used in this work.}
\label{j0_j1_operator_combos}
\end{table}

\begin{table}[ht]
\begin{tabular}{c|ccc}
\toprule
Type $(\psi \phi \theta \omega)$ & $I(J)^{P}$ & Diquark-Antidiquark & Dimeson \\
\hline
\multirow{3}{*}{$udcb/udsb/udsc$} & \multirow{3}{*}{$0(0)^+$}
& \multirow{3}{*}{$E(\gamma_5,\gamma_5),E(\gamma_t\gamma_5,\gamma_t\gamma_5)$} &
$M(\gamma_5,\gamma_5)-N(\gamma_5,\gamma_5)$\\
& & & $M(I,I)-N(I,I)$\\
& & & $M(\gamma_i,\gamma_i)$\\
\hline
\multirow{3}{*}{$uscb$} & \multirow{3}{*}{$\frac{1}{2}(0)^+$}
& \multirow{3}{*}{$E(\gamma_5,\gamma_5),\;E(\gamma_t\gamma_5,\gamma_t\gamma_5)$} &
$M(\gamma_5,\gamma_5),\;M(I,I)$\\
& & & $N(\gamma_5,\gamma_5),\;N(I,I)$\\
& & & $M(\gamma_i,\gamma_i)$\\
\botrule
\end{tabular}
\caption{$J^P=0^+$ operators used in this work.}
\end{table}

The lattice explicitly breaks rotational symmetry, 
breaking the continuum rotation group down
to the little group $O_h$. The operators listed above lie in either the $A_1$ or $T_1$
irrep, which map directly to the continuum 
$J=0$ and $J=1$ quantum numbers 
respectively. For the operators that are 
produced from the product
$T_1\bigotimes T_1 = A_1\bigoplus E \bigoplus T_1 \bigoplus T_2 $~\cite{Moore:2006ng} 
we pick out the scalar and vector components 
respectively~\cite{Leskovec:2019ioa}.

The operator combinations which in principle couple
to states in the $J^P=0^+$ and $1^+$
channels considered below are listed in 
Table~\ref{j0_j1_operator_combos}.
While the diquark-antidiquark operators, $D$ 
and $E$, were included in initial 
explorations of the $ud\bar{s}\bar{c}$
and $ud\bar{s}\bar{b}$ channels, they were
found to couple only weakly to the
corresponding ground states and hence 
not included in the final analyses of
these channels. They were 
also omitted from the final version of 
the analyses of the $ud\bar{c}\bar{b}$ 
channels where using the set of 
meson-meson operators was found to give 
more statistically precise results and
produce no significant change to the
energies of the lowest-lying states.

Throughout this work we will obtain our ground 
states from a generalised eigenvalue problem
(GEVP) approach~\cite{Michael:1982gb,Luscher:1990ck,Blossier:2009kd}. 
We use the solutions of the GEVP (sample by 
sample) to extract the ``optimised correlator''
\begin{equation}
C_i(t) = \sum_{j,k}V_{ij}(\tau)^\dagger C_{jk}(t) V_{ki}(\tau)
\label{optimizedcorrdefn}\end{equation}
where $V$ is the matrix with columns made of 
the eigenvector solutions of
\begin{equation}
C_{ij}(t)v_j(t) = \lambda_i C_{ij}(t+t_0)v_j(t)\ .
\end{equation}
$\tau$ (the ``diagonalization time''~\cite{Horz:2019rrn}) in 
Eq.~(\ref{optimizedcorrdefn}) is to be chosen 
large enough to produce improved projection onto 
the ground state. We then perform a correlated 
single-exponential fit to the resulting 
optimised correlators to obtain our final levels.
In the rest of this work we will quote results
obtained using $t_0/a=2$ and $\tau /a =4$, 
values for which the ground state masses are 
found to display good stability upon variation 
of these parameters. Larger values of $t_0$ and 
$\tau$ tend to make the solution statistically 
less precise, and eventually unstable.
\section{Wall sources and the box-sink construction}\label{sec:walls_boxsink}

Throughout this work we will use Coulomb-gauge-fixed wall sources \cite{Billoire:1985yn,Gupta:1990mr} (fixed to high-precision using the FACG 
algorithm of \cite{Hudspith:2014oja}). These sources have some benefits, as well as some 
peculiarities, which are discussed below.

A gauge-fixed wall source is a time-slice of point sources on a Coulomb gauge-fixed
background. Here we assume the gauge condition 
has been applied over the whole volume. The elements of the source do not need to be 
stochastically drawn as the gauge fixing 
condition connects colors appropriately; this means that these sources can be used for baryons~\cite{Daniel:1992ek,Aoki:1995bb,Antonio:2006px}.
Gauge-fixed wall source results can be 
considerably more statistically precise, especially at lighter pion masses, than point
sources at the same cost due to the volume-sampling. Like other wall sources, these 
automatically zero-momentum project, and if 
momentum is needed it must be put into the 
source explicitly as a momentum source or 
introduced via partial twisting~\cite{Bedaque:2004kc,Sachrajda:2004mi}, 
as was done for the NRQCD tuning (see App.\ref{app:NRQCD_tuning}).

Use of a guage-fixed wall source is much like 
that of a point source, though with some potential
complications for channels like those studied 
here where two-meson thresholds exist. When
combined with a local sink, one typically finds 
a negative sign for the amplitude of (at least)
the first excited state, and hence a ``Wall-Local'' correlator whose effective
mass approaches the ground state mass from below.
If one cannot measure the signal to large 
enough $t$ to ensure the ground state plateau 
has been reached, the ground state mass may 
then be under-estimated, and lead one 
to either conclude that a bound state exists 
when in fact one does not or over-estimate 
the binding in the case one does exist. This 
issue does not arise for point or stochastic 
wall sources, where the effective masses 
approach the ground-state mass from above and so 
offer a rigorous upper bound on that mass.

Often it useful to compute ``Wall-Wall'' correlators \cite{Aoki:2010dy}, defined by 
contracting objects in the usual way, with propagators that are summed over the spatial 
sites of the sink time-slice,
\begin{equation}
S^W(t) = \sum_x S(x,t).
\end{equation}
The resulting ``Wall-Wall'' correlation functions 
are symmetric under the exchange of operators 
at the source and sink
and have effective masses which 
approach the ground state from above. They 
are, however, very statistically noisy. Since 
Wall-Local correlator effective masses 
empirically approach their ground state 
plateaus from below (i.e., have negative 
excited state contamination) while those
of Wall-Wall correlators approach theirs 
from above (i.e., have positive excited state
contamination), sink combinations intermediate
between the two are expected to exist with 
small excited state contamination. The goal 
is to find, if possible, combinations of this
type which also suffer from only a limited loss
in statistical precision. One possibility would
be to use some form of sink smearing, though
the products of link matrices needed with 
the propagator solution would make that
computationally costly. Another would be
to simply take advantage of the gauge
condition and construct propagators by  
summing over points lying inside a sphere 
around each reference sink point $x$, 
\begin{equation}
S^B(x,t) = \frac{1}{N}\sum_{r^2\leq R^2} S(x+r,t).
\end{equation}
With this construction, varying $R^2$ between
$0$ and its maximum value, $3(L/2)^2$, produces a
continuous interpolation between the Wall-Local 
and Wall-Wall cases. We refer to sinks
constructed in this manner as ``Box-Sinks''. As 
the Box-Sink sum can be performed at the level 
of contractions and doesn't require extra 
inversions, the construction of Wall-Box 
correlators is reasonably cheap to implement. 
In channels where the ground state is a compact 
hadronic object, one would expect an optimal
Wall-Box improvement of the ground state plateau 
for Box-Sink radii roughly matching the physical 
ground-state hadron size. 

\begin{table}[ht]
\centering
    \begin{tabular}{c|cc|cc|cc}
    \toprule
    Operators & Flavor & $R^2$ & Flavor & $R^2$ & Flavor & $R^2$ \\
    \hline
    \multirow{3}{*}{Single Mesons} &
    $\bar{u}d$ & 0 & $\bar{s}u$ & 0 & $\bar{c}u$ & 49 \\
    & $\bar{c}s$ & 49 & $\bar{b}u$ & 49 & $\bar{b}s$ & 49 \\
    & $\bar{b}c$ & 20 & & & & \\
    \hline
    \multirow{2}{*}{Tetraquarks} &
    $ud\bar{s}\bar{c}$ & 49 & $ud\bar{s}\bar{b}$ & 49 & $ud\bar{c}\bar{b}$ & 49 \\
    & $us\bar{c}\bar{b}$ & 64 & $uc\bar{b}\bar{b}$ & 36 & $sc\bar{b}\bar{b}$ & 36 \\
    \botrule   
    \end{tabular}
    \caption{Values of the Box-Sink
    summation radii (in lattice units)
    used in this work. Values were chosen so 
    the corresponding effective mass plateaus 
    begin reasonably early.}
    \label{tab:R2choice}
\end{table}

\begin{figure}
    \centering
    \subfloat{
    \includegraphics[scale=.28]{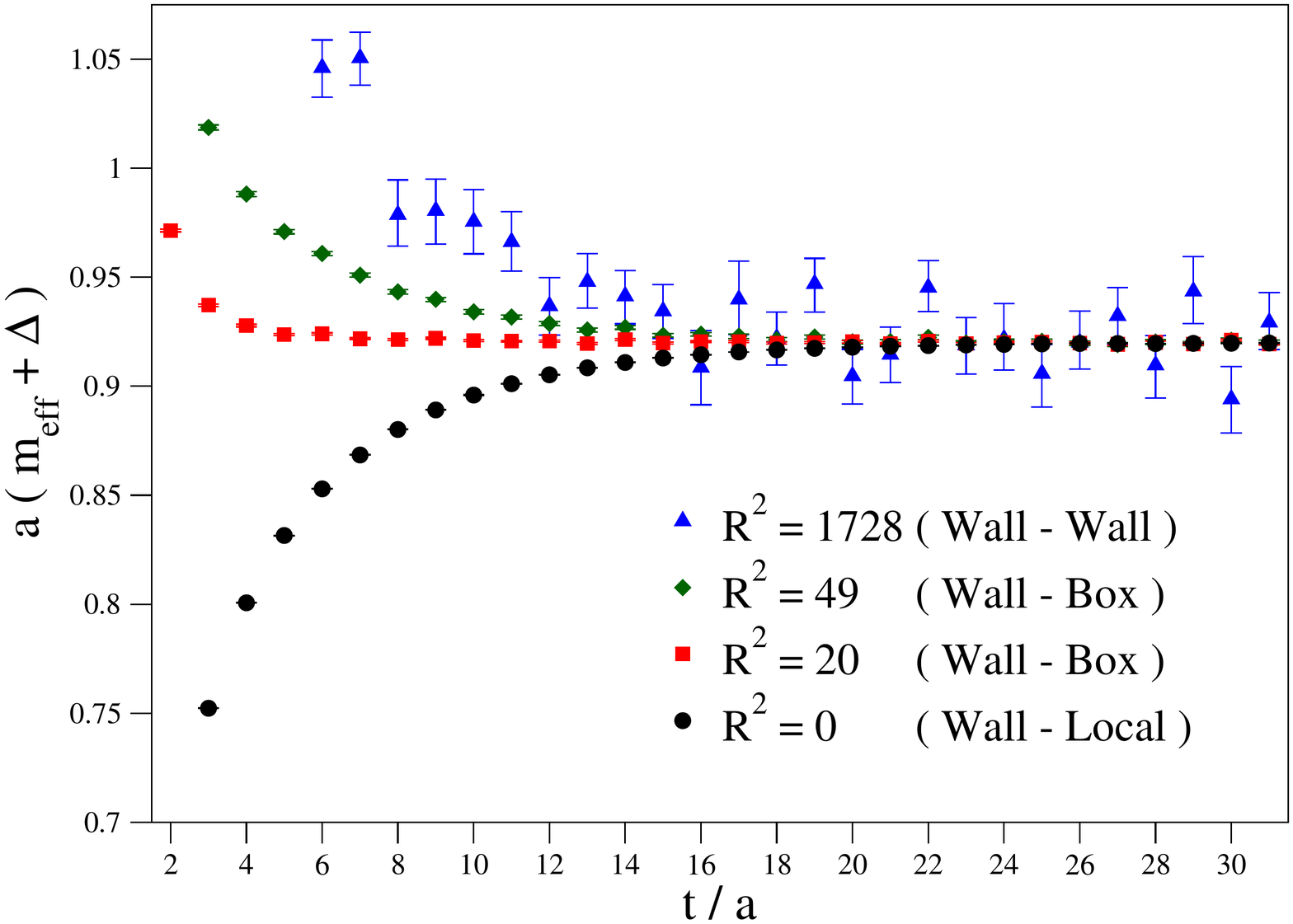}
    }
    \hspace{-24pt}
    \subfloat{
    \includegraphics[scale=0.28]{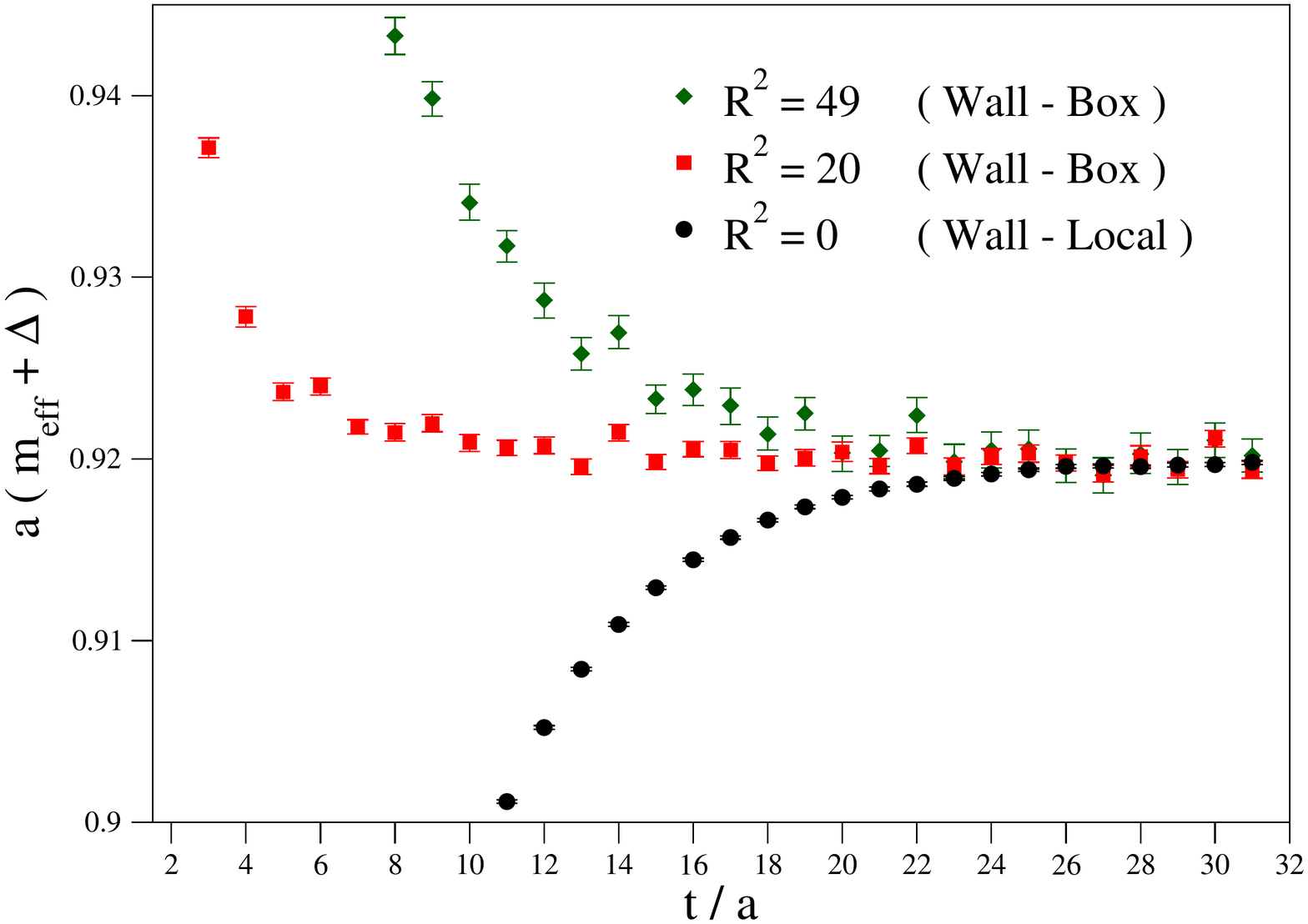}
    }
    \caption{An illustrative
    comparison of Box-Sink $B_c$ meson correlation function results for
    different Box-Sink radii. $R^2$ is
    in lattice units. The right panel 
    shows a zoom into the plateau region.}
    \label{fig:BCrComp}
\end{figure}

One should bear in mind that, in contrast to what happens in the Wall-Wall case, Wall-Box and 
Wall-Local correlation matrices are not symmetric \footnote{see App.~\ref{app:meson_amps} for more 
discussion of how to obtain physical amplitudes from these correlators}. The GEVP 
method, however, doesn't necessarily need this symmetry to obtain the underlying real
eigenvalues. Careful monitoring of the 
imaginary parts of the eigenvalues is needed 
and in practice the lack of symmetry was not
found to be an issue.

An illustration of the utility of the Box-Sink 
construction is provided, for the case of the 
$B_c$ meson channel, in Fig.~\ref{fig:BCrComp}. 
We see that intermediate values of $R^2$, indeed,
exist which extend the ground state plateau to 
significantly lower $t$ than is the case for the 
Wall-Local or Wall-Wall combinations. Results for
all Box-Sink radii must, of course, approach the same ground state 
mass at sufficiently large $t$. The point here is that the improvement in the ground state 
plateau occurs for small enough $R^2$ that the onset of the significant increase in noise that 
occurs as $R^2$ grows towards the Wall-Wall limit is avoided in the optimal $R^2$ plateau 
region. The figure also illustrates that a simple Wall-Local correlator can have 
significant excited state contamination and only approach the plateau at very large 
times. This is less of a problem in 
single-meson channels, like the $B_c$ channel,
where one can follow the signal 
out to large enough $t$, 
but becomes 
an issue in channels with larger
signal-to-noise problems, such as the tetraquark channels we study here, where we are typically 
only able to follow our signal out to $t/a\approx 20$ due to an exponential signal to noise problem. Such issues have been 
discussed in the literature for a long time in the context of the nucleon and $\rho$ meson, 
see, e.g., Refs.~\cite{Daniel:1992ek,Aoki:1995bb,Bhattacharya:1995fz}.

\begin{figure}
    \centering
    \subfloat[$ud\bar{b}\bar{b}$]{
    \includegraphics[scale=0.28]{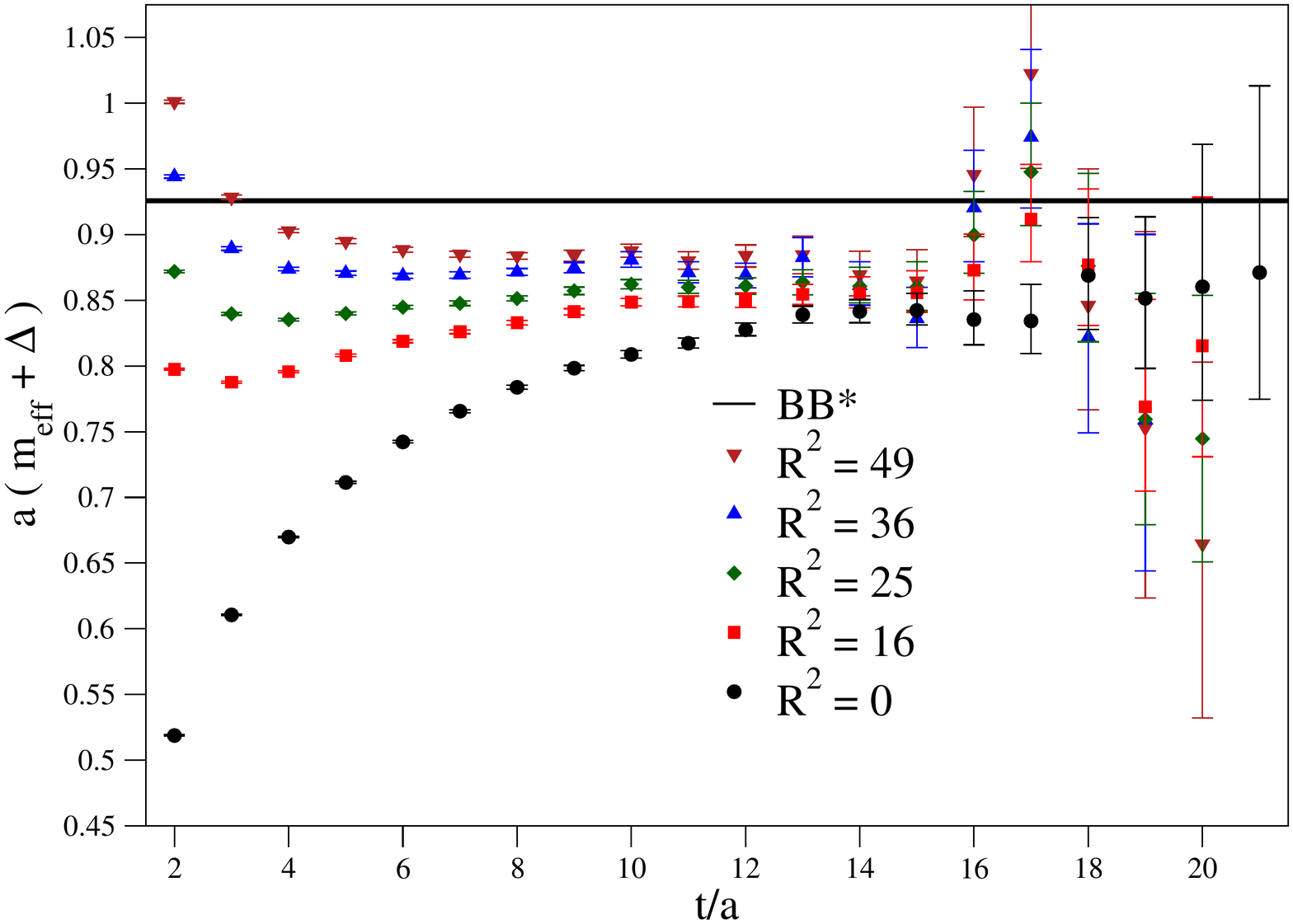}
    }
    \hspace{-24pt}
    \subfloat[$\ell s\bar{b}\bar{b}$]{
    \includegraphics[scale=0.28]{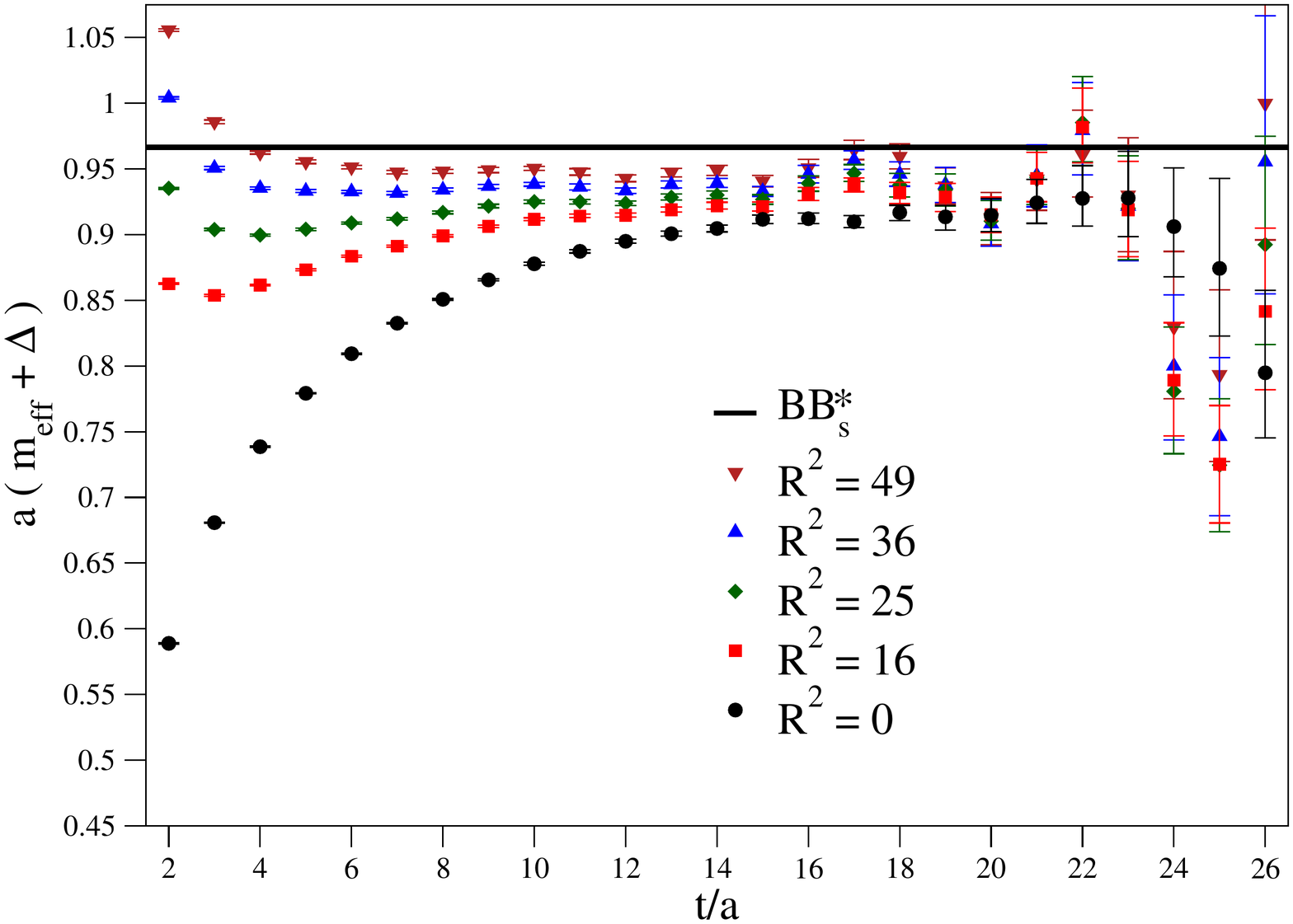}
    }
    \caption{Box-Sink effective mass plots for the lowest GEVP state in the $J^P=1^+$, $I=0$ $ud\bar{b}\bar{b}$ and $I=1/2$, $\ell s\bar{b}\bar{b}$ channels. Also shown is the lowest-lying two-meson threshold at rest.}
    \label{fig:effmass_udbb_lsbb}
\end{figure}

A further illustration of the utility of
the Box-Sink construction, now for
tetraquark signals, is provided by
the effective mass plots of the optimised ground-state correlators for the $J^P=1^+$,
$I=0$ $ud\bar{b}\bar{b}$ and $I=1/2$
$\ell s\bar{b}\bar{b}$ channels at
various Box-Sink radii $R^2$, shown
in Fig.~\ref{fig:effmass_udbb_lsbb}. While
the errors grow, as anticipated, as $R^2$ 
increases, the ground state signals
plateau much earlier, and 
in the region of much lower noise, than was 
the case for the corresponding Wall-Local 
plateaus shown in Ref.~\cite{Francis:2016hui}. 
This is especially true of the optimised
$R^2$ region, $R^2\simeq 36$ or $49$.
While the signals degrade quickly for larger
$t/a$ (and larger $R^2$), it is
clear that a ``window of consistency''
exists in the region of the Box-Sink
plateaus, for both channels. It is also
clear that, while the Wall-Local ($R^2=0$)
results reach the same plateaus, they do
so only at much later $t/a$, rather 
close to the region where the signals are lost.

The fact that enhanced excited-state 
contamination is seen in the Wall-Local results
of both the $B_c$ and tetraquark channels
suggests this behavior may be common for
Wall-Local correlators.
Fig.~\ref{fig:effmass_udbb_lsbb} illustrates,
in more detail, the potential danger this
creates. In the absence of the longer, more
reliable Box-Sink plateaus, it would be easy, 
with only the Wall-Local results, to mistakenly
identify the start of the Wall-Local plateaus 
as occurring around $t/a=14$, leading to 
an overestimate of the binding in both  
channels. The utility of the Box-Sink 
construction in avoiding this problem 
is clear. Given the results shown in
Fig.~\ref{fig:effmass_udbb_lsbb}, we expect 
our earlier Wall-Local-correlator-based estimates
for the physical point bindings in these
channels~\cite{Francis:2016hui} to be reduced
once the ongoing improved, multiple-$m_\pi$ 
Box-Sink analysis is completed. For now, the
results for these channels, at the single
$m_\pi$ of the current study, serve to 
establish the Box-Sink construction 
as an improved method for identifying channels
supporting deeply-bound tetraquark states and
quantifying the binding therein, and motivate 
our use of it in the studies of the other 
channels, reported below. 

\section{Lattice Details}\label{sec:lattice}

For all channels investigated in this work we 
have used a single $n_f=2+1$, $48^3\times64$ 
Wilson-Clover ensemble generated as an extension 
to the PACS-CS ensembles~\cite{Aoki:2009ix} 
used in our previous works~\cite{Francis:2016hui,Francis:2018jyb}. 
As this ensemble is generated with the same 
$\beta$ and $C_{SW}$, the lattice-spacing 
(for which we use $a^{-1}=2.194(10)$ GeV \cite{Namekawa:2013vu}) and the charm-quark
action parameters should be the same as for those previous ensembles.

This ensemble was generated using openQCD-1.6 \cite{openQCD} and all propagator inversions were performed using a version of this code (modified~\footnote{We implemented this 
heavy-quark action using AVX/FMA2 vector intrinsics directly in openQCD. Typical charm-quark 
propagator inversions are comparable to those of our strange-quark propagators} for 
the charm quarks) as well. For the light and strange quarks we use the DFL+SAP+GCR solver algorithm \cite{Luscher:2003qa,Luscher:2007es,Luscher:2007se}, and solve for multiple wall source positions as the same deflation subspace can be used. For the charm quarks, we just use the SAP+GCR solver as low-mode deflation is practically 
ineffective for very heavy quarks and there is a significant overhead in generating the deflation 
subspace compared to the solve time. For the charm-quark action we use the relativistic heavy 
quark action of 
Refs.~\cite{Aoki:2001ra,Aoki:2003dg} with the same parameters as listed in \cite{Namekawa:2011wt}. 
For the b-quark propagators we use tadpole-improved, 
tree-level NRQCD propagators generated on-the-fly from 
the gauge configuration. With several technical
improvements to our generation of NRQCD propagators the combination of partially-twisted boundary conditions 
and the beneficial statistical properties of gauge-fixed wall sources could be used to tune our bare-mass 
parameter to significantly higher precision than 
before. The partially-twisted boundary conditions,
in particular, allow for a denser selection of 
momenta in the $\eta_b$ and $\Upsilon$ dispersion relations within a lattice irrep than is possible 
with only Fourier modes, and thus improve the statistical resolution. Details of our
action and of the new tuning procedure we have used are provided in App.\ref{app:NRQCD_tuning}
Almost 1000 u, s, c and b propagators have been 
generated for each of the tetraquark channels 
considered here.

The new ensemble has some attractive features. 
In particular, its pion mass, $\approx 192\ {\text MeV}$, is reasonably close to physical 
while still maintaining $m_\pi L >4$. As for the previous PACS-CS ensembles, we have a strange
sea quark heavier than physical and thus have 
to use a partially-quenched strange (using the
$\kappa_s$-value of \cite{Mohler:2013rwa}) to get 
a respectable ($507\;\text{MeV}$) kaon mass. The 
Tsukuba tuning also gives close to physical $D$ 
and $D^*$ meson masses. The lattice-valued masses 
we obtain are presented in Table~\ref{tab:mestab} in 
App.~\ref{app:results_tabs}. Note that, for NRQCD, where there is an additive 
mass renormalisation, $\Delta$, only mass differences have physical meaning. We will 
present results in this lattice-valued form for the majority of this work. 

As we use NRQCD for the b quarks, channels 
with one or more b quarks do not have a
rigorous continuum limit. We do, however, 
include terms that should cancel up to $O(a^4)$
discretisation effects in our NRQCD 
Hamiltonian \cite{Lepage:1992tx}. The c quarks have been tuned to 
have a flat dispersion relation and so are 
expected to have a very mild approach 
to the continuum. The light-quarks are 
$O(a)$-improved non-perturbatively with the 
$C_{SW}$ term. In all, we expect discretisation effects to be relatively small.

\begin{table}[ht]
    \centering
    \begin{tabular}{cccccccc}
    \toprule
        $V$ & $\kappa_l$ & $\kappa_s^{\text{sea}}$ & $\kappa_s^{PQ}$ & $N_{\text{conf}}$ & $N_{\text{src}}$ & $m_\pi L$\\
        \hline
         $48^3\times 64$& $0.13777$ & $0.13640$ & $0.13666$ & 122 & 8 & 4.2 \\ 
         \botrule
    \end{tabular}
    \caption{Details of the ensemble 
    used in this work}
    \label{tab:ensemble_info}
\end{table}

Table~\ref{tab:ensemble_info} provides a
summary of the new ensemble parameters, 
the number of configurations, $N_{\text{conf}}$,
and the number of sources per configuration,
$N_{\text{src}}$, used in this study.

All vector correlators were averaged over the spatial index $i$, and all correlators were symmetrised with their backward-propagating states in order to improve statistical resolution.

\section{Results}\label{sec:results}

In this section, we present the measured spectra
(GEVP eigenvalues) and associated two-meson
thresholds for the channels discussed above. 
The eigenvalues are obtained from correlated,
single-exponential fits to the corresponding
optimised correlators. The windows of $t$ 
used in these fits are tabulated in
App.~\ref{app:results_tabs} and the corresponding
effective-mass plots, showing results for the 
ground states, as well as those excited states
for which some signal could be resolved,
collected in App.~\ref{app:effmassplots}. 
Reliable, early-onset ground-state plateaus 
are evident in all of the channels considered. 
Additional low-lying excited states are also
typically well determined. Results for a number
of higher-lying, less well determined excited 
states, obtained from fits to signals lost 
at relatively early times and/or with relatively 
short fit windows are also included in the 
tables and spectrum plots. These serve to 
identify, qualitatively, the parts of
higher-lying two-meson spectrum which 
couple most strongly to the operators 
under consideration.

\begin{figure}[h]
    \centering
    \subfloat[$ud\bar{c}\bar{b}$]
    {
    \includegraphics[scale=0.28]{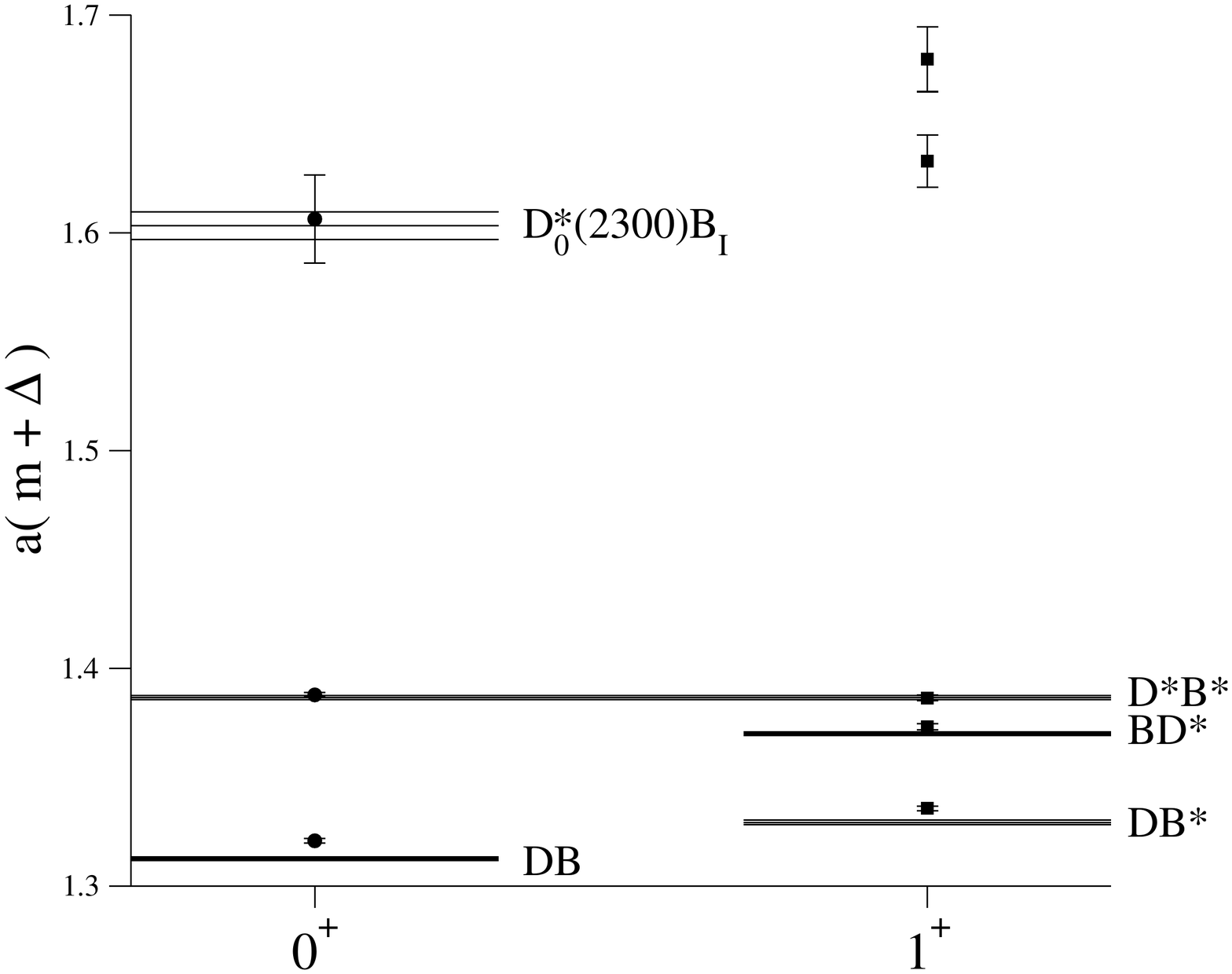}
    \label{fig:udcb}
    }
    \hspace{-32pt}
    \subfloat[$\ell s\bar{c}\bar{b}$]
    {
    \includegraphics[scale=0.28]{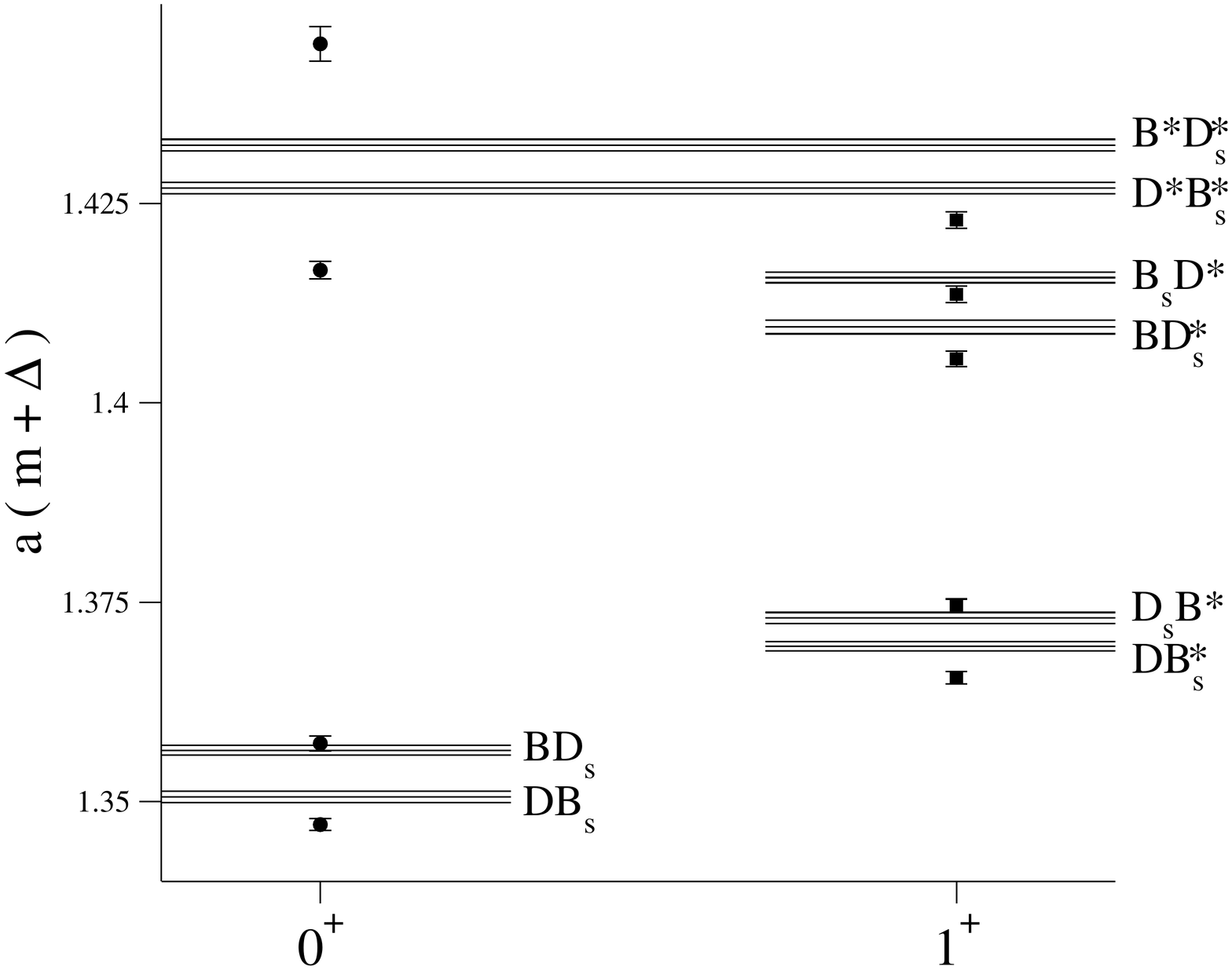}
    \label{fig:uscb}
    }
    \caption{Lowest-lying GEVP 
    eigenvalues and associated
    two-meson thresholds for the $I=0$, $J^P=0^+$ and $1^+$ $ud\bar{c}\bar{b}$ (left) and $J^P=0^+$ and $1^+$ $\ell s\bar{c}\bar{b}$ (right) channels.}
\end{figure}

Fig.~\ref{fig:udcb} shows the spectrum 
results for the $I=0$, $J^P=0^+$ and $1^+$ 
$ud\bar{c}\bar{b}$ channels. The ground states 
in both cases are found to be roughly consistent
with the lowest two-meson threshold, with 
no further states near threshold. There is thus no
sign of an extra, bound tetraquark state in 
either channel.

The results for the $J^P=0^+$ and $1^+$ 
$\ell s\bar{c}\bar{b}$ channels are shown in
Fig.~\ref{fig:uscb}. Here we see the ground 
states in both channels pushed marginally below 
the corresponding lowest two-meson thresholds,
$DB_s$ and $DB_s^*$, respectively. With no other 
states in the vicinity, however, we suspect this
reflects either a residual systematic (from 
either the GEVP procedure or our fitting to the 
optimised ground-state correlators) or a small
($< 10$ MeV) FV shift, and that
the ground states in these channels correspond 
to the lowest two-meson thresholds. With only 
a single volume, the possibility of weakly 
bound ground states cannot, however, be
unambiguously ruled out. Interestingly, the 
three highest excited states in the $J^P=1^+$
channel all lie just below higher-lying 
two-meson thresholds, with splittings 
consistent with those between these 
thresholds. This ``pushing down'' of the
eigenvalues is of the order of a few MeV at 
most. This pattern is not, however, observed 
for the $J^P=0^+$ channel, where the third
excited state lies much higher than the third
excited two-meson threshold, $D^*B_s^*$. 

\begin{figure}[h!]
    \centering
    \subfloat[$ud\bar{s}\bar{b}$]
    {
    \includegraphics[scale=0.28]{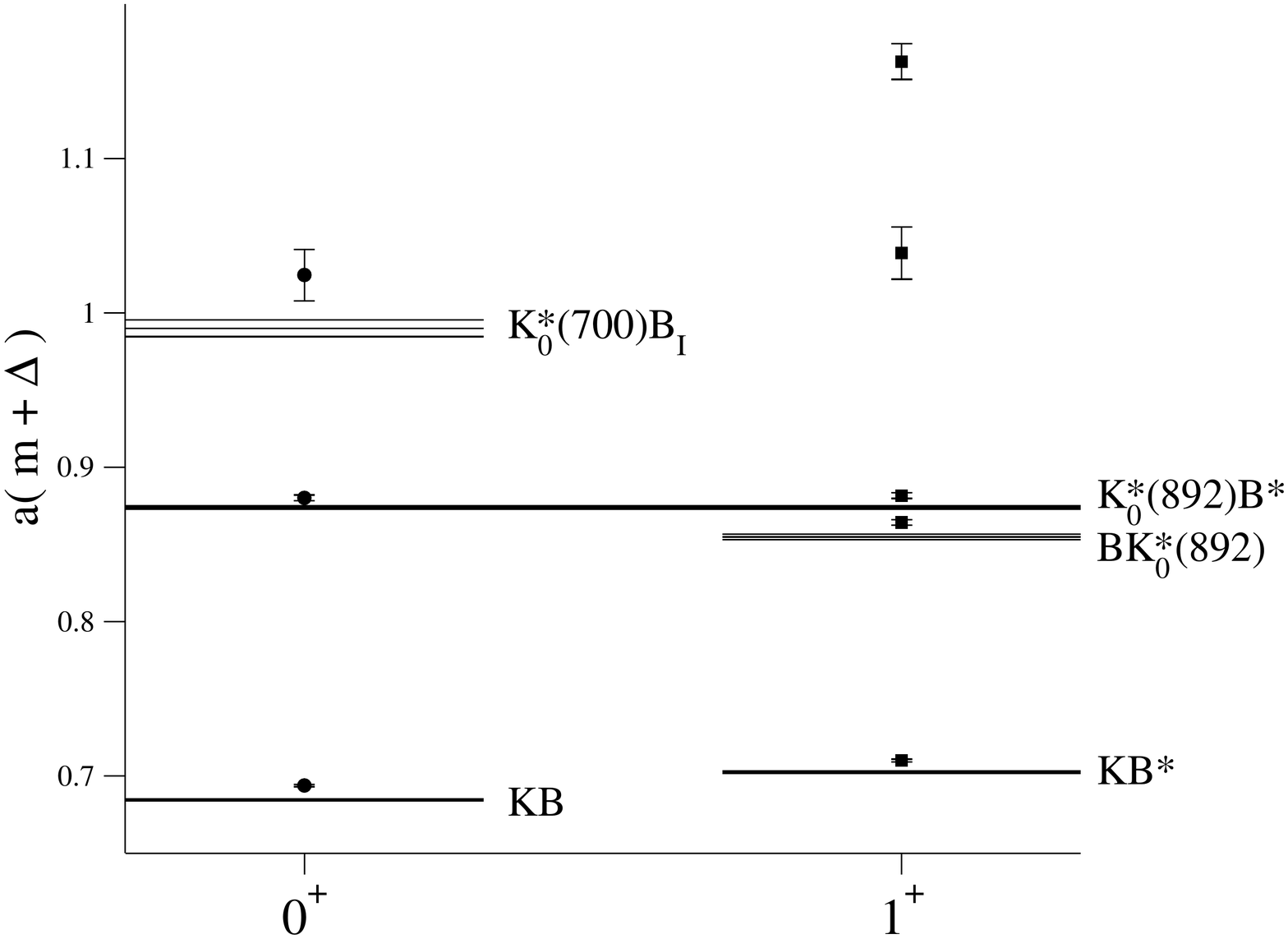}
    \label{fig:udsb}
    }
    \hspace{-32pt}
    \subfloat[$ud\bar{s}\bar{c}$]
    {
    \includegraphics[scale=0.28]{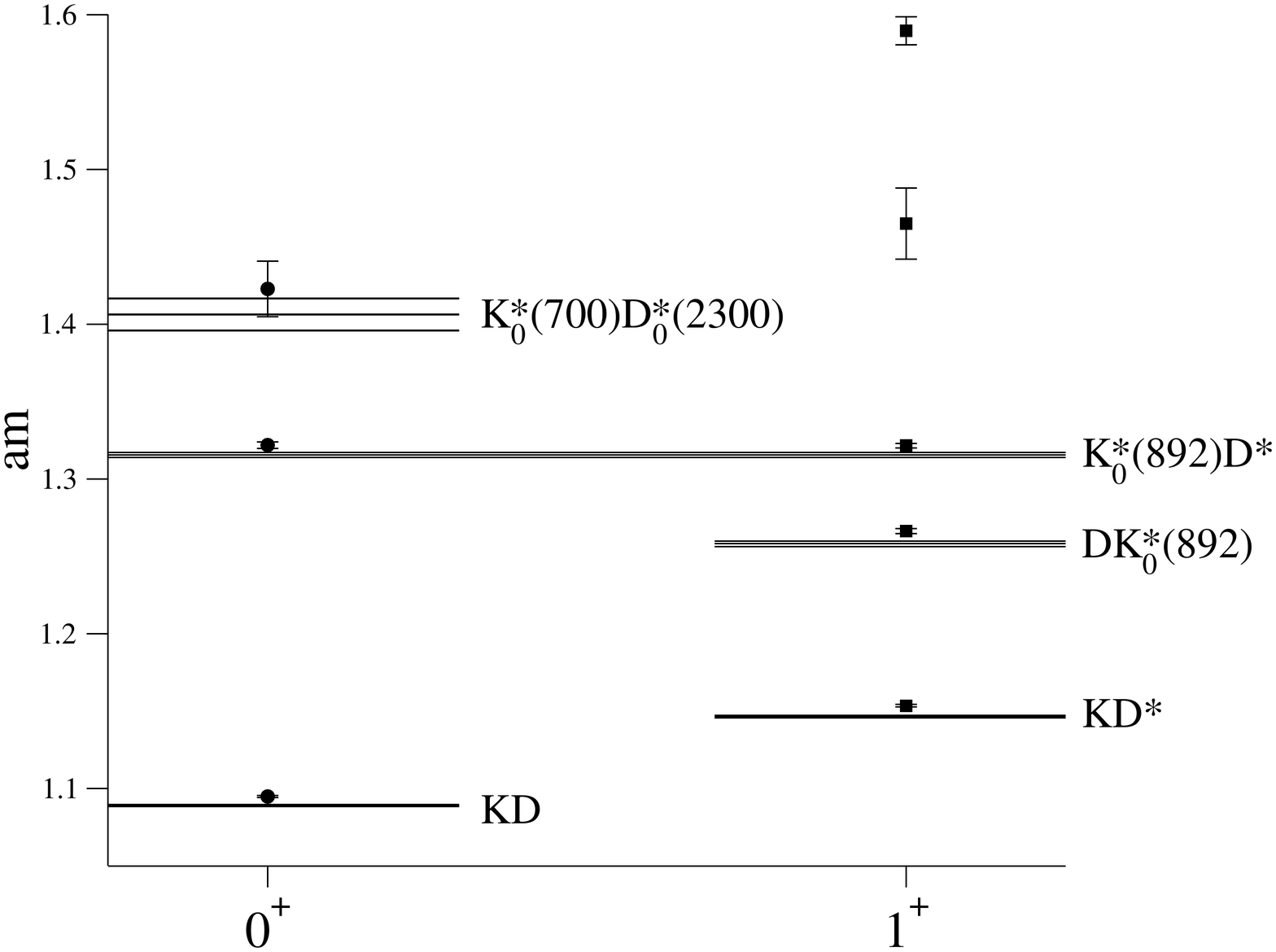}
    \label{fig:udsc}
    }
    \caption{Lowest-lying GEVP 
    eigenvalues and associated
    two-meson thresholds for the $I=0$, $J^P=0^+$ and
    $1^+$ $ud\bar{s}\bar{b}$ (left) and $ud\bar{s}\bar{c}$ (right) channels
    }
\end{figure}

Fig.~\ref{fig:udsb} shows our results for the
$I=0$, $J^P=0^+$ and $1^+$ $ud\bar{s}\bar{b}$ 
channels. The ground states again come out
consistent with the corresponding lowest
two-meson thresholds, with no sign of an extra 
state below, or even near, these thresholds,
suggesting the observed ground states once more 
correspond to the two-meson threshold states. Results for the analogous $I=0$, $J^P=0^+$ and
$1^+$ $ud\bar{s}\bar{c}$ channels are shown in
Fig.~\ref{fig:udsc}. The ground states lie
slightly above the lowest two-meson thresholds, 
with, once more, no sign of an additional, 
bound state below threshold in either channel.

\begin{figure}
    \subfloat[$uc\bar{b}\bar{b}$]
    {
    \includegraphics[scale=0.28]{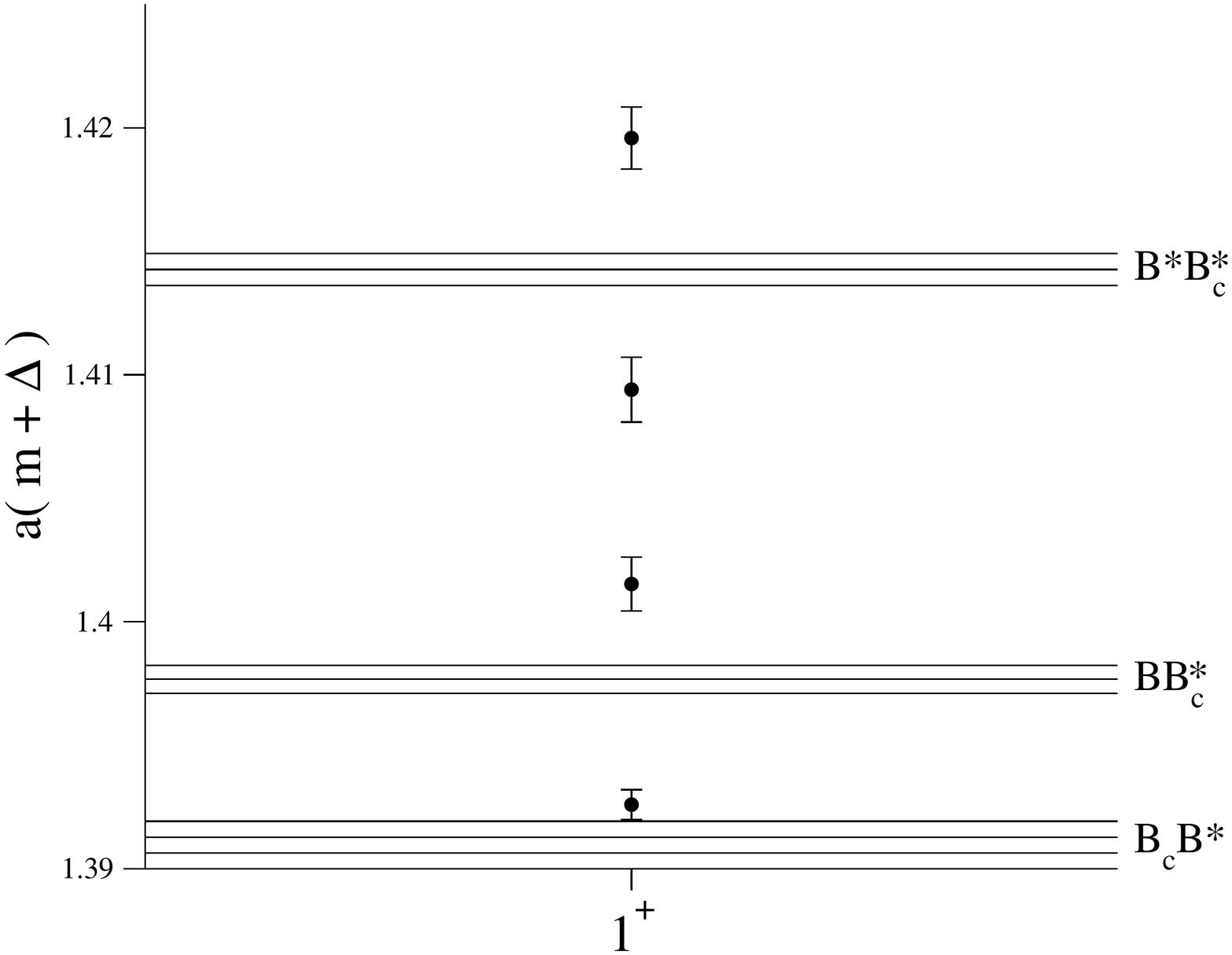}
    }
    \hspace{-32pt}
    \subfloat[$sc\bar{b}\bar{b}$]
    {
    \includegraphics[scale=0.28]{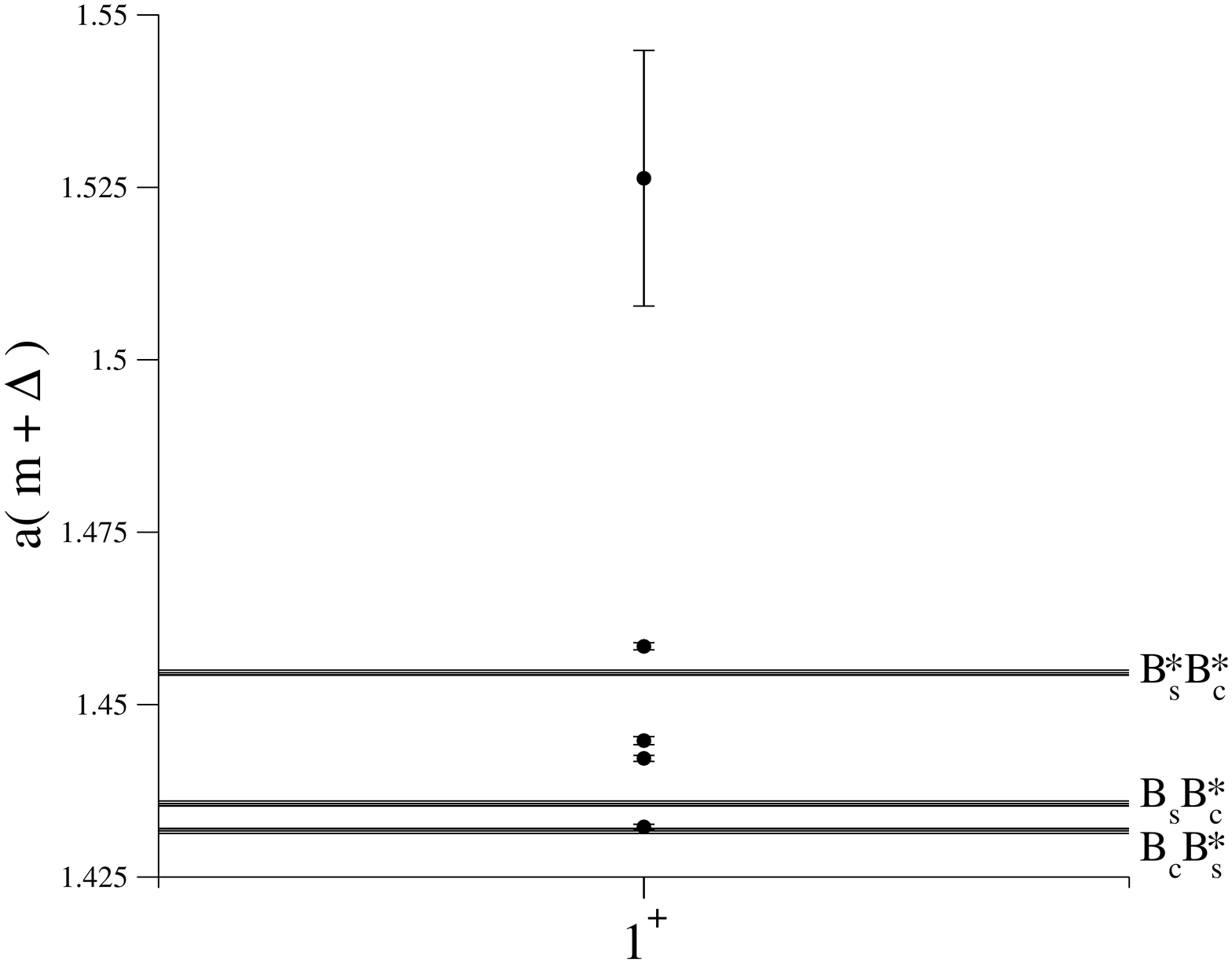}
    }
    \caption{Lowest-lying GEVP 
    eigenvalues and associated
    two-meson thresholds for the $J^P=1^+$ $uc\bar{b}\bar{b}$ and $sc\bar{b}\bar{b}$ channels.
    }
    \label{fig:scbb}
\end{figure}

Finally, in Fig.~\ref{fig:scbb} we show our 
results for the triply-heavy $J^P=1^+$,
$uc\bar{b}\bar{b}$ and $sc\bar{b}\bar{b}$ 
channels. The ground states are again consistent 
with the corresponding two-meson thresholds
($BB_c^*$ and $B_sB_c^*$, respectively),
confirming the earlier results for these channels
reported in Ref.~\cite{Junnarkar:2018twb}. 

\section{Discussion and Conclusions}\label{sec:conclusions}

In this work we have used lattice simulations
to investigate a number of exotic tetraquark 
channels in which model and/or QCD sum rule
studies and/or heavy-quark symmetry arguments
predict bound, strong-interaction-stable 
tetraquark states to exist. We have also 
revisited our earlier lattice studies of the 
doubly bottom, $J^P=1^+$, $\bar{3}_F$
$ud\bar{b}\bar{b}$ and $\ell s\bar{b}\bar{b}$ 
and bottom-charm,
$J^P=1^+$, $I=0$ $ud\bar{c}\bar{b}$ 
channels~\cite{Francis:2016hui,Francis:2018jyb},
introducing a number of significant improvements
over those earlier works. In particular, we 
have (i) generated a new ensemble with 
close-to-physical ($\simeq 192$ MeV) pion mass
and significantly larger volume 
($m_\pi L =4.2\, >\, 4$) to better control 
possible finite volume effects; (ii) added
additional operators to improve our ability to
resolve the tower of states contributing to each
correlator, and (iii) made major improvements
to our ground-state signals via the Box-Sink
construction. We have also, as described in more detail in
App.~\ref{app:NRQCD_tuning}, re-tuned our NRQCD bare mass to a much more precise and reliable value, using
partially-twisted boundary conditions. All of these  represent dramatic improvements over our 
previous work.

The only channels in which we find evidence
for deeply bound, strong-interaction-stable
tetraquark states are the doubly bottom
$J^P=1^+$, $\bar{3}_F$, $ud\bar{b}\bar{b}$
and $\ell s\bar{b}\bar{b}$ channels already
identified as supporting such bound states 
in earlier lattice studies, including our own.
Our new results for these channels, albeit
so far only at a single $m_\pi \simeq 192$ MeV,
display significantly
improved ground-state plateaus, largely due to 
the improved, Box-Sink construction. With the 
bindings at this $m_\pi$ reduced compared to 
those implied for the same $m_\pi$ by the fits 
to the $m_\pi$ dependences reported in
Ref.~\cite{Francis:2016hui}, we expect to 
find similarly reduced physical-$m_\pi$
results once we complete our ongoing
multiple-$m_\pi$ analysis of these channels
and are able to perform the physical-point 
extrapolation. With the binding in these channels
expected to increase with decreasing $m_\pi$, 
the final physical-point bindings will be deeper
than those of the current study. The current,
single-$m_\pi$ results, thus provide further
confirmation of the conclusions reached in
previous lattice studies regarding the existence
of a $\bar{3}_F$ of strong-interaction-stable, 
$J^P=1^+$, doubly bottom tetraquark states.
Once the multiple-$m_\pi$ analysis has been
completed, the improved ground state signals
produced by the Box-Sink construction should 
also significantly improve the reliability 
of the extrapolated physical-$m_\pi$ results.

In contrast to the doubly bottom, $\bar{3}_F$,
$J^P=1^+$ channels, no bound, non-molecular
ground states are found in any of the other ten
channels considered here. States with bindings 
$\gtrsim 10-20$ MeV, as predicted by many model
and QCD sum rule studies, should have been 
easily detectable in these analyses. 

Among the channels where current results rule out
the possibility of a bound, non-molecular ground
state is the $I=0$, $J^P=1^+$ $ud\bar{c}\bar{b}$
channel. This conclusion stands in contrast to
that reached in our earlier study of this
channel~\cite{Francis:2018jyb}, where indications
for possible modest tetraquark binding were
found, albeit based on rather short, late-time
plateau signals, and with worries about possible
FV effects for the ensemble with $m_\pi =164$
MeV, where $m_\pi L=2.4$. The desirability of
studying this channel at larger volumes, in
order to more strongly test the bound state
interpretation of the ground state results, was
stressed already in Ref.~\cite{Francis:2018jyb},
and the current study provides precisely such a 
larger-volume extension, one which, moreover,
has the advantage of the significantly improved
Box-Sink analysis methodology. The indications 
of binding seen in the earlier study do not
survive the improved, larger-volume, Box-Sink
analysis: the ground-state mass is found to 
lie, clearly slightly above, rather 
than below, $DB^*$ threshold. The very 
good associated ground-state effective-mass 
plateau is shown in the right-hand panel of 
Fig.~\ref{fig:effmass_udcb} in 
App.~\ref{app:effmassplots}. We conclude that 
there is, in fact, no non-molecular bound 
tetraquark state in this channel, and that the
late-time behavior of the earlier Wall-Local
results was likely affected by the
late-Wall-Local plateauing problem discussed 
in detail for the $B_c$ and doubly bottom 
$\bar{3}_F$, $J^P=1^+$ channels in Sec.~\ref{sec:walls_boxsink} above.

Turning to a comparison to the results of other
approaches, we note first that the predictions
of the heavy-quark-symmetry 
approach~\cite{Eichten:2017ffp,Braaten:2020nwp}
for the binding energies in the 
$I=0$, $J^P=1^+$ $ud\bar{b}\bar{b}$ and 
$J^P=1^+$ $\ell s\bar{b}\bar{b}$ channels, as
well as for an absence of binding in the $J^P=0^+$ and 
$1^+$, $I=0$ $ud\bar{c}\bar{b}$ and $J^P=0^+$ and 
$1^+$ $\ell s\bar{c}\bar{b}$ channels, are all 
compatible with current lattice results.

We turn next to tests of QCD sum rule and model
predictions. As stressed
by the authors of the various model tetraquark
studies, with the parameters of the models already
fully determined in earlier non-exotic sector
studies, the predictions for tetraquark binding 
are parameter free, and hence allow for highly
non-trivial tests of the underlying models.
In cases where multiple studies of the 
same model exist, we focus,
where possible, on those in which numerical
convergence of the prediction for the ground 
state energy has been demonstrated, taking
these to supercede earlier, less complete 
studies, whether by the same or different authors.

The earliest of the chiral quark model (ChQM)
studies, Ref.~\cite{Pepin:1996id}, employs a
model with pseudoscalar nonet
exchange potentials acting between light quarks
only and colour Coulomb and pairwise linear
confinement potentials acting between all quarks.
The resulting $I=0$, $J^P=1^+$ $ud\bar{b}\bar{b}$ 
binding, $497$ MeV, far exceeds that allowed 
by lattice results, thus ruling out this
implementation of the ChQM approach. Apart from
Ref.~\cite{Huang:2019otd}, all other
ChQM studies we are aware of
include octet (or nonet) pseudoscalar exchange 
potentials and scalar exchange potentials, 
both acting between light quarks only, pairwise 
or multi-body confinement potentials, acting 
between all quarks, and, in
addition to the colour Coulomb part of the 
effective OGE interaction, 
the associated colour-spin-dependent hyperfine 
interaction, again acting between all quarks.
Specific implementations differ, especially with
regard to the form of the effective 
scalar exchange(s). In Ref.~\cite{Zhang:2007mu} 
$u,\, d$ and $s$ quarks are assumed to interact 
via the exchange of a full nonet of scalar mesons.
Other implementations include only the exchange 
of the $\sigma$, either in ``SU(3)'' form, 
where the exchange is restricted to $u,\, d$ and 
$s$ quarks~\cite{Vijande:2003ki,Vijande:2006jf,Vijande:2009kj,Carames:2011zz,Chen:2018hts,Deng:2018kly,Yang:2019itm}, or ``SU(2)'' form, where it is
restricted further to $u$ and $d$ quarks only~\cite{Yang:2009zzp,Chen:2018hts,Tan:2020ldi}.
A final ChQM implementation, used to
study the $I=0$, $J^P=0^+$ and $1^+$ 
$ud\bar{s}\bar{b}$ channels~\cite{Huang:2019otd},
is the quark delocalization color screening 
model (QDCSM), which retains pseudoscalar exchange
between light quarks and OGE
between all quarks, but includes no scalar meson
exchange, instead employing a modified 
confinement form that allows two-center
cluster configurations with quark delocalization between the clusters. 
The scalar-nonet implementation is ruled out
by its $32$ MeV prediction for the
$I=0$, $J^P=1^+$ $ud\bar{b}\bar{b}$ binding,
which lies well below the range, $\sim 100-180$
MeV, allowed by current lattice results.
The SU(2) implementation is also ruled out, this 
time by significant {\it over-bindings} relative 
to lattice results: $318$~\cite{Tan:2020ldi} or 
$404$ MeV~\cite{Yang:2009zzp} for the
$I=0$, $J^P=1^+$ $ud\bar{b}\bar{b}$ channel
and $178$ and $199$ MeV~\cite{Tan:2020ldi} for the
$I=0$, $J^P=0^+$ and $1^+$ $ud\bar{c}\bar{b}$
channels (which our results show to be at most
weakly unbound). The SU(3) implementation
is similarly ruled out, with predicted
$I=0$, $J^P=1+$ $ud\bar{b}\bar{b}$ bindings 
of $341$~\cite{Vijande:2003ki}, 
$214$~\cite{Vijande:2009kj}, 
$217$~\cite{Carames:2011zz}, 
$278$~\cite{Deng:2018kly} or
$359$~\cite{Yang:2019itm} MeV, 
$I=0$, $J^P=0^+$ and $1^+$ $ud\bar{c}\bar{b}$ 
bindings of $136\pm 12$ and 
$171\pm 12$ MeV~\cite{Deng:2018kly} or
$178$ and $217$ MeV~\cite{Yang:2019itm},
and $I=0$, $J^P=0^+$ and $1^+$ $ud\bar{s}\bar{b}$
bindings of $70$ and $68$ MeV~\cite{Chen:2018hts}
(our lattice results again show the ground
states to be at most weakly bound in these 
channels). Finally, the QCDSM implementation 
predicts bindings of $74$ and $58$ MeV for the
$I=0$, $J^P=0^+$ and $1^+$ $ud\bar{s}\bar{b}$
channels, and hence is also ruled out by the
absence of such deep binding in our lattice 
results. We conclude that all implementations
of the ChQM approach are ruled out by current
lattice results, and hence that the ChQM framework
does not provide a reliable phenomenological
representation of QCD in the low-energy regime.

The results of non-chiral models fare much better 
when tested against lattice results. Multiple
non-relativistic quark model studies 
of tetraquark channels exist using the
AL1~\cite{Semay:1994ht} and Bhaduri (or BCN)~\cite{Bhaduri:1981pn} potentials. These
potentials are characterized by pairwise linear 
confinement, colour Coulomb and regularized OGE
hyperfine interactions. The AL1 model is a variation of
the BCN model in which the strict OGE relation 
between the strengths of the colour Coulomb 
and hyperfine interactions has been relaxed. 
Ref.~\cite{Semay:1994ht} also investigates three
additional models (the AL2, AP1 and AP2 models),
characterized by alterations to
the radial dependence of the AL1 confinement
potential and/or the regularization of the AL1 
OGE potentials. Predictions for binding in
the $I=0$, $J^P=1^+$ $ud\bar{b}\bar{b}$;
$J^P=1^+$ $\ell s\bar{b}\bar{b}$; $I=0$,
$J^P=0^+$ and $1^+$, $ud\bar{c}\bar{b}$; 
and $I=0$, $J^P=1^+$ $ud\bar{c}\bar{c}$ channels
exist for all these models, as well as for the
Godfrey-Isgur-Capstick (GIC) 
model~\cite{Godfrey:1985xj,Capstick:1986bm}, 
a ``relativized'' variant of this class of  model.
BCN results may be found in 
Refs.~\cite{Zouzou:1986qh,SilvestreBrac:1993ss,Semay:1994ht,Janc:2004qn,Vijande:2007rf,Vijande:2009kj,Yang:2009zzp}, AL1 results in 
Refs.~\cite{Semay:1994ht,Janc:2004qn,Caramees:2018oue,Hernandez:2019eox}, AL2,
AP1 and AP2 results in Ref.~\cite{Semay:1994ht}, and
GIC results in Ref.~\cite{Lu:2020rog}. 
Predictions for binding in the 
$I=0$, $J^P=1^+$ $ud\bar{b}\bar{b}$; 
$J^P=1^+$ $\ell s\bar{b}\bar{b}$; 
$I=0$, $J^P=0^+$ $ud\bar{c}\bar{b}$; 
and $I=0$, $J^P=1^+$ $ud\bar{c}\bar{b}$ channels,
obtained from the non-relativistic BCN, AL1,
AL2, AP1 and AP2 models, range from $131$ to $167$ MeV~\cite{SilvestreBrac:1993ss,Semay:1994ht,Janc:2004qn,Vijande:2007rf,Vijande:2009kj}, $40$ to $61$
MeV~\cite{Semay:1994ht}, $0$ to $23$
MeV~\cite{Semay:1994ht,Caramees:2018oue} and $0$
to $23$ MeV~\cite{Semay:1994ht,Caramees:2018oue},
respectively. Taking the $\sim 20-35$ MeV 
variations with model choice for these channels 
seen in Ref.~\cite{Semay:1994ht} as a measure of
the uncertainties associated with the specifics 
of the implementations of the interactions in 
this class of model, one sees that the 
predictions are compatible within errors
with current lattice results. Future lattice
studies, using larger-volume ensembles to
better quantify possible residual FV effects,
will be useful for sharpening these tests for
the $J^P=0^+$ and $1^+$, $I=0$ $ud\bar{c}\bar{b}$ 
channels, where the results of this paper
show no evidence for binding while the most
recent AL1 model study~\cite{Caramees:2018oue}
finds modest $23$ MeV bindings for both.
The relativized GIC model, in contrast, 
predicts only $54$ MeV binding in the
$I=0$, $J^P=1^+$ $ud\bar{b}\bar{b}$ channel
and no binding in the $J^P=1^+$ $\ell s\bar{b}\bar{b}$
channel~\cite{Lu:2020rog}, and hence is ruled out by 
lattice results.

Another model approach ruled out by lattice 
results, specifically those for the $I=0$
$ud\bar{c}\bar{b}$ channels, is that of the 
LAMP model. In this approach, an ansatz
for a relativistic confinement
potential is employed in the multiquark
sector. Taking the $BD$ interaction studied
in Ref.~\cite{Silbar:2013dda} to be specific,
linearly rising mean-field confinement 
potentials for the $B$ and $D$ are merged
into a one-body, two-well, W-shaped confinement 
potential for the $BD$ system by truncating  
the individual $B$ and $D$ wells at the
midpoint between their centers. Such a 
potential allows for the delocalization of 
an initially single-well light-quark 
wavefunction into the second of the two 
potential wells. The optimal well-center 
separation and degree of delocalization
are determined by a variational minimization. 
The result, in which the effect of the OGE
hyperfine interaction has been neglected, is 
a $BD$ binding of $155$ MeV, clearly 
incompatible with the lattice results above.
This rules out the LAMP ansatz for the effective 
confining interaction.

Turning to tests of QCD sum rule results, 
we see that recent predictions for very 
deep tetraquark binding (by $569\pm 260$
MeV~\cite{Agaev:2018khe} in the $I=0$, $J^P=1^+$
$ud\bar{b}\bar{b}$ channel, $477\pm 250$ 
MeV~\cite{Agaev:2020zag} in the $J^P=1^+$ 
$\ell s\bar{b}\bar{b}$ channel, $485\pm 150$ 
MeV~\cite{Agaev:2018khe} in the $I=0$, $J^P=0^+$
$ud\bar{c}\bar{b}$ channel, $407\pm 160$ 
MeV~\cite{Agaev:2019lwh}, $467\pm 150$ 
MeV~\cite{Agaev:2020zag} and $200\pm 130$ 
MeV~\cite{Wang:2020jgb} in the $J^P=0^+$ 
$\ell s\bar{c}\bar{b}$ channel,
$180\pm 110$ MeV~\cite{Wang:2020jgb} in the 
$J^P=1^+$ $\ell s\bar{c}\bar{b}$ channel,
and $394\pm 170$ MeV~\cite{Sundu:2019feu} in the
$I=0$, $J^P=0^+$ $ud\bar{s}\bar{b}$ channel)
are not compatible, within quoted uncertainties,
with current lattice results. Other sum
sum rule predictions ($400\pm 300$ 
MeV~\cite{Navarra:2007yw,Du:2012wp}, 
$80\pm 80$ MeV~\cite{Wang:2017uld} 
and $240\pm 150$ MeV~\cite{Tang:2019nwv} in the
$I=0$, $J^P=1^+$ $ud\bar{b}\bar{b}$ channel,
$290\pm 300$ MeV~\cite{Du:2012wp}, 
$140\pm 80$ MeV~\cite{Wang:2017uld} and
$180\pm 160$ MeV~\cite{Tang:2019nwv} in
the $J^P=1^+$ $\ell s\bar{b}\bar{b}$ channel,
$5\pm 100$ and $60\pm 100$ MeV ~\cite{Chen:2013aba}
in the $I=0$, $J^P=0^+$ and $J^P=1^+$ 
$ud\bar{c}\bar{b}$ channels), are compatible with
lattice results within errors, though, with the 
exception of the $I=0$, $J^P=1^+$
$ud\bar{b}\bar{b}$ prediction of 
Ref.~\cite{Wang:2017uld}, with significantly 
higher central values. It is not clear whether 
the above disagreements with lattice results 
are due to shortcomings associated with 
the use of the SVZ ansatz, inaccuracies in the
factorization approximation for some of the
higher dimension condensates, a need for updates 
to OPE input (e.g., the quark condensate, where 
precise results from FLAG are now  
available~\cite{Aoki:2019cca}), or other aspects 
of the sum rule implementations. 
On this last point it is worth noting the 
discrepancy between the predictions, $80\pm 80$ 
MeV~\cite{Wang:2017uld} and $569\pm 260$ 
MeV~\cite{Agaev:2018khe}, for binding in the
$I=0$, $J^P=1^+$ $ud\bar{b}\bar{b}$ channel
obtained from analyses employing the same 
interpolating operator, essentially identical 
OPE input, and both truncating the OPE at 
the same dimension, $D=10$. The predictions,
$142\pm 80$ MeV~\cite{Wang:2017uld} and 
$477\pm 250$ MeV~\cite{Agaev:2020zag}, for
binding in the $J^P=1^+$ $\ell s\bar{b}\bar{b}$ 
channel also come from analyses using the same
interpolating operator, essentially identical
OPE input and truncation of the OPE at the same
dimension, $D=10$. A similar discrepancy exists in
the $I=0$, $J^P=0^+$ $ud\bar{c}\bar{b}$ channel
where very different ground state masses, 
$7140\pm 100$ MeV~\cite{Chen:2013aba}
and $6660\pm 150$ MeV~\cite{Agaev:2018khe},
are predicted by analyses using the same
interpolating operator and similar OPE input, 
differing only in the dimension at which the OPE 
was truncated ($D=8$ in Ref.~\cite{Chen:2013aba}
and $D=10$ in Ref.~\cite{Agaev:2018khe}). These
observation make clear that significant, 
yet-to-be-fully-quantified systematic 
uncertainties exist for applications of 
the conventional implementation of the QCD sum 
rule framework to the multiquark sector. The
targets provided by current (and future) lattice
results for tetraquark binding (or lack thereof) 
in the channels investigated in this paper 
should help in further investigations of these
issues, and have the potential to aid in 
the development of improved sum rule 
implementation strategies.

We close with a reminder of improvements/extensions
to the present study either already in progress 
or planned for the near future. The first is the 
completion of the extension of the 
$m_\pi\simeq 192$ MeV Box-Sink analysis reported 
here to additional $48^3\times 64$ ensembles 
having both lighter and heavier $m_\pi$. This 
will allow us to update and sharpen our previous
physical-point extrapolations for the bindings 
of the doubly bottom, $J^P=1^+$, $\bar{3}_F$ 
states, and further test non-chiral model
predictions. This work is ongoing. The next 
step after that is to generate a new set of ensembles with even larger volume covering a
similar range of $m_\pi$. This will allow us 
to carry out quantitative FV studies, and 
extend the current analysis to channels, such 
as the doubly charmed channels, where shallow 
binding remains a possibility and experimental 
detection would be less challenging than in 
the doubly bottom sector. 

\section*{Acknowledgements}
BC, RL and KM are supported by grants from
the Natural Sciences and Engineering Council
of Canada, RJH by the European Research Council (ERC) under the European Union’s Horizon 2020 research and innovation programme through grant 
agreement 771971-SIMDAMA.
All computations were carried out using a
crucial allocation from Compute Canada on the
Niagara supercomputer at Scinet.

\begin{appendix}
\section{Tuning NRQCD}\label{app:NRQCD_tuning}

We use the tree-level ($c_i=1$), tadpole-improved, NRQCD action of \cite{Lepage:1992tx}, defined by the symmetric time evolution \cite{Lewis:2008fu},
\begin{equation}
\begin{aligned}
G(x,t+1) &= \left( 1 - \frac{\delta H}{2} \right) \left( 1 - \frac{H_0}{2n}\right)^n U_t(x,t_0)^\dagger \left( 1 - \frac{H_0}{2n} \right)^n \left( 1 - \frac{\delta H}{2} \right)  G(t),
\end{aligned}
\end{equation}
where 
\begin{equation}
\begin{aligned}
H_0 = &-c_0\frac{1}{2 M_0} \Delta^2,\\
H_I = &\left(-c_1\frac{1}{8 M_0^2}-c_6 \frac{1}{16 n M_0^2}\right)\left(\Delta^2\right)^2 +c_2 \frac{i}{8 M_0^2}( \tilde\Delta\cdot \tilde{E} - \tilde{E} \cdot \tilde\Delta) + c_5 \frac{\Delta^4}{24 M_0}\\
H_D = &-c_3 \frac{1}{8 M_0^2}\sigma\cdot\left( \tilde\Delta\times \tilde{E} - \tilde{E} \times \tilde\Delta \right) -c_4 \frac{1}{8 M_0} \sigma\cdot \tilde{B}\\
 \delta H = &H_I + H_D.
\end{aligned}
\end{equation}
A tilde indicates that an improved version has been used (see e.g. \cite{Groote:2000jd} for the tadpole-improved derivative terms) and $M_0$ is a bare parameter that we tune to recover the spin-averaged kinetic mass of the connected $\eta_b$ and $\Upsilon$. We use the fourth root of the plaquette for the tadpole improvement and a value of $n=4$ for the stability parameter.

\begin{figure}[ht]
\centering
\subfloat[Effective masses]
{
	\includegraphics[scale=0.28]{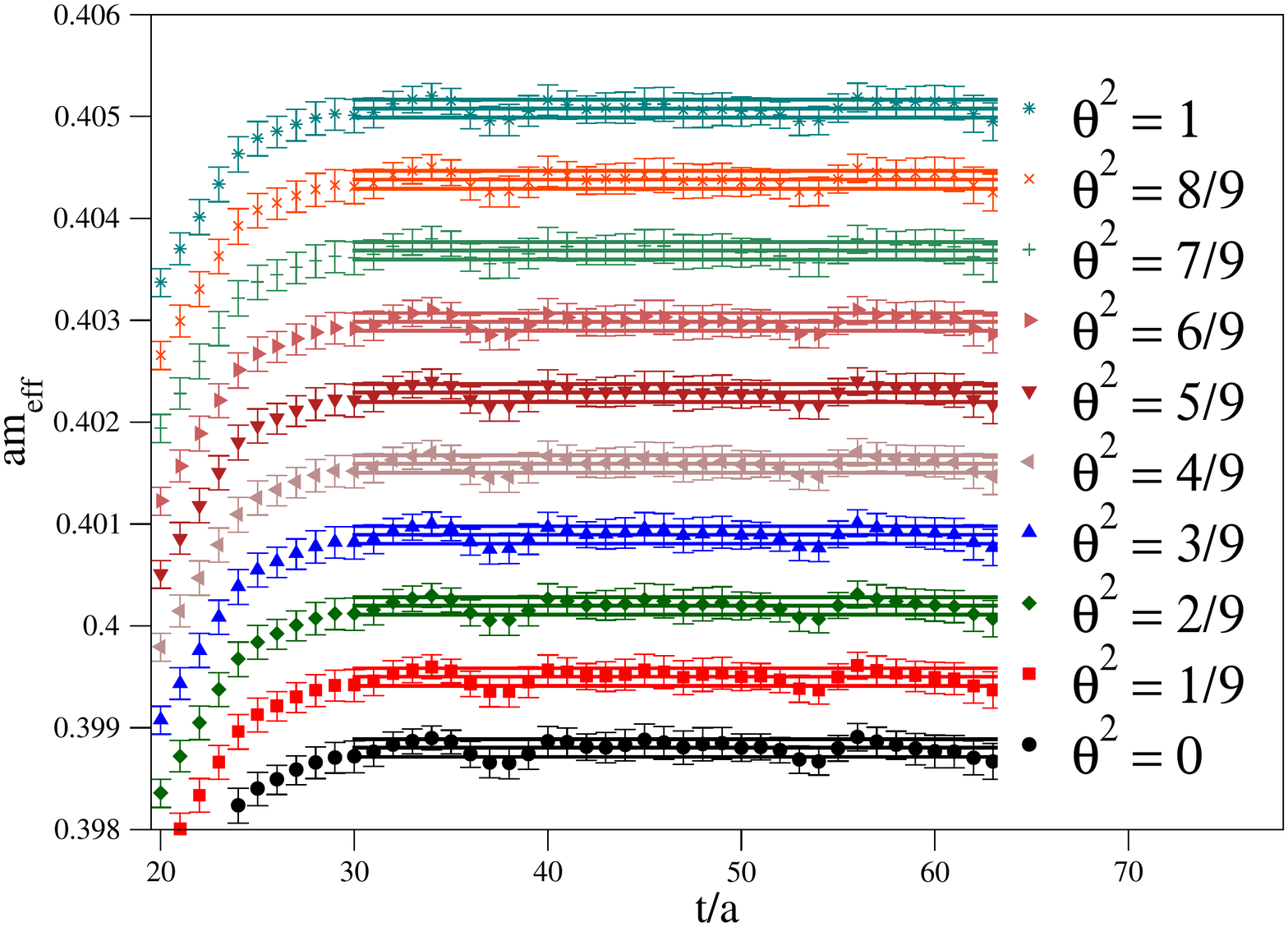}
}
\subfloat[]
{
\includegraphics[scale=0.28]{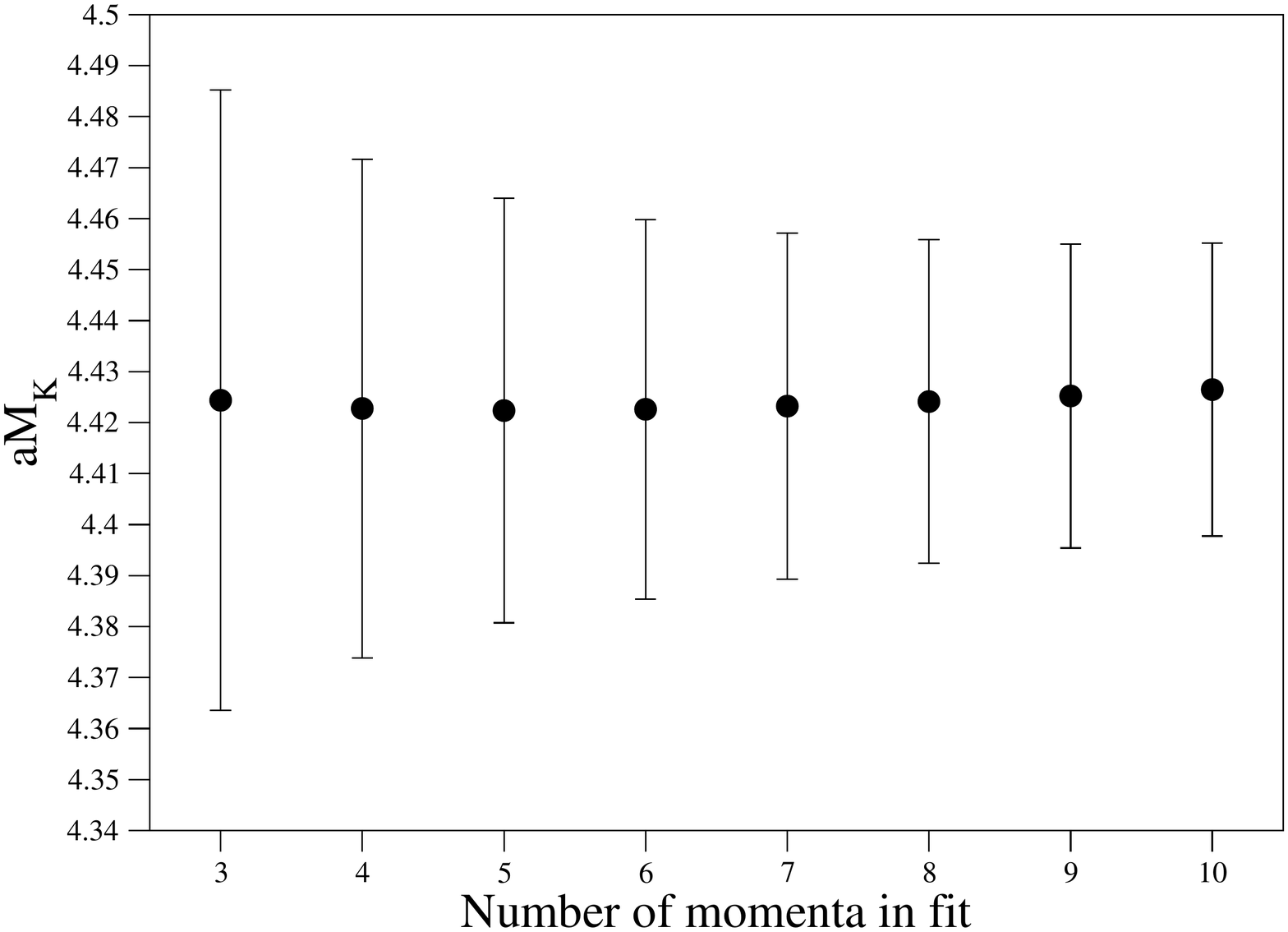}
}
\caption{(Left) Effective masses for each twist angle, and corresponding fit for $aM_0=1.88$. (Right) Stability and error reduction as the number of momenta is increased.}\label{fig:twistfits}
\end{figure}

For the tuning we use Coulomb gauge-fixed wall-sources and we implement the momenta by solving one of the b propagators on a partially-twisted boundary \cite{Bedaque:2004kc,Sachrajda:2004mi}. This is practically implemented by applying a $U(1)$ phase to the gauge field \cite{Nguyen:2011ek},
\begin{equation}
U_\mu(x) \rightarrow e^{i2\pi \theta_\mu/L_\mu }U_\mu(x)
\end{equation}
which yields $p_\mu=\frac{2\pi}{L_\mu}\theta_\mu$.
The benefit of this approach is that one can stay within the same irrep for arbitrary momenta, smoothly interpolating discretisation and finite-volume effects. We choose to twist equally in all directions to minimise the rotation-breaking hypercubic artifacts and the effects introduced by the twisting. The momenta we use is listed in Table~\ref{tab:moms} and the effective masses for these twists is shown in Fig.~\ref{fig:twistfits}.

\begin{table}[ht]
\begin{tabular}{c|c|c|c}
\toprule
$(\theta,\theta,\theta)$ & $(ap)^2$ & $(\theta,\theta,\theta)$ & $(ap)^2$ \\
\hline
$(0,0,0)$ & $0$ & $(\frac{\sqrt{5}}{3},\frac{\sqrt{5}}{3},\frac{\sqrt{5}}{3})$ & $0.064255$\\
$(\frac{1}{3},\frac{1}{3},\frac{1}{3})$ & $0.012851$ & $(\frac{\sqrt{6}}{3},\frac{\sqrt{6}}{3},\frac{\sqrt{6}}{3})$ & $0.077106$ \\
$(\frac{\sqrt{2}}{3},\frac{\sqrt{2}}{3},\frac{\sqrt{2}}{3})$ & $0.025702$ & $(\frac{\sqrt{7}}{3},\frac{\sqrt{7}}{3},\frac{\sqrt{7}}{3})$ & $0.089957$ \\
$(\frac{\sqrt{3}}{3},\frac{\sqrt{3}}{3},\frac{\sqrt{3}}{3})$ & $0.038553$ & $(\frac{\sqrt{8}}{3},\frac{\sqrt{8}}{3},\frac{\sqrt{8}}{3})$ & $0.102808$ \\
$(\frac{2}{3},\frac{2}{3},\frac{2}{3})$ & $0.051404$ & $(1,1,1)$ & $0.115659$\\
\botrule
\end{tabular}
\caption{Twist angles, chosen so the $(ap)^2$ values are evenly spaced between $(0,0,0)$ and $(1,1,1)$.}\label{tab:moms}
\end{table}

Once we have computed the correlation functions along all our directions we simultaneously fit them to the Ansatz,
\begin{equation}
A(1+(ap)^2 B+C((ap)^2)^2)\exp\left[-t\;\left( (aM_1) + \frac{1}{2aM_K}(ap)^2\right) \right],
\end{equation}
where $aM_1$ is the mass of the ground state and $aM_K$ is the kinetic mass. This enables us to describe our data simply by five fit parameters.

\begin{figure}
\centering
\includegraphics[scale=0.4]{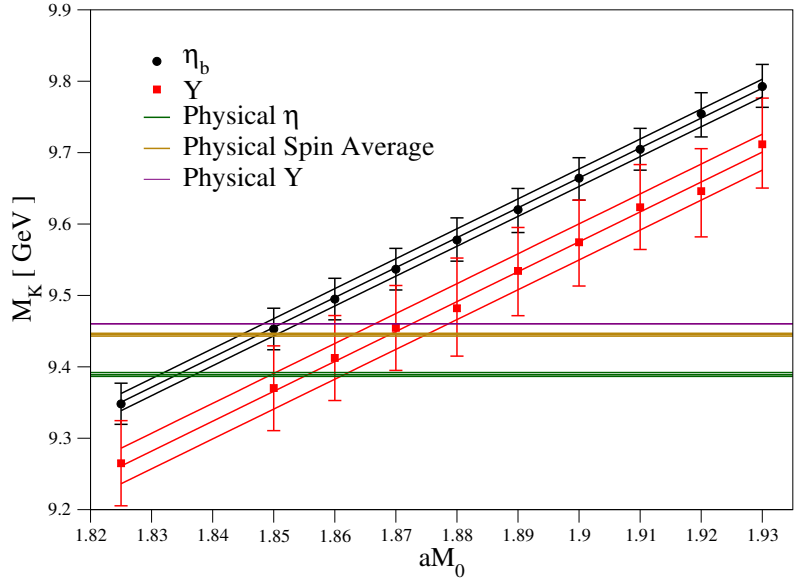}
\caption{Kinetic mass dependence on the bare mass $aM_0$. The fit is correlated and constrained to have the same slope for the $\eta_b$, $\Upsilon$ and spin average.}\label{fig:M0_determination}
\end{figure}

We note that, as in Ref.~\cite{Dowdall:2011wh},
the kinetic mass ordering is inverted compared to 
nature. This situation can be 
improved by the inclusion of yet-higher-order 
($1/M^3$) terms in the NRQCD 
Hamiltonian~\cite{Manohar:1997qy}, specifically
the parameter $c_7$. However, we follow the 
approach of \cite{Dowdall:2011wh} and tune to 
the spin average of these kinetic masses instead.
Using HPQCD's estimate~\cite{Dowdall:2011wh} for the spin-averaged mass, $9.445(2)$ GeV, and the 
value $a^{-1}=2.194$ GeV~\cite{Namekawa:2013vu} for our lattice spacing, we find the physical 
b quark corresponds to a bare mass $aM_0=1.864$. This tuning is obtained by measuring the kinetic masses for several bare masses and fitting their dependence linearly, as is shown in Fig.~\ref{fig:M0_determination}.

When we compute our heavy-light mesons we pack the backward and forward propagating NRQCD propagators into one as the NRQCD propagator has only 2 Dirac indices whereas the full light, strange, and charm propagators have four,
\begin{equation}
S(x) = 
\begin{pmatrix}
S^{\text{fwd}}(x) & 0 \\
0 & S^{\text{bwd}}(x) \\
\end{pmatrix}
\end{equation}
We also apply anti-periodic boundary conditions in time to the NRQCD propagators.

\section{Meson masses}\label{app:meson_amps}

As we use Coulomb gauge-fixed wall sources our amplitudes are partially related to physical matrix elements, but are polluted by the wall source \cite{Daniel:1992ek}. If we consider for example the pseudoscalar $J^P=0^{-}$ meson channel, we can use any of the following 8 correlation functions at large asymptotic times,
\begin{equation}\label{eq:simfit_mes}
\begin{aligned}
C^{WL}_{PP}(t) = \frac{P^WP^L}{2m_P}\left(e^{-m_Pt}+e^{-m_P(L_t-t)}\right),\\
C^{WL}_{AA}(t) = \frac{A^WA^L}{2m_P}\left(e^{-m_Pt}+e^{-m_P(L_t-t)}\right),\\
C^{WL}_{AP}(t) = \frac{A^WP^L}{2m_P}\left(e^{-m_Pt}-e^{-m_P(L_t-t)}\right),\\
C^{WL}_{PA}(t) = \frac{P^WA^L}{2m_P}\left(e^{-m_Pt}-e^{-m_P(L_t-t)}\right),\\
C^{WW}_{PP}(t) = \frac{P^WP^W}{2m_P}\left(e^{-m_Pt}+e^{-m_P(L_t-t)}\right),\\
C^{WW}_{AA}(t) = \frac{A^WA^W}{2m_P}\left(e^{-m_Pt}+e^{-m_P(L_t-t)}\right),\\
C^{WW}_{AP}(t) = \frac{A^WP^W}{2m_P}\left(e^{-m_Pt}-e^{-m_P(L_t-t)}\right),\\
C^{WW}_{PA}(t) = \frac{P^WA^W}{2m_P}\left(e^{-m_Pt}-e^{-m_P(L_t-t)}\right).
\end{aligned}
\end{equation}
Here we use $P$ to denote the operator with a $\gamma_5$-insertion and $A$ to denote the temporal axial current operator. Using these sources means,
for example, that $C^{WL}_{PA}$ is not equal to $C^{WL}_{AP}$. One can, however, clearly fit all 8 
correlators simultaneously with 5 fit parameters. One can similarly analyse correlators involving the 
$V_i=\gamma_i$ and $T_{it}=\gamma_i\gamma_t$ currents for $J^P=1^{-}$ states or $I$ and 
$\gamma_t$ for $J^P=0^+$ states. The same fit form
can be used when the sink is non-local too, although neither matrix element will correspond to a physical one. We 
use these fits to determine the masses of mesons 
not containing a b quark. This fit is not possible for 
mesons containing b quarks as $C_{AA}=C_{PP}$ at 
our level of heavy-light approximation.

\section{Tables of results}\label{app:results_tabs}

In this work we use simple, local operators of the form
\begin{equation}
M = (\bar{\psi}\Gamma \phi)
\end{equation}
to determine the masses of ordinary meson states.
The results, in lattice units, and with the 
additive offset in the case of mesons containing 
a b quark, are given in Tab.~\ref{tab:mestab}. Where there are several $\Gamma$ insertions listed, 
a simultaneous fit of the form of 
Eq.~\ref{eq:simfit_mes} was used, otherwise a correlated single-cosh fit was performed 
on the one channel.
As some of the particles listed in the table are resonances, the ground-state masses obtained from
the associated local-operator correlators should not be directly compared to the masses of these 
resonances in nature. To simplify notation, we have, nonetheless, used the name of the 
lowest-lying resonance in the channel to label the channel itself in these cases, trusting this will 
not lead to any confusion. Throughout this work we use the shorthand name $B_I$ for the scalar, $J^P=0^+$ B meson.

\begin{table}[ht]
    \centering
    \begin{tabular}{cc|c|cc|c}
    \toprule
    State & $\Gamma$ & $am$ & State & $\Gamma$ & $a(m+\Delta)$ \\
    \hline
    $\pi$ & $\gamma_5,\gamma_t\gamma_5$ & 0.0880(8) & $B$ & $\gamma_5$ & 0.4546(6) \\
    $K$ & $\gamma_5,\gamma_t\gamma_5$ & 0.2313(2) & $B^*$& $\gamma_i$ & 0.4712(6) \\
    $K_0^*(700)$ & $I,\gamma_t$ & 0.402(3) & $B_I$ & $I$ & 0.590(5) \\
    $K^*(892)$ & $\gamma_i,\gamma_i\gamma_t$ & 0.401(9) & $B_s$& $\gamma_5$ & 0.4926(3) \\
    $D$& $\gamma_5,\gamma_t\gamma_5$ & 0.8585(6) & $B_s^*$& $\gamma_i$ & 0.5116(3) \\
    $D^*$ & $\gamma_i,\gamma_i\gamma_t$ & 0.9155(6) & $B_c$& $\gamma_5$ & 0.92011(8) \\
    $D^*_0(2300)$ & $I,\gamma_t$ & 1.016(5) & $B_c^*$ & $\gamma_i$ & 0.9429(1)\\
    \botrule
    \end{tabular}
    \caption{Table of results for the ground-state
    effective masses associated with the simple, local meson operators used in this work.}
    \label{tab:mestab}
\end{table}

Tables~\ref{tab:udcb_evtab} through
\ref{tab:scbb_evtab} present the results
for the lowest resolvable GEVP eigenvalues 
in the ten tetraquark channels specified in the main text, obtained 
from correlated, single-exponential fits
to the plateau regions of the corresponding 
optimised correlators. Entries in column 1
specify the quantum numbers of the state. Those in
columns 2, 3 and 4 give the corresponding 
eigenvalue (in lattice units), the fit window
interval (in lattice units) and the associated  
$\chi^2/dof$. Fits for higher-lying
excited states having reasonable $\chi^2/dof$ 
have been included even in cases where the fit
window is relatively short since such results 
help identify those parts of the higher-lying
two-meson spectrum having the largest couplings
to the operators used in constructing the
optimised correlators.

\begin{table}[ht]
    \centering
    \begin{tabular}{c|ccc}
    \toprule
    $J^P$ & $a(m+\Delta)$ & fit (low,high) & $\chi^2/dof$ \\
    \hline
    \multirow{3}{*}{$0^+$} 
    & 1.3208(10) & (7,18) & 1.2 \\
    & 1.3878(11) & (5,18) & 1.1 \\
    & 1.606(20) & (6,12) & 0.9 \\
    \hline
    \multirow{5}{*}{$1^+$} 
    & 1.3356(10) & (7,17) & 1.2 \\
    & 1.3731(13) & (7,17) & 0.8 \\
    & 1.3864(12) & (6,18) & 0.9 \\
    & 1.632(12) & (5,9) & 1.1 \\
    & 1.680(15) & (5,10) & 0.6 \\
    \botrule
    \end{tabular}
    \caption{Eigenvalues and fit parameters for the $I=0$, $J^P=0^+$ and $1^+$ $ud\bar{c}\bar{b}$ channels}
    \label{tab:udcb_evtab}
\end{table}

\begin{table}[h]
    \centering
    \begin{tabular}{c|ccc}
    \toprule
    $J^P$ & $a(m+\Delta)$ & fit (low,high) & $\chi^2/dof$ \\
    \hline
    \multirow{4}{*}{$0^+$}
    & 1.3471(8) & (8,27) & 1.2 \\
    & 1.3573(9) & (10,19) & 1.0 \\
    & 1.417(1) & (8,18) & 0.8 \\
    & 1.445(2) & (10,15) & 1.5 \\
    \hline
    \multirow{6}{*}{$1^+$}
    & 1.3656(8) & (7,20) & 1.2 \\
    & 1.3746(8) & (8,18) & 1.1 \\
    & 1.406(1) & (8,18) & 1.2 \\
    & 1.414(1) & (7,20) & 1.2 \\
    & 1.423(1) & (8,18) & 1.0 \\
    & 1.59(1) & (8,18) & 1.0 \\
    \botrule
    \end{tabular}
    \caption{Eigenvalues and fit parameters for the $J^P=0^+$ and $1^+$ 
    $\ell s\bar{c}\bar{b}$ channels}
    \label{tab:uscb_evtab}
\end{table}

\newpage

\begin{table}[ht]
    \centering
    \begin{tabular}{c|ccc}
    \toprule
    $J^P$ & $a(m+\Delta)$ & fit (low,high) & $\chi^2/dof$ \\
    \hline
    \multirow{3}{*}{$0^+$}
     & 0.6939(7) & (7,25) & 1.1 \\
     & 0.8802(19) & (7,16) & 1.1 \\
     & 1.023(17) & (6,11) & 0.6 \\
    \hline
    \multirow{5}{*}{$1^+$}
    & 0.7100(9) & (7,20) & 1.0 \\
    & 0.8639(19) & (7,16) & 1.0 \\
    & 0.8695(52) & (10,16) & 1.0 \\
    & 1.038(17) & (6,12) & 0.5 \\
    & 1.163(12) & (5,12) & 1 \\
    \botrule
    \end{tabular}
    \caption{Eigenvalues and fit parameters for the $I=0$, $J^P=0^+$ and $1^+$ $ud\bar{s}\bar{b}$ channels}
    \label{tab:udsb_evtab}
\end{table}

\begin{table}[h]
    \centering
    \begin{tabular}{c|ccc}
    \toprule
    $J^P$ & $a(m+\Delta)$ & fit (low,high) & $\chi^2/dof$ \\
    \hline
    \multirow{3}{*}{$0^+$}
    & 1.0949(6) & (7,17) & 1.3 \\
    & 1.321(2) & (4,13) & 0.7 \\
    & 1.423(2) & (3,9) & 0.8 \\
    \hline
    \multirow{5}{*}{$1^+$}
    & 1.1536(7) & (8,19) & 1.0 \\
    & 1.266(2) & (8,22) & 1.0 \\
    & 1.321(2) & (6,18) & 1.0 \\
    & 1.465(2) & (6,14) & 0.7 \\
    & 1.590(1) & (4,13) & 0.5 \\
    \botrule
    \end{tabular}
    \caption{Eigenvalues and fit parameters for the $I=0$, $J^P=0^+$ and $1^+$ $ud\bar{s}\bar{c}$ channels}
    \label{tab:udsc_evtab}
\end{table}

\begin{table}[h]
    \centering
    \begin{tabular}{c|ccc}
     \toprule
    $J^P$ & $a(m+\Delta)$ & fit (low,high) & $\chi^2/dof$ \\
    \hline
     \multirow{4}{*}{$1^+$}
    & 1.3926(6) & (6,25) & 0.7 \\
    & 1.4015(11) & (10,25) & 1.2 \\
    & 1.4094(13) & (10,17) & 1.3 \\
    & 1.4196(13) & (10,17) & 1.3 \\
    \botrule
    \end{tabular}
    \caption{Eigenvalues and fit parameters for the $I=0$, $J^P=1^+$ $uc\bar{b}\bar{b}$ channel}
    \label{tab:ucbb_evtab}
\end{table}

\begin{table}[ht]
    \centering
    \begin{tabular}{c|ccc}
    \toprule
    $J^P$ & $a(m+\Delta)$ & fit (low,high) & $\chi^2/dof$ \\
    \hline
     \multirow{5}{*}{$1^+$}
    & 1.4323(4) & (8,22) & 1.2 \\
    & 1.4422(5) & (13,22) & 0.8 \\  
    & 1.4448(6) & (16,25) & 0.6 \\
    & 1.4585(5) & (13,22) & 0.8\\
    & 1.5263(184) & (14,22) & 0.6 \\
    \botrule
    \end{tabular}
    \caption{Eigenvalues and fit parameters for the $I=0$, $J^P=1^+$ $sc\bar{b}\bar{b}$ channel}
    \label{tab:scbb_evtab}
\end{table}

\newpage

\section{Effective mass plots}\label{app:effmassplots}


If we consider our correlators $C(t)$ to behave 
like single exponentials at large $t$ (a reasonable 
expectation for heavy states) we can define an 
effective mass via
\begin{equation}
am_\text{eff}(t) = \tanh^{-1}\left( \frac{C(t-1)-C(t+1)}{C(t-1)+C(t+1)}\right).
\end{equation}
This has the desirable property that 
$\tanh^{-1}(x)$ is defined for all $x$ in 
$[-1,1]$, and provides a definition that is, empirically, better behaved for very noisy data 
than the more common $\log$ or $\cosh^{-1}$ definitions, $\log\left(\frac{C(t)}{C(t+1)}\right)$ or $\cosh^{-1}{\left(\frac{C(t+1)+C(t-1)}{2C(t)}\right)}$,
where fluctuations can lead to negative values of 
the ratios appearing in the arguments of the 
$\log$ or $\cosh^{-1}$, causing these definitions 
to break down.

\begin{figure}[h]
    \centering
    \subfloat{
        \includegraphics[scale=0.27]{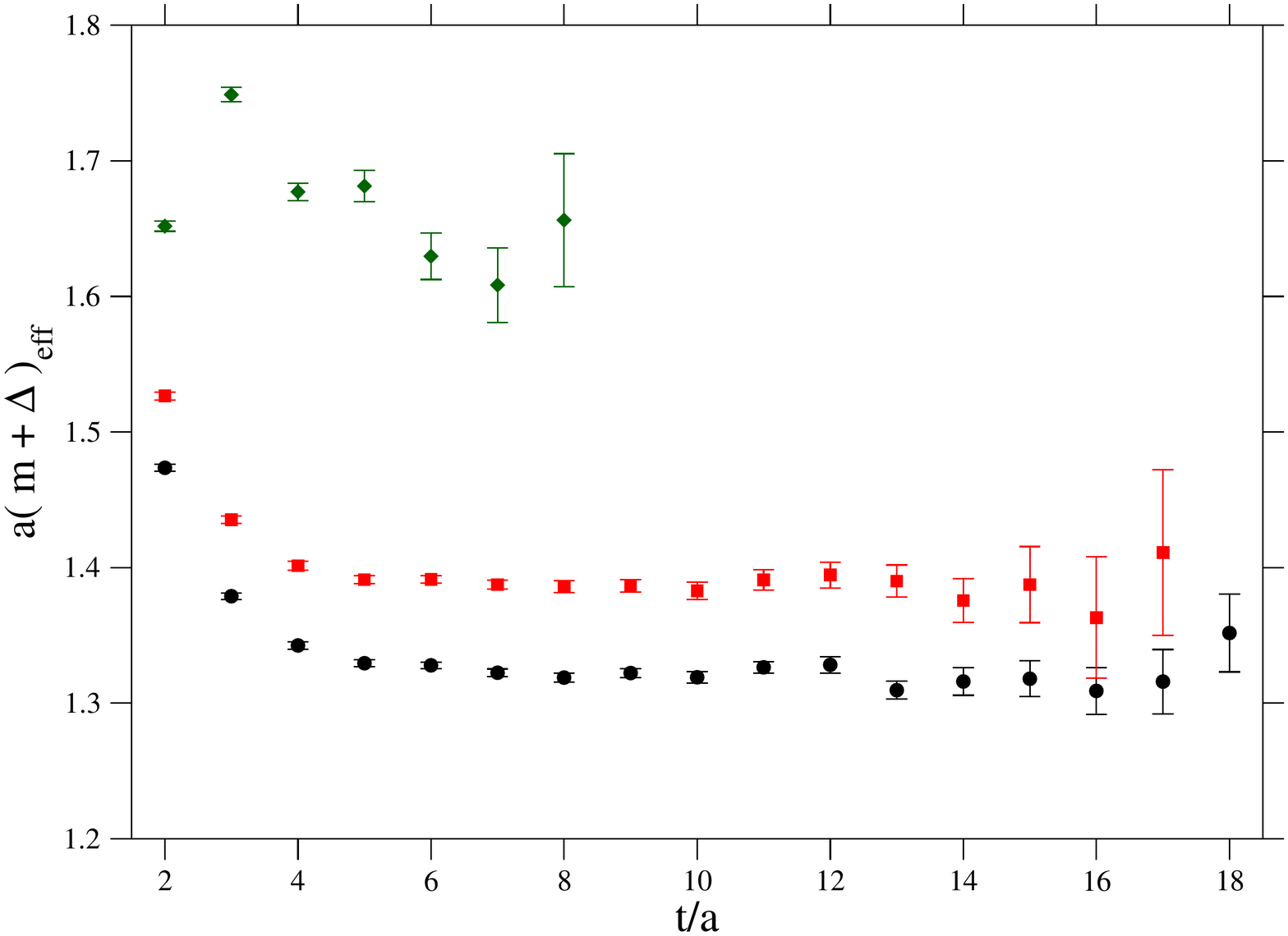}
    }
    \subfloat{
        \includegraphics[scale=0.27]{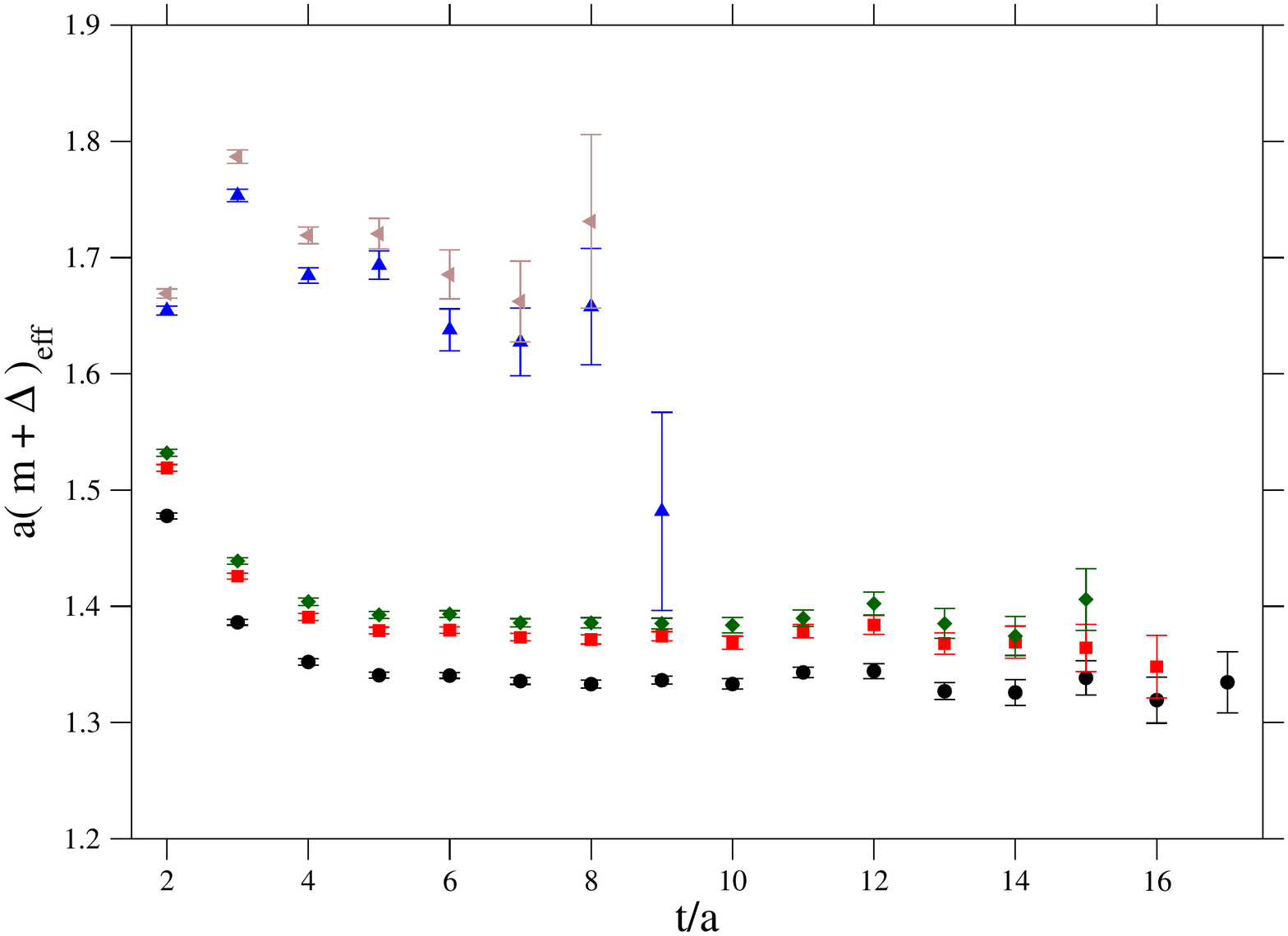}
    }
    \vspace{-16pt}
    \caption{Effective mass plots for the $I=0$, $J^P=0^+$ (left) and $1^+$ (right) $ud\bar{c}\bar{b}$ channels.}
    \label{fig:effmass_udcb}
\end{figure}

\begin{figure}[h!]
    \centering
    \subfloat{
        \includegraphics[scale=0.27]{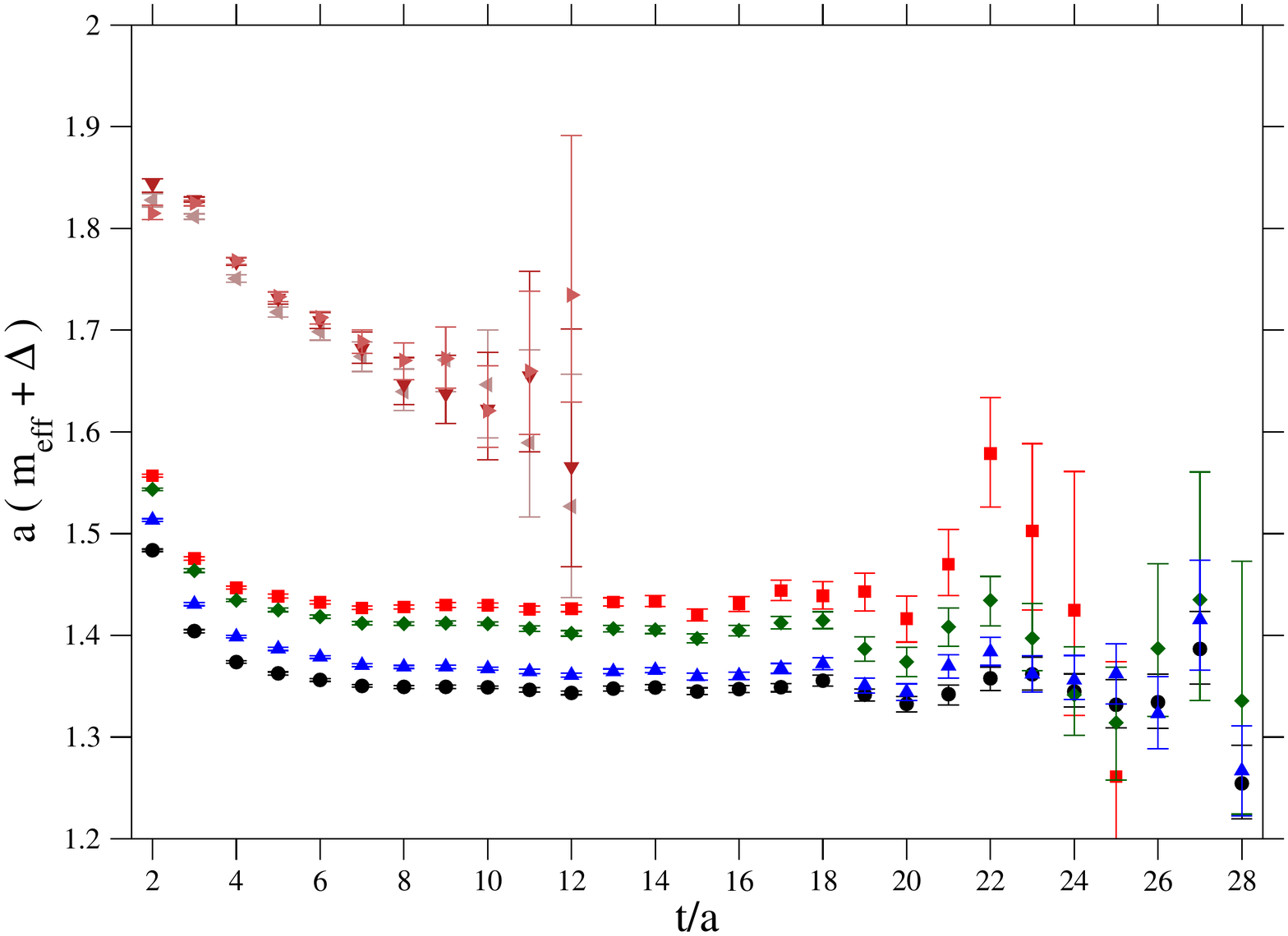}
    }
    \hspace{-24pt}
    \subfloat{
        \includegraphics[scale=0.27]{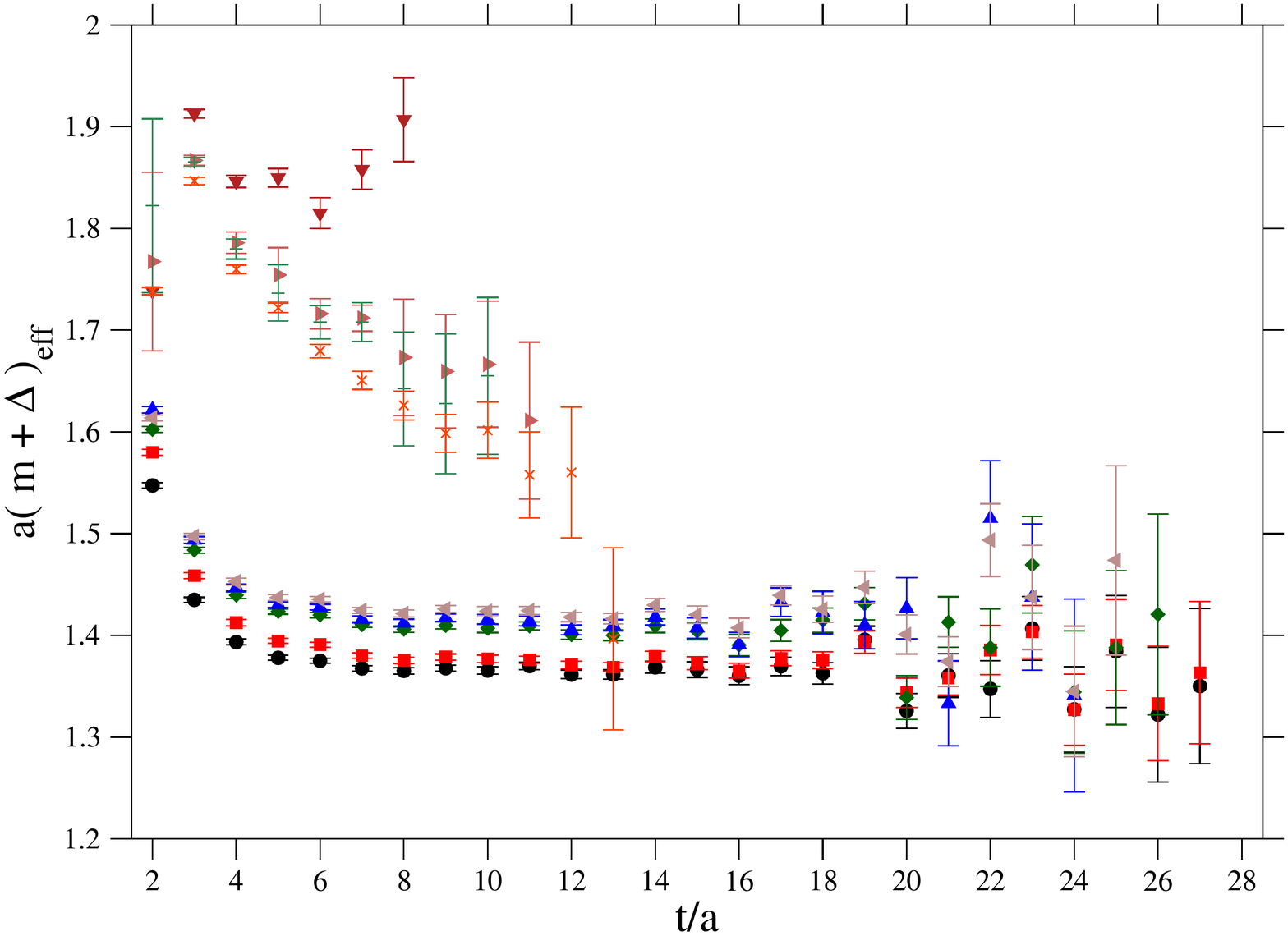}
    }
    \vspace{-16pt}
    \caption{Effective mass plots for the $J^P=0^+$ (left) and $1^+$ (right) $\ell s\bar{c}\bar{b}$ channels.}
    \label{fig:effmass_uscb}
\end{figure}

\begin{figure}[h!]
    \centering
    \subfloat{
        \includegraphics[scale=0.27]{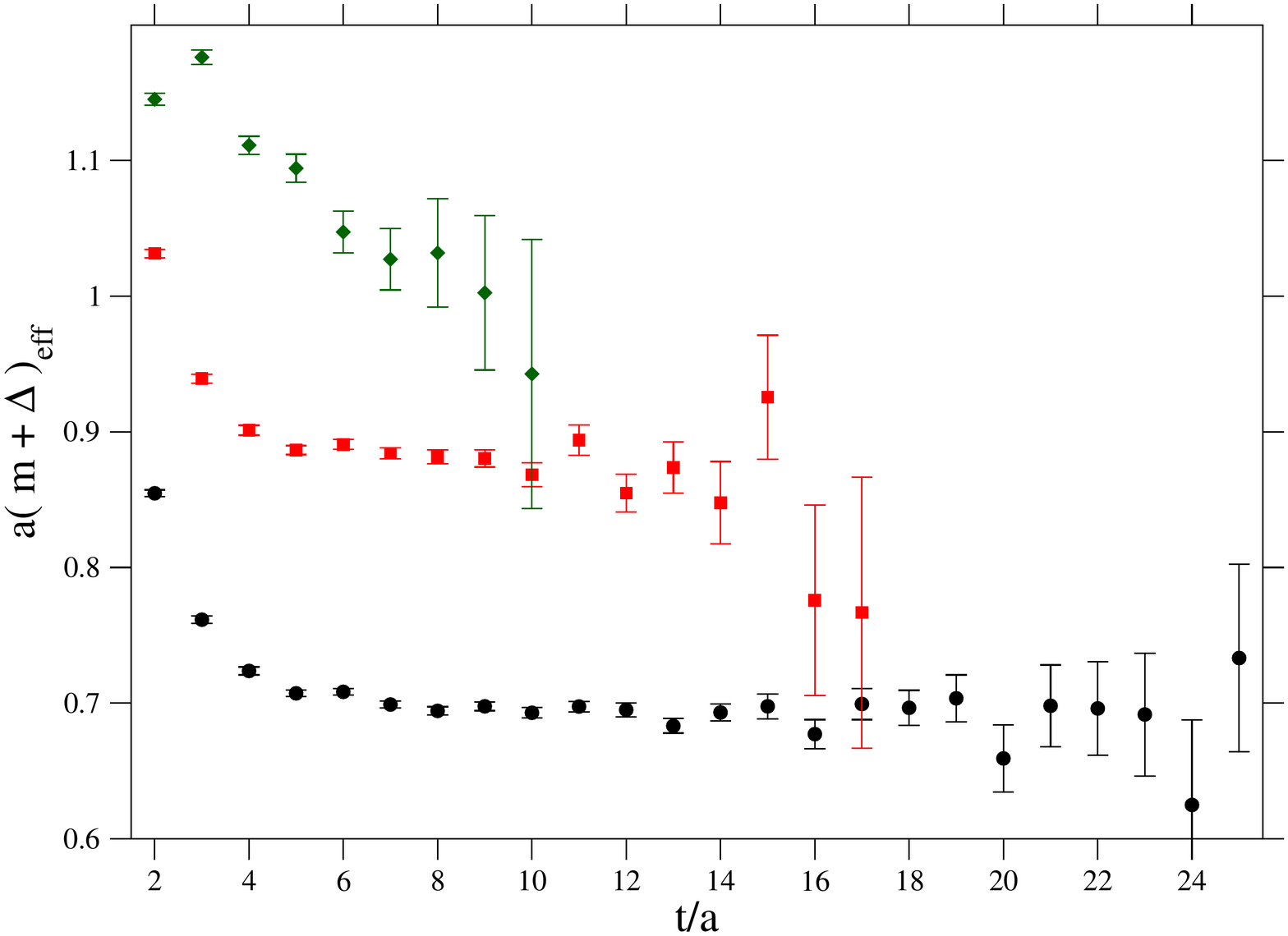}
    }
    \subfloat{
        \includegraphics[scale=0.27]{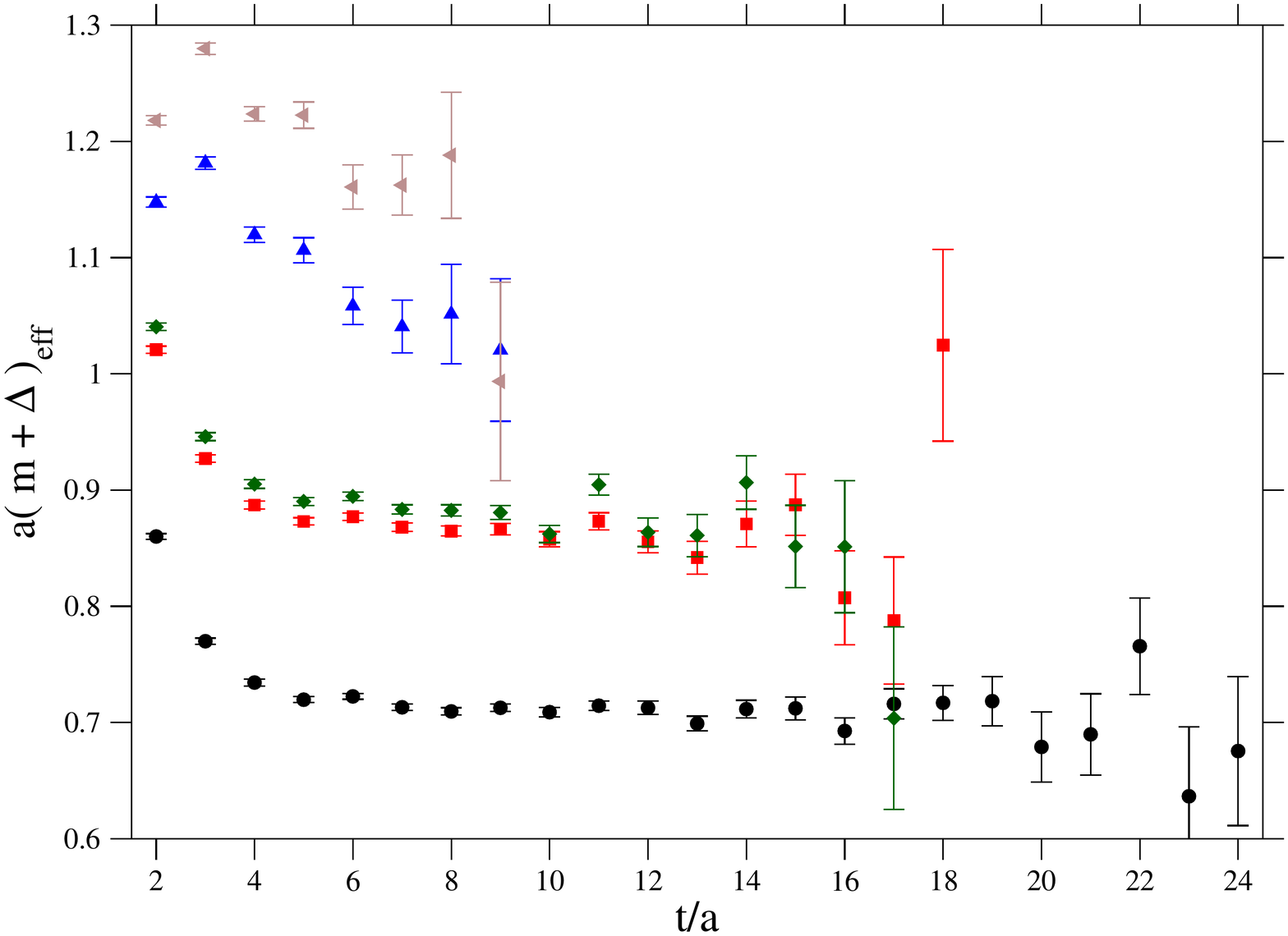}
    }
    \vspace{-16pt}
    \caption{Effective mass plots for the $I=0$, $J^P=0^+$ (left) and $1^+$ (right) $ud\bar{s}\bar{b}$ channels.}
    \label{fig:effmass_udsb}
\end{figure}

\begin{figure}[h!]
    \centering
    \subfloat{
        \includegraphics[scale=0.27]{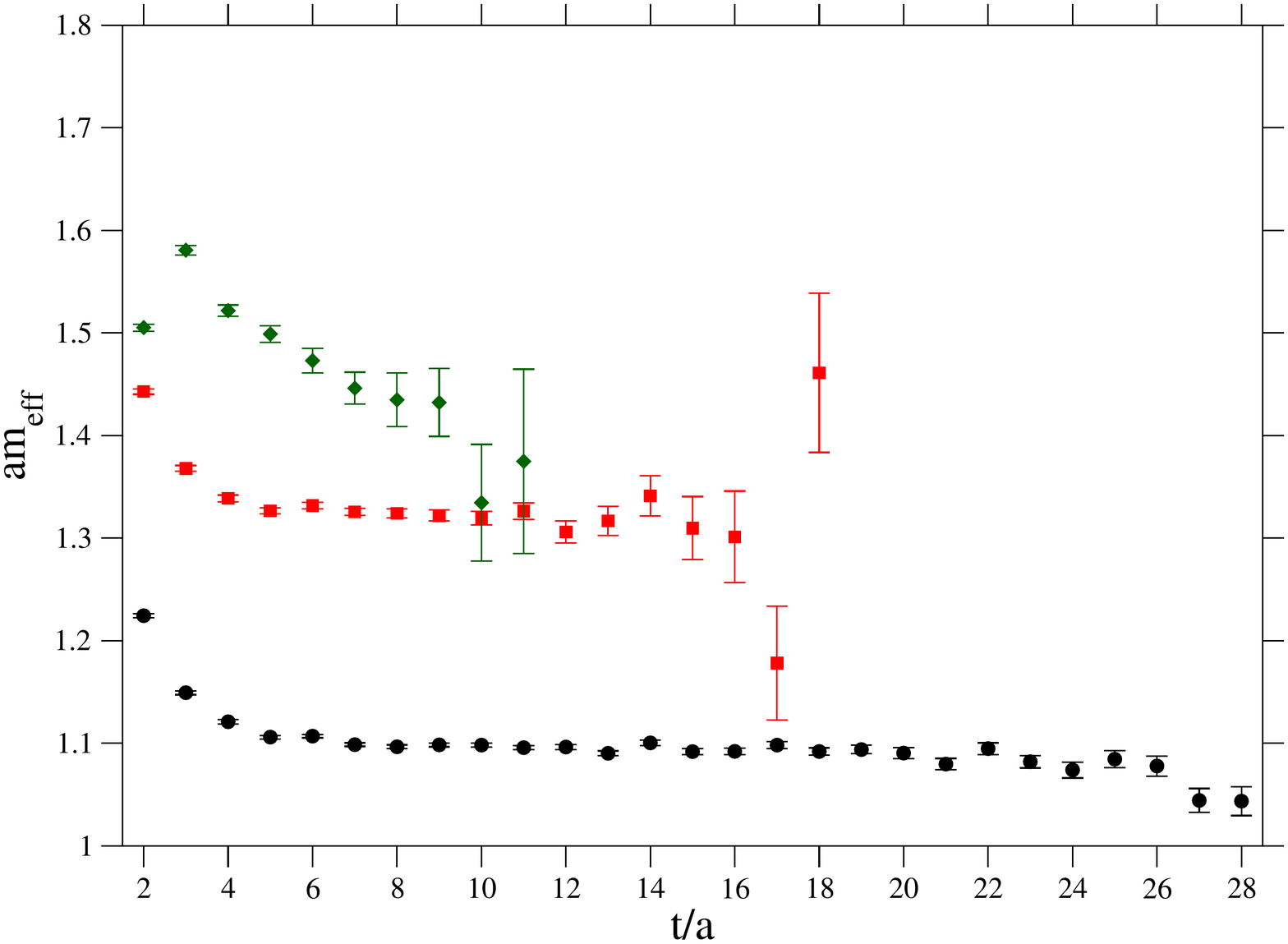}
    }
    \subfloat{
        \includegraphics[scale=0.27]{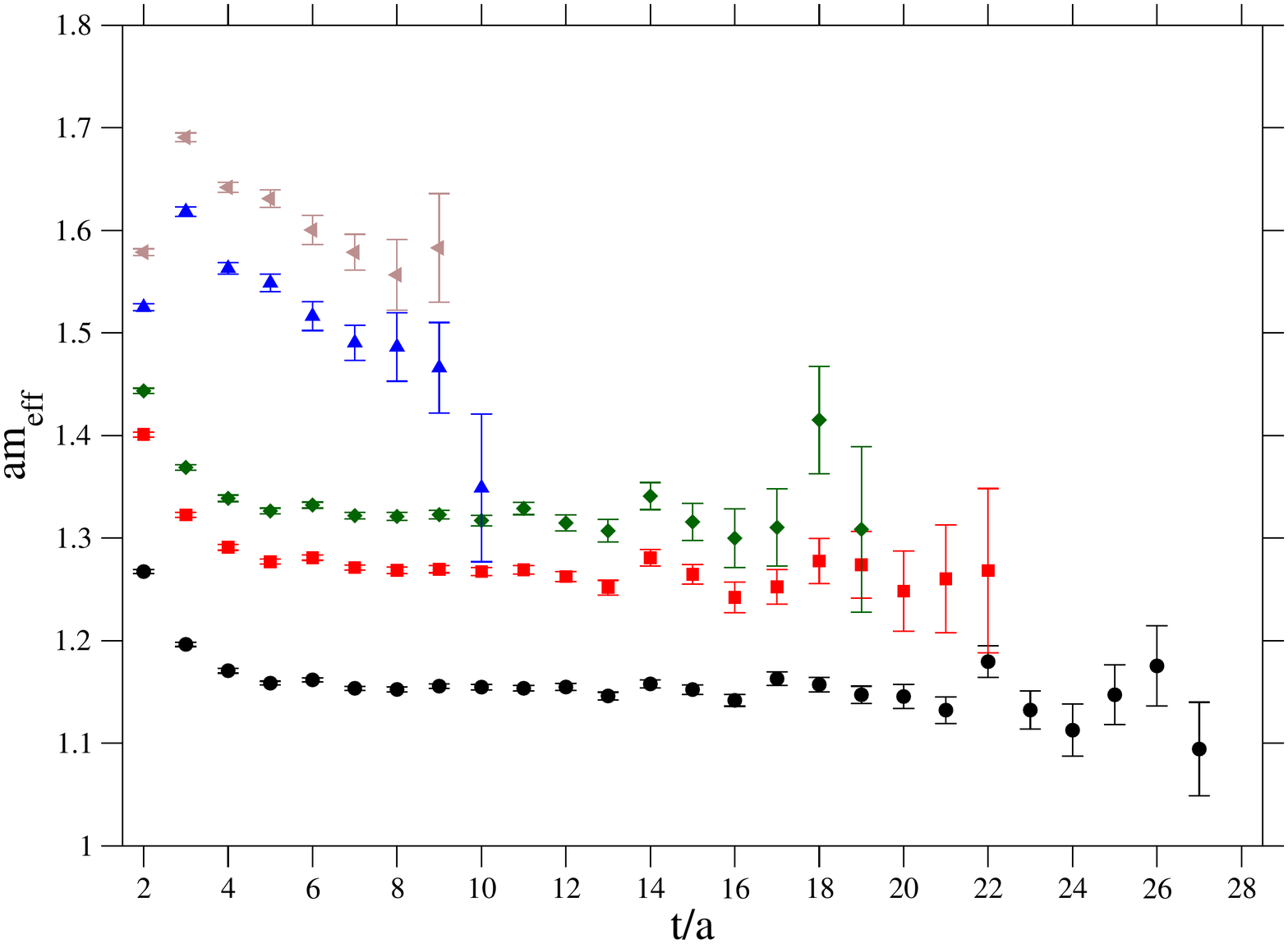}
    }
    \vspace{-16pt}
    \caption{Effective mass plots for the $I=0$, $J^P=0^+$ (left) and $1^+$ (right) $ud\bar{s}\bar{c}$ channels.}
    \label{fig:effmass_udsc}
\end{figure}

\begin{figure}[ht]
    \centering
    \subfloat{
    \includegraphics[scale=0.27]{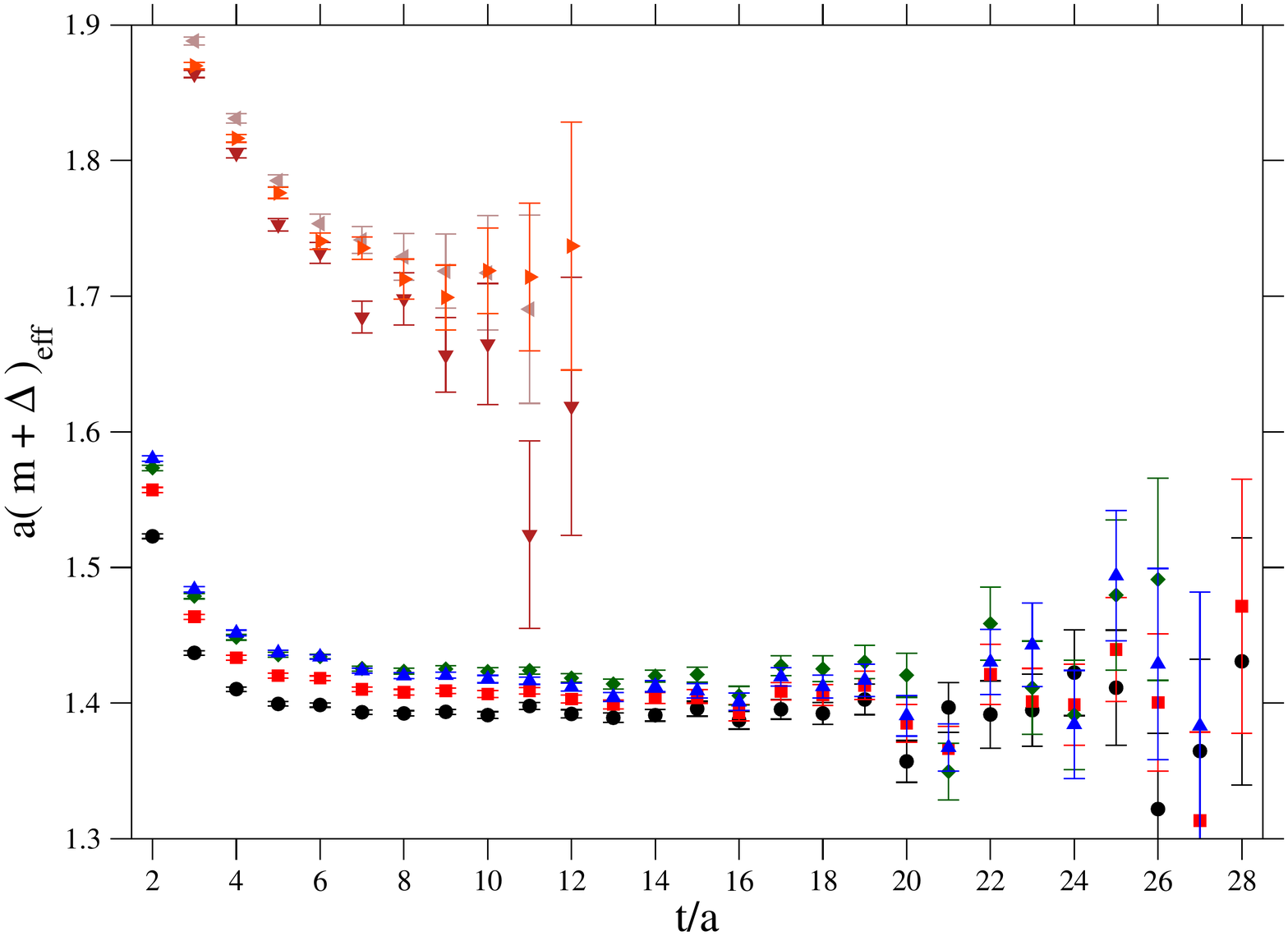}
    }
    \subfloat{
        \includegraphics[scale=0.27]{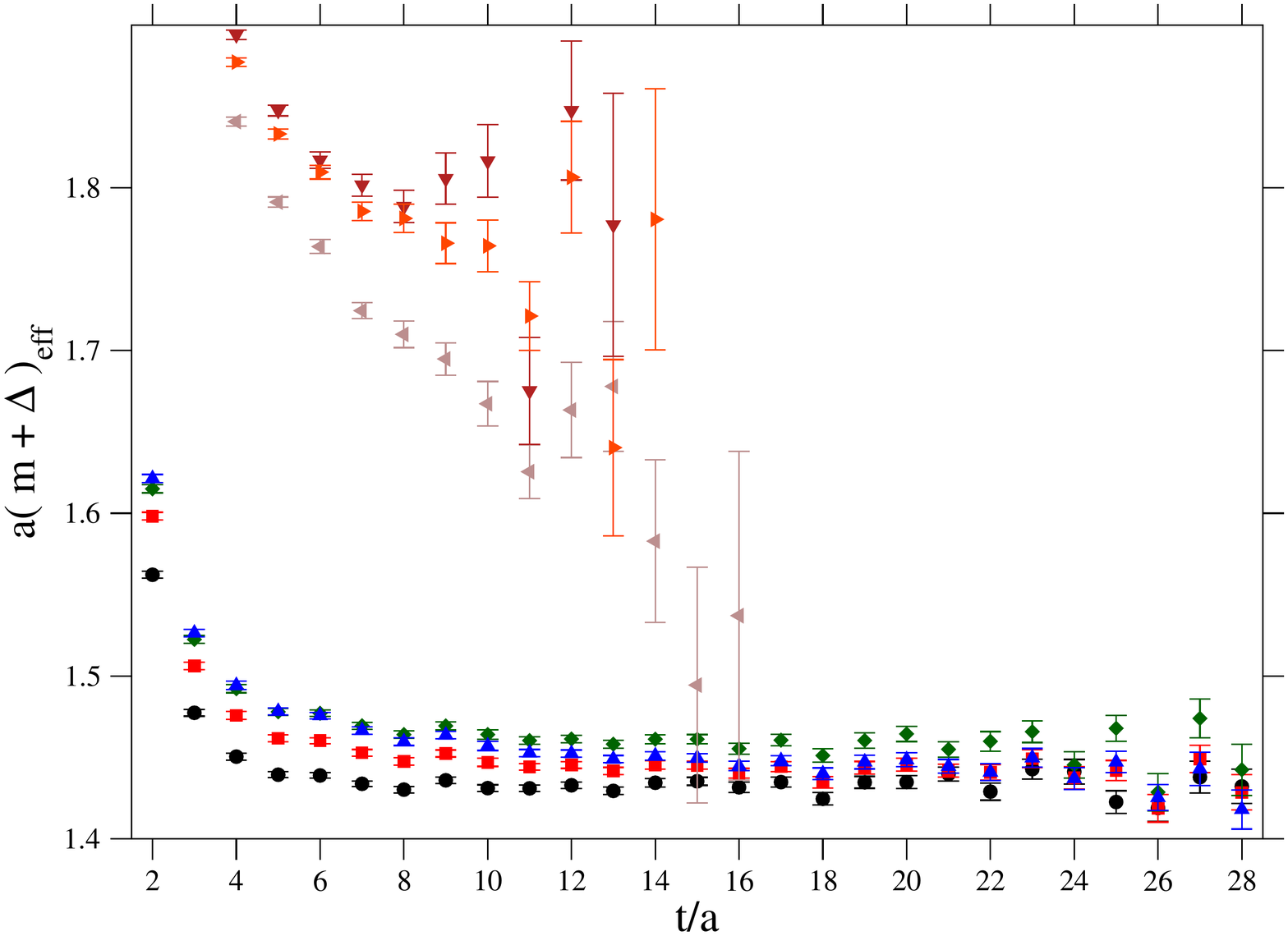}
    }
    \vspace{-16pt}
    \caption{Effective mass plots for the $J^P=1^+$ $uc\bar{b}\bar{b}$ (left) and $sc\bar{b}\bar{b}$ (right) channels.}
    \label{fig:effmass_ucbb}
\end{figure}

Figs.~\ref{fig:effmass_udcb}, 
\ref{fig:effmass_uscb},
\ref{fig:effmass_udsb} and \ref{fig:effmass_udsc}
show the resulting $J^P=0^+$ and $1^+$ effective
masses for the $ud\bar{c}\bar{b}$, 
$\ell s\bar{c}\bar{b}$, $ud\bar{s}\bar{b}$ 
and $ud\bar{s}\bar{c}$ channels,
respectively. Fig.~\ref{fig:effmass_ucbb}, 
similarly, shows the effective masses for 
the $J^P=1^+$ $uc\bar{b}\bar{b}$ and 
$sc\bar{b}\bar{b}$ channels.

\newpage

\end{appendix}

\bibliography{uscb}{}

\begin{thebibliography}{107}%
\makeatletter
\providecommand \@ifxundefined [1]{%
 \@ifx{#1\undefined}
}%
\providecommand \@ifnum [1]{%
 \ifnum #1\expandafter \@firstoftwo
 \else \expandafter \@secondoftwo
 \fi
}%
\providecommand \@ifx [1]{%
 \ifx #1\expandafter \@firstoftwo
 \else \expandafter \@secondoftwo
 \fi
}%
\providecommand \natexlab [1]{#1}%
\providecommand \enquote  [1]{``#1''}%
\providecommand \bibnamefont  [1]{#1}%
\providecommand \bibfnamefont [1]{#1}%
\providecommand \citenamefont [1]{#1}%
\providecommand \href@noop [0]{\@secondoftwo}%
\providecommand \href [0]{\begingroup \@sanitize@url \@href}%
\providecommand \@href[1]{\@@startlink{#1}\@@href}%
\providecommand \@@href[1]{\endgroup#1\@@endlink}%
\providecommand \@sanitize@url [0]{\catcode `\\12\catcode `\$12\catcode
  `\&12\catcode `\#12\catcode `\^12\catcode `\_12\catcode `\%12\relax}%
\providecommand \@@startlink[1]{}%
\providecommand \@@endlink[0]{}%
\providecommand \url  [0]{\begingroup\@sanitize@url \@url }%
\providecommand \@url [1]{\endgroup\@href {#1}{\urlprefix }}%
\providecommand \urlprefix  [0]{URL }%
\providecommand \Eprint [0]{\href }%
\providecommand \doibase [0]{http://dx.doi.org/}%
\providecommand \selectlanguage [0]{\@gobble}%
\providecommand \bibinfo  [0]{\@secondoftwo}%
\providecommand \bibfield  [0]{\@secondoftwo}%
\providecommand \translation [1]{[#1]}%
\providecommand \BibitemOpen [0]{}%
\providecommand \bibitemStop [0]{}%
\providecommand \bibitemNoStop [0]{.\EOS\space}%
\providecommand \EOS [0]{\spacefactor3000\relax}%
\providecommand \BibitemShut  [1]{\csname bibitem#1\endcsname}%
\let\auto@bib@innerbib\@empty
\bibitem [{\citenamefont {Bicudo}\ \emph {et~al.}(2016)\citenamefont {Bicudo},
  \citenamefont {Cichy}, \citenamefont {Peters},\ and\ \citenamefont
  {Wagner}}]{Bicudo:2015kna}%
  \BibitemOpen
  \bibfield  {author} {\bibinfo {author} {\bibfnamefont {P.}~\bibnamefont
  {Bicudo}}, \bibinfo {author} {\bibfnamefont {K.}~\bibnamefont {Cichy}},
  \bibinfo {author} {\bibfnamefont {A.}~\bibnamefont {Peters}}, \ and\ \bibinfo
  {author} {\bibfnamefont {M.}~\bibnamefont {Wagner}},\ }\href {\doibase
  10.1103/PhysRevD.93.034501} {\bibfield  {journal} {\bibinfo  {journal} {Phys.
  Rev. D}\ }\textbf {\bibinfo {volume} {93}},\ \bibinfo {pages} {034501}
  (\bibinfo {year} {2016})},\ \Eprint {http://arxiv.org/abs/1510.03441}
  {arXiv:1510.03441 [hep-lat]} \BibitemShut {NoStop}%
\bibitem [{\citenamefont {Bicudo}\ \emph {et~al.}(2017)\citenamefont {Bicudo},
  \citenamefont {Scheunert},\ and\ \citenamefont {Wagner}}]{Bicudo:2016ooe}%
  \BibitemOpen
  \bibfield  {author} {\bibinfo {author} {\bibfnamefont {P.}~\bibnamefont
  {Bicudo}}, \bibinfo {author} {\bibfnamefont {J.}~\bibnamefont {Scheunert}}, \
  and\ \bibinfo {author} {\bibfnamefont {M.}~\bibnamefont {Wagner}},\ }\href
  {\doibase 10.1103/PhysRevD.95.034502} {\bibfield  {journal} {\bibinfo
  {journal} {Phys. Rev. D}\ }\textbf {\bibinfo {volume} {95}},\ \bibinfo
  {pages} {034502} (\bibinfo {year} {2017})},\ \Eprint
  {http://arxiv.org/abs/1612.02758} {arXiv:1612.02758 [hep-lat]} \BibitemShut
  {NoStop}%
\bibitem [{\citenamefont {Francis}\ \emph {et~al.}(2017)\citenamefont
  {Francis}, \citenamefont {Hudspith}, \citenamefont {Lewis},\ and\
  \citenamefont {Maltman}}]{Francis:2016hui}%
  \BibitemOpen
  \bibfield  {author} {\bibinfo {author} {\bibfnamefont {A.}~\bibnamefont
  {Francis}}, \bibinfo {author} {\bibfnamefont {R.~J.}\ \bibnamefont
  {Hudspith}}, \bibinfo {author} {\bibfnamefont {R.}~\bibnamefont {Lewis}}, \
  and\ \bibinfo {author} {\bibfnamefont {K.}~\bibnamefont {Maltman}},\ }\href
  {\doibase 10.1103/PhysRevLett.118.142001} {\bibfield  {journal} {\bibinfo
  {journal} {Phys. Rev. Lett.}\ }\textbf {\bibinfo {volume} {118}},\ \bibinfo
  {pages} {142001} (\bibinfo {year} {2017})},\ \Eprint
  {http://arxiv.org/abs/1607.05214} {arXiv:1607.05214 [hep-lat]} \BibitemShut
  {NoStop}%
\bibitem [{\citenamefont {Junnarkar}\ \emph {et~al.}(2019)\citenamefont
  {Junnarkar}, \citenamefont {Mathur},\ and\ \citenamefont
  {Padmanath}}]{Junnarkar:2018twb}%
  \BibitemOpen
  \bibfield  {author} {\bibinfo {author} {\bibfnamefont {P.}~\bibnamefont
  {Junnarkar}}, \bibinfo {author} {\bibfnamefont {N.}~\bibnamefont {Mathur}}, \
  and\ \bibinfo {author} {\bibfnamefont {M.}~\bibnamefont {Padmanath}},\ }\href
  {\doibase 10.1103/PhysRevD.99.034507} {\bibfield  {journal} {\bibinfo
  {journal} {Phys. Rev. D}\ }\textbf {\bibinfo {volume} {99}},\ \bibinfo
  {pages} {034507} (\bibinfo {year} {2019})},\ \Eprint
  {http://arxiv.org/abs/1810.12285} {arXiv:1810.12285 [hep-lat]} \BibitemShut
  {NoStop}%
\bibitem [{\citenamefont {Leskovec}\ \emph {et~al.}(2019)\citenamefont
  {Leskovec}, \citenamefont {Meinel}, \citenamefont {Pflaumer},\ and\
  \citenamefont {Wagner}}]{Leskovec:2019ioa}%
  \BibitemOpen
  \bibfield  {author} {\bibinfo {author} {\bibfnamefont {L.}~\bibnamefont
  {Leskovec}}, \bibinfo {author} {\bibfnamefont {S.}~\bibnamefont {Meinel}},
  \bibinfo {author} {\bibfnamefont {M.}~\bibnamefont {Pflaumer}}, \ and\
  \bibinfo {author} {\bibfnamefont {M.}~\bibnamefont {Wagner}},\ }\href
  {\doibase 10.1103/PhysRevD.100.014503} {\bibfield  {journal} {\bibinfo
  {journal} {Phys. Rev.}\ }\textbf {\bibinfo {volume} {D100}},\ \bibinfo
  {pages} {014503} (\bibinfo {year} {2019})},\ \Eprint
  {http://arxiv.org/abs/1904.04197} {arXiv:1904.04197 [hep-lat]} \BibitemShut
  {NoStop}%
\bibitem [{\citenamefont {Eichten}\ and\ \citenamefont
  {Quigg}(2017)}]{Eichten:2017ffp}%
  \BibitemOpen
  \bibfield  {author} {\bibinfo {author} {\bibfnamefont {E.~J.}\ \bibnamefont
  {Eichten}}\ and\ \bibinfo {author} {\bibfnamefont {C.}~\bibnamefont
  {Quigg}},\ }\href {\doibase 10.1103/PhysRevLett.119.202002} {\bibfield
  {journal} {\bibinfo  {journal} {Phys. Rev. Lett.}\ }\textbf {\bibinfo
  {volume} {119}},\ \bibinfo {pages} {202002} (\bibinfo {year} {2017})},\
  \Eprint {http://arxiv.org/abs/1707.09575} {arXiv:1707.09575 [hep-ph]}
  \BibitemShut {NoStop}%
\bibitem [{\citenamefont {Braaten}\ \emph {et~al.}(2020)\citenamefont
  {Braaten}, \citenamefont {He},\ and\ \citenamefont
  {Mohapatra}}]{Braaten:2020nwp}%
  \BibitemOpen
  \bibfield  {author} {\bibinfo {author} {\bibfnamefont {E.}~\bibnamefont
  {Braaten}}, \bibinfo {author} {\bibfnamefont {L.-P.}\ \bibnamefont {He}}, \
  and\ \bibinfo {author} {\bibfnamefont {A.}~\bibnamefont {Mohapatra}},\
  }\href@noop {} {\  (\bibinfo {year} {2020})},\ \Eprint
  {http://arxiv.org/abs/2006.08650} {arXiv:2006.08650 [hep-ph]} \BibitemShut
  {NoStop}%
\bibitem [{Note1()}]{Note1}%
  \BibitemOpen
  \bibinfo {note} {Other arguments favoring the possibility of bound doubly
  bottom tetraquarks, may be found in Refs.~\cite {Manohar:1992nd} and \cite
  {Czarnecki:2017vco}.}\BibitemShut {Stop}%
\bibitem [{\citenamefont {Francis}\ \emph {et~al.}(2019)\citenamefont
  {Francis}, \citenamefont {Hudspith}, \citenamefont {Lewis},\ and\
  \citenamefont {Maltman}}]{Francis:2018jyb}%
  \BibitemOpen
  \bibfield  {author} {\bibinfo {author} {\bibfnamefont {A.}~\bibnamefont
  {Francis}}, \bibinfo {author} {\bibfnamefont {R.~J.}\ \bibnamefont
  {Hudspith}}, \bibinfo {author} {\bibfnamefont {R.}~\bibnamefont {Lewis}}, \
  and\ \bibinfo {author} {\bibfnamefont {K.}~\bibnamefont {Maltman}},\ }\href
  {\doibase 10.1103/PhysRevD.99.054505} {\bibfield  {journal} {\bibinfo
  {journal} {Phys. Rev. D}\ }\textbf {\bibinfo {volume} {99}},\ \bibinfo
  {pages} {054505} (\bibinfo {year} {2019})},\ \Eprint
  {http://arxiv.org/abs/1810.10550} {arXiv:1810.10550 [hep-lat]} \BibitemShut
  {NoStop}%
\bibitem [{\citenamefont {Aaij}\ \emph {et~al.}(2017)\citenamefont {Aaij} \emph
  {et~al.}}]{Aaij:2017ueg}%
  \BibitemOpen
  \bibfield  {author} {\bibinfo {author} {\bibfnamefont {R.}~\bibnamefont
  {Aaij}} \emph {et~al.} (\bibinfo {collaboration} {LHCb}),\ }\href {\doibase
  10.1103/PhysRevLett.119.112001} {\bibfield  {journal} {\bibinfo  {journal}
  {Phys. Rev. Lett.}\ }\textbf {\bibinfo {volume} {119}},\ \bibinfo {pages}
  {112001} (\bibinfo {year} {2017})},\ \Eprint
  {http://arxiv.org/abs/1707.01621} {arXiv:1707.01621 [hep-ex]} \BibitemShut
  {NoStop}%
\bibitem [{\citenamefont {Aaij}\ \emph {et~al.}(2018)\citenamefont {Aaij} \emph
  {et~al.}}]{Aaij:2018gfl}%
  \BibitemOpen
  \bibfield  {author} {\bibinfo {author} {\bibfnamefont {R.}~\bibnamefont
  {Aaij}} \emph {et~al.} (\bibinfo {collaboration} {LHCb}),\ }\href {\doibase
  10.1103/PhysRevLett.121.162002} {\bibfield  {journal} {\bibinfo  {journal}
  {Phys. Rev. Lett.}\ }\textbf {\bibinfo {volume} {121}},\ \bibinfo {pages}
  {162002} (\bibinfo {year} {2018})},\ \Eprint
  {http://arxiv.org/abs/1807.01919} {arXiv:1807.01919 [hep-ex]} \BibitemShut
  {NoStop}%
\bibitem [{\citenamefont {Khachatryan}\ \emph {et~al.}(2017)\citenamefont
  {Khachatryan} \emph {et~al.}}]{Khachatryan:2016ydm}%
  \BibitemOpen
  \bibfield  {author} {\bibinfo {author} {\bibfnamefont {V.}~\bibnamefont
  {Khachatryan}} \emph {et~al.} (\bibinfo {collaboration} {CMS}),\ }\href
  {\doibase 10.1007/JHEP05(2017)013} {\bibfield  {journal} {\bibinfo  {journal}
  {JHEP}\ }\textbf {\bibinfo {volume} {05}},\ \bibinfo {pages} {013} (\bibinfo
  {year} {2017})},\ \Eprint {http://arxiv.org/abs/1610.07095} {arXiv:1610.07095
  [hep-ex]} \BibitemShut {NoStop}%
\bibitem [{\citenamefont {Gershon}\ and\ \citenamefont
  {Poluektov}(2019)}]{Gershon:2018gda}%
  \BibitemOpen
  \bibfield  {author} {\bibinfo {author} {\bibfnamefont {T.}~\bibnamefont
  {Gershon}}\ and\ \bibinfo {author} {\bibfnamefont {A.}~\bibnamefont
  {Poluektov}},\ }\href {\doibase 10.1007/JHEP01(2019)019} {\bibfield
  {journal} {\bibinfo  {journal} {JHEP}\ }\textbf {\bibinfo {volume} {01}},\
  \bibinfo {pages} {019} (\bibinfo {year} {2019})},\ \Eprint
  {http://arxiv.org/abs/1810.06657} {arXiv:1810.06657 [hep-ph]} \BibitemShut
  {NoStop}%
\bibitem [{\citenamefont {Ader}\ \emph {et~al.}(1982)\citenamefont {Ader},
  \citenamefont {Richard},\ and\ \citenamefont {Taxil}}]{Ader:1981db}%
  \BibitemOpen
  \bibfield  {author} {\bibinfo {author} {\bibfnamefont {J.}~\bibnamefont
  {Ader}}, \bibinfo {author} {\bibfnamefont {J.}~\bibnamefont {Richard}}, \
  and\ \bibinfo {author} {\bibfnamefont {P.}~\bibnamefont {Taxil}},\ }\href
  {\doibase 10.1103/PhysRevD.25.2370} {\bibfield  {journal} {\bibinfo
  {journal} {Phys. Rev. D}\ }\textbf {\bibinfo {volume} {25}},\ \bibinfo
  {pages} {2370} (\bibinfo {year} {1982})}\BibitemShut {NoStop}%
\bibitem [{\citenamefont {Heller}\ and\ \citenamefont
  {Tjon}(1987)}]{Heller:1986bt}%
  \BibitemOpen
  \bibfield  {author} {\bibinfo {author} {\bibfnamefont {L.}~\bibnamefont
  {Heller}}\ and\ \bibinfo {author} {\bibfnamefont {J.}~\bibnamefont {Tjon}},\
  }\href {\doibase 10.1103/PhysRevD.35.969} {\bibfield  {journal} {\bibinfo
  {journal} {Phys. Rev. D}\ }\textbf {\bibinfo {volume} {35}},\ \bibinfo
  {pages} {969} (\bibinfo {year} {1987})}\BibitemShut {NoStop}%
\bibitem [{\citenamefont {Carlson}\ \emph {et~al.}(1988)\citenamefont
  {Carlson}, \citenamefont {Heller},\ and\ \citenamefont
  {Tjon}}]{Carlson:1987hh}%
  \BibitemOpen
  \bibfield  {author} {\bibinfo {author} {\bibfnamefont {J.}~\bibnamefont
  {Carlson}}, \bibinfo {author} {\bibfnamefont {L.}~\bibnamefont {Heller}}, \
  and\ \bibinfo {author} {\bibfnamefont {J.}~\bibnamefont {Tjon}},\ }\href
  {\doibase 10.1103/PhysRevD.37.744} {\bibfield  {journal} {\bibinfo  {journal}
  {Phys. Rev. D}\ }\textbf {\bibinfo {volume} {37}},\ \bibinfo {pages} {744}
  (\bibinfo {year} {1988})}\BibitemShut {NoStop}%
\bibitem [{\citenamefont {Zouzou}\ \emph {et~al.}(1986)\citenamefont {Zouzou},
  \citenamefont {Silvestre-Brac}, \citenamefont {Gignoux},\ and\ \citenamefont
  {Richard}}]{Zouzou:1986qh}%
  \BibitemOpen
  \bibfield  {author} {\bibinfo {author} {\bibfnamefont {S.}~\bibnamefont
  {Zouzou}}, \bibinfo {author} {\bibfnamefont {B.}~\bibnamefont
  {Silvestre-Brac}}, \bibinfo {author} {\bibfnamefont {C.}~\bibnamefont
  {Gignoux}}, \ and\ \bibinfo {author} {\bibfnamefont {J.}~\bibnamefont
  {Richard}},\ }\href {\doibase 10.1007/BF01557611} {\bibfield  {journal}
  {\bibinfo  {journal} {Z. Phys. C}\ }\textbf {\bibinfo {volume} {30}},\
  \bibinfo {pages} {457} (\bibinfo {year} {1986})}\BibitemShut {NoStop}%
\bibitem [{\citenamefont {Lipkin}(1986)}]{Lipkin:1986dw}%
  \BibitemOpen
  \bibfield  {author} {\bibinfo {author} {\bibfnamefont {H.~J.}\ \bibnamefont
  {Lipkin}},\ }\href {\doibase 10.1016/0370-2693(86)90843-9} {\bibfield
  {journal} {\bibinfo  {journal} {Phys. Lett. B}\ }\textbf {\bibinfo {volume}
  {172}},\ \bibinfo {pages} {242} (\bibinfo {year} {1986})}\BibitemShut
  {NoStop}%
\bibitem [{\citenamefont {Silvestre-Brac}\ and\ \citenamefont
  {Semay}(1993)}]{SilvestreBrac:1993ss}%
  \BibitemOpen
  \bibfield  {author} {\bibinfo {author} {\bibfnamefont {B.}~\bibnamefont
  {Silvestre-Brac}}\ and\ \bibinfo {author} {\bibfnamefont {C.}~\bibnamefont
  {Semay}},\ }\href {\doibase 10.1007/BF01565058} {\bibfield  {journal}
  {\bibinfo  {journal} {Z. Phys. C}\ }\textbf {\bibinfo {volume} {57}},\
  \bibinfo {pages} {273} (\bibinfo {year} {1993})}\BibitemShut {NoStop}%
\bibitem [{\citenamefont {Semay}\ and\ \citenamefont
  {Silvestre-Brac}(1994)}]{Semay:1994ht}%
  \BibitemOpen
  \bibfield  {author} {\bibinfo {author} {\bibfnamefont {C.}~\bibnamefont
  {Semay}}\ and\ \bibinfo {author} {\bibfnamefont {B.}~\bibnamefont
  {Silvestre-Brac}},\ }\href {\doibase 10.1007/BF01413104} {\bibfield
  {journal} {\bibinfo  {journal} {Z. Phys. C}\ }\textbf {\bibinfo {volume}
  {61}},\ \bibinfo {pages} {271} (\bibinfo {year} {1994})}\BibitemShut
  {NoStop}%
\bibitem [{\citenamefont {Pepin}\ \emph {et~al.}(1997)\citenamefont {Pepin},
  \citenamefont {Stancu}, \citenamefont {Genovese},\ and\ \citenamefont
  {Richard}}]{Pepin:1996id}%
  \BibitemOpen
  \bibfield  {author} {\bibinfo {author} {\bibfnamefont {S.}~\bibnamefont
  {Pepin}}, \bibinfo {author} {\bibfnamefont {F.}~\bibnamefont {Stancu}},
  \bibinfo {author} {\bibfnamefont {M.}~\bibnamefont {Genovese}}, \ and\
  \bibinfo {author} {\bibfnamefont {J.}~\bibnamefont {Richard}},\ }\href
  {\doibase 10.1016/S0370-2693(96)01597-3} {\bibfield  {journal} {\bibinfo
  {journal} {Phys. Lett. B}\ }\textbf {\bibinfo {volume} {393}},\ \bibinfo
  {pages} {119} (\bibinfo {year} {1997})},\ \Eprint
  {http://arxiv.org/abs/hep-ph/9609348} {arXiv:hep-ph/9609348} \BibitemShut
  {NoStop}%
\bibitem [{\citenamefont {Brink}\ and\ \citenamefont
  {Stancu}(1998)}]{Brink:1998as}%
  \BibitemOpen
  \bibfield  {author} {\bibinfo {author} {\bibfnamefont {D.}~\bibnamefont
  {Brink}}\ and\ \bibinfo {author} {\bibfnamefont {F.}~\bibnamefont {Stancu}},\
  }\href {\doibase 10.1103/PhysRevD.57.6778} {\bibfield  {journal} {\bibinfo
  {journal} {Phys. Rev. D}\ }\textbf {\bibinfo {volume} {57}},\ \bibinfo
  {pages} {6778} (\bibinfo {year} {1998})}\BibitemShut {NoStop}%
\bibitem [{\citenamefont {Vijande}\ \emph {et~al.}(2004)\citenamefont
  {Vijande}, \citenamefont {Fernandez}, \citenamefont {Valcarce},\ and\
  \citenamefont {Silvestre-Brac}}]{Vijande:2003ki}%
  \BibitemOpen
  \bibfield  {author} {\bibinfo {author} {\bibfnamefont {J.}~\bibnamefont
  {Vijande}}, \bibinfo {author} {\bibfnamefont {F.}~\bibnamefont {Fernandez}},
  \bibinfo {author} {\bibfnamefont {A.}~\bibnamefont {Valcarce}}, \ and\
  \bibinfo {author} {\bibfnamefont {B.}~\bibnamefont {Silvestre-Brac}},\ }\href
  {\doibase 10.1140/epja/i2003-10128-9} {\bibfield  {journal} {\bibinfo
  {journal} {Eur. Phys. J. A}\ }\textbf {\bibinfo {volume} {19}},\ \bibinfo
  {pages} {383} (\bibinfo {year} {2004})},\ \Eprint
  {http://arxiv.org/abs/hep-ph/0310007} {arXiv:hep-ph/0310007} \BibitemShut
  {NoStop}%
\bibitem [{\citenamefont {Janc}\ and\ \citenamefont
  {Rosina}(2004)}]{Janc:2004qn}%
  \BibitemOpen
  \bibfield  {author} {\bibinfo {author} {\bibfnamefont {D.}~\bibnamefont
  {Janc}}\ and\ \bibinfo {author} {\bibfnamefont {M.}~\bibnamefont {Rosina}},\
  }\href {\doibase 10.1007/s00601-004-0068-9} {\bibfield  {journal} {\bibinfo
  {journal} {Few Body Syst.}\ }\textbf {\bibinfo {volume} {35}},\ \bibinfo
  {pages} {175} (\bibinfo {year} {2004})},\ \Eprint
  {http://arxiv.org/abs/hep-ph/0405208} {arXiv:hep-ph/0405208} \BibitemShut
  {NoStop}%
\bibitem [{\citenamefont {Vijande}\ \emph {et~al.}(2006)\citenamefont
  {Vijande}, \citenamefont {Valcarce},\ and\ \citenamefont
  {Tsushima}}]{Vijande:2006jf}%
  \BibitemOpen
  \bibfield  {author} {\bibinfo {author} {\bibfnamefont {J.}~\bibnamefont
  {Vijande}}, \bibinfo {author} {\bibfnamefont {A.}~\bibnamefont {Valcarce}}, \
  and\ \bibinfo {author} {\bibfnamefont {K.}~\bibnamefont {Tsushima}},\ }\href
  {\doibase 10.1103/PhysRevD.74.054018} {\bibfield  {journal} {\bibinfo
  {journal} {Phys. Rev. D}\ }\textbf {\bibinfo {volume} {74}},\ \bibinfo
  {pages} {054018} (\bibinfo {year} {2006})},\ \Eprint
  {http://arxiv.org/abs/hep-ph/0608316} {arXiv:hep-ph/0608316} \BibitemShut
  {NoStop}%
\bibitem [{\citenamefont {Vijande}\ \emph {et~al.}(2007)\citenamefont
  {Vijande}, \citenamefont {Weissman}, \citenamefont {Valcarce},\ and\
  \citenamefont {Barnea}}]{Vijande:2007rf}%
  \BibitemOpen
  \bibfield  {author} {\bibinfo {author} {\bibfnamefont {J.}~\bibnamefont
  {Vijande}}, \bibinfo {author} {\bibfnamefont {E.}~\bibnamefont {Weissman}},
  \bibinfo {author} {\bibfnamefont {A.}~\bibnamefont {Valcarce}}, \ and\
  \bibinfo {author} {\bibfnamefont {N.}~\bibnamefont {Barnea}},\ }\href
  {\doibase 10.1103/PhysRevD.76.094027} {\bibfield  {journal} {\bibinfo
  {journal} {Phys. Rev. D}\ }\textbf {\bibinfo {volume} {76}},\ \bibinfo
  {pages} {094027} (\bibinfo {year} {2007})},\ \Eprint
  {http://arxiv.org/abs/0710.2516} {arXiv:0710.2516 [hep-ph]} \BibitemShut
  {NoStop}%
\bibitem [{\citenamefont {Ebert}\ \emph {et~al.}(2007)\citenamefont {Ebert},
  \citenamefont {Faustov}, \citenamefont {Galkin},\ and\ \citenamefont
  {Lucha}}]{Ebert:2007rn}%
  \BibitemOpen
  \bibfield  {author} {\bibinfo {author} {\bibfnamefont {D.}~\bibnamefont
  {Ebert}}, \bibinfo {author} {\bibfnamefont {R.}~\bibnamefont {Faustov}},
  \bibinfo {author} {\bibfnamefont {V.}~\bibnamefont {Galkin}}, \ and\ \bibinfo
  {author} {\bibfnamefont {W.}~\bibnamefont {Lucha}},\ }\href {\doibase
  10.1103/PhysRevD.76.114015} {\bibfield  {journal} {\bibinfo  {journal} {Phys.
  Rev. D}\ }\textbf {\bibinfo {volume} {76}},\ \bibinfo {pages} {114015}
  (\bibinfo {year} {2007})},\ \Eprint {http://arxiv.org/abs/0706.3853}
  {arXiv:0706.3853 [hep-ph]} \BibitemShut {NoStop}%
\bibitem [{\citenamefont {Zhang}\ \emph {et~al.}(2008)\citenamefont {Zhang},
  \citenamefont {Zhang},\ and\ \citenamefont {Zhang}}]{Zhang:2007mu}%
  \BibitemOpen
  \bibfield  {author} {\bibinfo {author} {\bibfnamefont {M.}~\bibnamefont
  {Zhang}}, \bibinfo {author} {\bibfnamefont {H.}~\bibnamefont {Zhang}}, \ and\
  \bibinfo {author} {\bibfnamefont {Z.}~\bibnamefont {Zhang}},\ }\href
  {\doibase 10.1088/0253-6102/50/2/31} {\bibfield  {journal} {\bibinfo
  {journal} {Commun. Theor. Phys.}\ }\textbf {\bibinfo {volume} {50}},\
  \bibinfo {pages} {437} (\bibinfo {year} {2008})},\ \Eprint
  {http://arxiv.org/abs/0711.1029} {arXiv:0711.1029 [nucl-th]} \BibitemShut
  {NoStop}%
\bibitem [{\citenamefont {Vijande}\ \emph {et~al.}(2009)\citenamefont
  {Vijande}, \citenamefont {Valcarce},\ and\ \citenamefont
  {Barnea}}]{Vijande:2009kj}%
  \BibitemOpen
  \bibfield  {author} {\bibinfo {author} {\bibfnamefont {J.}~\bibnamefont
  {Vijande}}, \bibinfo {author} {\bibfnamefont {A.}~\bibnamefont {Valcarce}}, \
  and\ \bibinfo {author} {\bibfnamefont {N.}~\bibnamefont {Barnea}},\ }\href
  {\doibase 10.1103/PhysRevD.79.074010} {\bibfield  {journal} {\bibinfo
  {journal} {Phys. Rev. D}\ }\textbf {\bibinfo {volume} {79}},\ \bibinfo
  {pages} {074010} (\bibinfo {year} {2009})},\ \Eprint
  {http://arxiv.org/abs/0903.2949} {arXiv:0903.2949 [hep-ph]} \BibitemShut
  {NoStop}%
\bibitem [{\citenamefont {Yang}\ \emph {et~al.}(2009)\citenamefont {Yang},
  \citenamefont {Deng}, \citenamefont {Ping},\ and\ \citenamefont
  {Goldman}}]{Yang:2009zzp}%
  \BibitemOpen
  \bibfield  {author} {\bibinfo {author} {\bibfnamefont {Y.}~\bibnamefont
  {Yang}}, \bibinfo {author} {\bibfnamefont {C.}~\bibnamefont {Deng}}, \bibinfo
  {author} {\bibfnamefont {J.}~\bibnamefont {Ping}}, \ and\ \bibinfo {author}
  {\bibfnamefont {T.}~\bibnamefont {Goldman}},\ }\href {\doibase
  10.1103/PhysRevD.80.114023} {\bibfield  {journal} {\bibinfo  {journal} {Phys.
  Rev. D}\ }\textbf {\bibinfo {volume} {80}},\ \bibinfo {pages} {114023}
  (\bibinfo {year} {2009})}\BibitemShut {NoStop}%
\bibitem [{\citenamefont {Carames}\ \emph {et~al.}(2011)\citenamefont
  {Carames}, \citenamefont {Valcarce},\ and\ \citenamefont
  {Vijande}}]{Carames:2011zz}%
  \BibitemOpen
  \bibfield  {author} {\bibinfo {author} {\bibfnamefont {T.}~\bibnamefont
  {Carames}}, \bibinfo {author} {\bibfnamefont {A.}~\bibnamefont {Valcarce}}, \
  and\ \bibinfo {author} {\bibfnamefont {J.}~\bibnamefont {Vijande}},\ }\href
  {\doibase 10.1016/j.physletb.2011.04.023} {\bibfield  {journal} {\bibinfo
  {journal} {Phys. Lett. B}\ }\textbf {\bibinfo {volume} {699}},\ \bibinfo
  {pages} {291} (\bibinfo {year} {2011})}\BibitemShut {NoStop}%
\bibitem [{\citenamefont {Silbar}\ and\ \citenamefont
  {Goldman}(2014)}]{Silbar:2013dda}%
  \BibitemOpen
  \bibfield  {author} {\bibinfo {author} {\bibfnamefont {R.~R.}\ \bibnamefont
  {Silbar}}\ and\ \bibinfo {author} {\bibfnamefont {T.}~\bibnamefont
  {Goldman}},\ }\href {\doibase 10.1142/S0218301314500918} {\bibfield
  {journal} {\bibinfo  {journal} {Int. J. Mod. Phys. E}\ }\textbf {\bibinfo
  {volume} {23}},\ \bibinfo {pages} {1450091} (\bibinfo {year} {2014})},\
  \Eprint {http://arxiv.org/abs/1304.5480} {arXiv:1304.5480 [nucl-th]}
  \BibitemShut {NoStop}%
\bibitem [{\citenamefont {Karliner}\ and\ \citenamefont
  {Rosner}(2017)}]{Karliner:2017qjm}%
  \BibitemOpen
  \bibfield  {author} {\bibinfo {author} {\bibfnamefont {M.}~\bibnamefont
  {Karliner}}\ and\ \bibinfo {author} {\bibfnamefont {J.~L.}\ \bibnamefont
  {Rosner}},\ }\href {\doibase 10.1103/PhysRevLett.119.202001} {\bibfield
  {journal} {\bibinfo  {journal} {Phys. Rev. Lett.}\ }\textbf {\bibinfo
  {volume} {119}},\ \bibinfo {pages} {202001} (\bibinfo {year} {2017})},\
  \Eprint {http://arxiv.org/abs/1707.07666} {arXiv:1707.07666 [hep-ph]}
  \BibitemShut {NoStop}%
\bibitem [{\citenamefont {Caramés}\ \emph {et~al.}(2019)\citenamefont
  {Caramés}, \citenamefont {Vijande},\ and\ \citenamefont
  {Valcarce}}]{Caramees:2018oue}%
  \BibitemOpen
  \bibfield  {author} {\bibinfo {author} {\bibfnamefont {T.~F.}\ \bibnamefont
  {Caramés}}, \bibinfo {author} {\bibfnamefont {J.}~\bibnamefont {Vijande}}, \
  and\ \bibinfo {author} {\bibfnamefont {A.}~\bibnamefont {Valcarce}},\ }\href
  {\doibase 10.1103/PhysRevD.99.014006} {\bibfield  {journal} {\bibinfo
  {journal} {Phys. Rev. D}\ }\textbf {\bibinfo {volume} {99}},\ \bibinfo
  {pages} {014006} (\bibinfo {year} {2019})},\ \Eprint
  {http://arxiv.org/abs/1812.08991} {arXiv:1812.08991 [hep-ph]} \BibitemShut
  {NoStop}%
\bibitem [{\citenamefont {Chen}\ and\ \citenamefont
  {Ping}(2018)}]{Chen:2018hts}%
  \BibitemOpen
  \bibfield  {author} {\bibinfo {author} {\bibfnamefont {X.}~\bibnamefont
  {Chen}}\ and\ \bibinfo {author} {\bibfnamefont {J.}~\bibnamefont {Ping}},\
  }\href {\doibase 10.1103/PhysRevD.98.054022} {\bibfield  {journal} {\bibinfo
  {journal} {Phys. Rev. D}\ }\textbf {\bibinfo {volume} {98}},\ \bibinfo
  {pages} {054022} (\bibinfo {year} {2018})},\ \Eprint
  {http://arxiv.org/abs/1807.10505} {arXiv:1807.10505 [hep-ph]} \BibitemShut
  {NoStop}%
\bibitem [{\citenamefont {Deng}\ \emph {et~al.}(2020)\citenamefont {Deng},
  \citenamefont {Chen},\ and\ \citenamefont {Ping}}]{Deng:2018kly}%
  \BibitemOpen
  \bibfield  {author} {\bibinfo {author} {\bibfnamefont {C.}~\bibnamefont
  {Deng}}, \bibinfo {author} {\bibfnamefont {H.}~\bibnamefont {Chen}}, \ and\
  \bibinfo {author} {\bibfnamefont {J.}~\bibnamefont {Ping}},\ }\href {\doibase
  10.1140/epja/s10050-019-00012-y} {\bibfield  {journal} {\bibinfo  {journal}
  {Eur. Phys. J. A}\ }\textbf {\bibinfo {volume} {56}},\ \bibinfo {pages} {9}
  (\bibinfo {year} {2020})},\ \Eprint {http://arxiv.org/abs/1811.06462}
  {arXiv:1811.06462 [hep-ph]} \BibitemShut {NoStop}%
\bibitem [{\citenamefont {Park}\ \emph {et~al.}(2019)\citenamefont {Park},
  \citenamefont {Noh},\ and\ \citenamefont {Lee}}]{Park:2018wjk}%
  \BibitemOpen
  \bibfield  {author} {\bibinfo {author} {\bibfnamefont {W.}~\bibnamefont
  {Park}}, \bibinfo {author} {\bibfnamefont {S.}~\bibnamefont {Noh}}, \ and\
  \bibinfo {author} {\bibfnamefont {S.~H.}\ \bibnamefont {Lee}},\ }\href
  {\doibase 10.1016/j.nuclphysa.2018.12.019} {\bibfield  {journal} {\bibinfo
  {journal} {Nucl. Phys. A}\ }\textbf {\bibinfo {volume} {983}},\ \bibinfo
  {pages} {1} (\bibinfo {year} {2019})},\ \Eprint
  {http://arxiv.org/abs/1809.05257} {arXiv:1809.05257 [nucl-th]} \BibitemShut
  {NoStop}%
\bibitem [{\citenamefont {Yang}\ \emph {et~al.}(2020)\citenamefont {Yang},
  \citenamefont {Ping},\ and\ \citenamefont {Segovia}}]{Yang:2019itm}%
  \BibitemOpen
  \bibfield  {author} {\bibinfo {author} {\bibfnamefont {G.}~\bibnamefont
  {Yang}}, \bibinfo {author} {\bibfnamefont {J.}~\bibnamefont {Ping}}, \ and\
  \bibinfo {author} {\bibfnamefont {J.}~\bibnamefont {Segovia}},\ }\href
  {\doibase 10.1103/PhysRevD.101.014001} {\bibfield  {journal} {\bibinfo
  {journal} {Phys. Rev. D}\ }\textbf {\bibinfo {volume} {101}},\ \bibinfo
  {pages} {014001} (\bibinfo {year} {2020})},\ \Eprint
  {http://arxiv.org/abs/1911.00215} {arXiv:1911.00215 [hep-ph]} \BibitemShut
  {NoStop}%
\bibitem [{\citenamefont {Huang}\ and\ \citenamefont
  {Ping}(2019)}]{Huang:2019otd}%
  \BibitemOpen
  \bibfield  {author} {\bibinfo {author} {\bibfnamefont {H.}~\bibnamefont
  {Huang}}\ and\ \bibinfo {author} {\bibfnamefont {J.}~\bibnamefont {Ping}},\
  }\href {\doibase 10.1140/epjc/s10052-019-7065-0} {\bibfield  {journal}
  {\bibinfo  {journal} {Eur. Phys. J. C}\ }\textbf {\bibinfo {volume} {79}},\
  \bibinfo {pages} {556} (\bibinfo {year} {2019})},\ \Eprint
  {http://arxiv.org/abs/1902.05778} {arXiv:1902.05778 [hep-ph]} \BibitemShut
  {NoStop}%
\bibitem [{\citenamefont {Hernández}\ \emph {et~al.}(2020)\citenamefont
  {Hernández}, \citenamefont {Vijande}, \citenamefont {Valcarce},\ and\
  \citenamefont {Richard}}]{Hernandez:2019eox}%
  \BibitemOpen
  \bibfield  {author} {\bibinfo {author} {\bibfnamefont {E.}~\bibnamefont
  {Hernández}}, \bibinfo {author} {\bibfnamefont {J.}~\bibnamefont {Vijande}},
  \bibinfo {author} {\bibfnamefont {A.}~\bibnamefont {Valcarce}}, \ and\
  \bibinfo {author} {\bibfnamefont {J.-M.}\ \bibnamefont {Richard}},\ }\href
  {\doibase 10.1016/j.physletb.2019.135073} {\bibfield  {journal} {\bibinfo
  {journal} {Phys. Lett. B}\ }\textbf {\bibinfo {volume} {800}},\ \bibinfo
  {pages} {135073} (\bibinfo {year} {2020})},\ \Eprint
  {http://arxiv.org/abs/1910.13394} {arXiv:1910.13394 [hep-ph]} \BibitemShut
  {NoStop}%
\bibitem [{\citenamefont {Tan}\ \emph {et~al.}(2020)\citenamefont {Tan},
  \citenamefont {Lu},\ and\ \citenamefont {Ping}}]{Tan:2020ldi}%
  \BibitemOpen
  \bibfield  {author} {\bibinfo {author} {\bibfnamefont {Y.}~\bibnamefont
  {Tan}}, \bibinfo {author} {\bibfnamefont {W.}~\bibnamefont {Lu}}, \ and\
  \bibinfo {author} {\bibfnamefont {J.}~\bibnamefont {Ping}},\ }\href@noop {}
  {\  (\bibinfo {year} {2020})},\ \Eprint {http://arxiv.org/abs/2004.02106}
  {arXiv:2004.02106 [hep-ph]} \BibitemShut {NoStop}%
\bibitem [{\citenamefont {Lü}\ \emph {et~al.}(2020)\citenamefont {Lü},
  \citenamefont {Chen},\ and\ \citenamefont {Dong}}]{Lu:2020rog}%
  \BibitemOpen
  \bibfield  {author} {\bibinfo {author} {\bibfnamefont {Q.-F.}\ \bibnamefont
  {Lü}}, \bibinfo {author} {\bibfnamefont {D.-Y.}\ \bibnamefont {Chen}}, \
  and\ \bibinfo {author} {\bibfnamefont {Y.-B.}\ \bibnamefont {Dong}},\
  }\href@noop {} {\  (\bibinfo {year} {2020})},\ \Eprint
  {http://arxiv.org/abs/2006.08087} {arXiv:2006.08087 [hep-ph]} \BibitemShut
  {NoStop}%
\bibitem [{\citenamefont {Navarra}\ \emph {et~al.}(2007)\citenamefont
  {Navarra}, \citenamefont {Nielsen},\ and\ \citenamefont
  {Lee}}]{Navarra:2007yw}%
  \BibitemOpen
  \bibfield  {author} {\bibinfo {author} {\bibfnamefont {F.~S.}\ \bibnamefont
  {Navarra}}, \bibinfo {author} {\bibfnamefont {M.}~\bibnamefont {Nielsen}}, \
  and\ \bibinfo {author} {\bibfnamefont {S.~H.}\ \bibnamefont {Lee}},\ }\href
  {\doibase 10.1016/j.physletb.2007.04.010} {\bibfield  {journal} {\bibinfo
  {journal} {Phys. Lett. B}\ }\textbf {\bibinfo {volume} {649}},\ \bibinfo
  {pages} {166} (\bibinfo {year} {2007})},\ \Eprint
  {http://arxiv.org/abs/hep-ph/0703071} {arXiv:hep-ph/0703071} \BibitemShut
  {NoStop}%
\bibitem [{\citenamefont {Du}\ \emph {et~al.}(2013)\citenamefont {Du},
  \citenamefont {Chen}, \citenamefont {Chen},\ and\ \citenamefont
  {Zhu}}]{Du:2012wp}%
  \BibitemOpen
  \bibfield  {author} {\bibinfo {author} {\bibfnamefont {M.-L.}\ \bibnamefont
  {Du}}, \bibinfo {author} {\bibfnamefont {W.}~\bibnamefont {Chen}}, \bibinfo
  {author} {\bibfnamefont {X.-L.}\ \bibnamefont {Chen}}, \ and\ \bibinfo
  {author} {\bibfnamefont {S.-L.}\ \bibnamefont {Zhu}},\ }\href {\doibase
  10.1103/PhysRevD.87.014003} {\bibfield  {journal} {\bibinfo  {journal} {Phys.
  Rev. D}\ }\textbf {\bibinfo {volume} {87}},\ \bibinfo {pages} {014003}
  (\bibinfo {year} {2013})},\ \Eprint {http://arxiv.org/abs/1209.5134}
  {arXiv:1209.5134 [hep-ph]} \BibitemShut {NoStop}%
\bibitem [{\citenamefont {Chen}\ \emph {et~al.}(2014)\citenamefont {Chen},
  \citenamefont {Steele},\ and\ \citenamefont {Zhu}}]{Chen:2013aba}%
  \BibitemOpen
  \bibfield  {author} {\bibinfo {author} {\bibfnamefont {W.}~\bibnamefont
  {Chen}}, \bibinfo {author} {\bibfnamefont {T.}~\bibnamefont {Steele}}, \ and\
  \bibinfo {author} {\bibfnamefont {S.-L.}\ \bibnamefont {Zhu}},\ }\href
  {\doibase 10.1103/PhysRevD.89.054037} {\bibfield  {journal} {\bibinfo
  {journal} {Phys. Rev. D}\ }\textbf {\bibinfo {volume} {89}},\ \bibinfo
  {pages} {054037} (\bibinfo {year} {2014})},\ \Eprint
  {http://arxiv.org/abs/1310.8337} {arXiv:1310.8337 [hep-ph]} \BibitemShut
  {NoStop}%
\bibitem [{\citenamefont {Wang}(2018)}]{Wang:2017uld}%
  \BibitemOpen
  \bibfield  {author} {\bibinfo {author} {\bibfnamefont {Z.-G.}\ \bibnamefont
  {Wang}},\ }\href {\doibase 10.5506/APhysPolB.49.1781} {\bibfield  {journal}
  {\bibinfo  {journal} {Acta Phys. Polon. B}\ }\textbf {\bibinfo {volume}
  {49}},\ \bibinfo {pages} {1781} (\bibinfo {year} {2018})},\ \Eprint
  {http://arxiv.org/abs/1708.04545} {arXiv:1708.04545 [hep-ph]} \BibitemShut
  {NoStop}%
\bibitem [{\citenamefont {Chen}\ \emph {et~al.}(2017)\citenamefont {Chen},
  \citenamefont {Chen}, \citenamefont {Liu}, \citenamefont {Steele},\ and\
  \citenamefont {Zhu}}]{Chen:2017rhl}%
  \BibitemOpen
  \bibfield  {author} {\bibinfo {author} {\bibfnamefont {W.}~\bibnamefont
  {Chen}}, \bibinfo {author} {\bibfnamefont {H.-X.}\ \bibnamefont {Chen}},
  \bibinfo {author} {\bibfnamefont {X.}~\bibnamefont {Liu}}, \bibinfo {author}
  {\bibfnamefont {T.}~\bibnamefont {Steele}}, \ and\ \bibinfo {author}
  {\bibfnamefont {S.-L.}\ \bibnamefont {Zhu}},\ }\href {\doibase
  10.1103/PhysRevD.95.114005} {\bibfield  {journal} {\bibinfo  {journal} {Phys.
  Rev. D}\ }\textbf {\bibinfo {volume} {95}},\ \bibinfo {pages} {114005}
  (\bibinfo {year} {2017})},\ \Eprint {http://arxiv.org/abs/1705.10088}
  {arXiv:1705.10088 [hep-ph]} \BibitemShut {NoStop}%
\bibitem [{\citenamefont {Agaev}\ \emph
  {et~al.}(2019{\natexlab{a}})\citenamefont {Agaev}, \citenamefont {Azizi},
  \citenamefont {Barsbay},\ and\ \citenamefont {Sundu}}]{Agaev:2018khe}%
  \BibitemOpen
  \bibfield  {author} {\bibinfo {author} {\bibfnamefont {S.}~\bibnamefont
  {Agaev}}, \bibinfo {author} {\bibfnamefont {K.}~\bibnamefont {Azizi}},
  \bibinfo {author} {\bibfnamefont {B.}~\bibnamefont {Barsbay}}, \ and\
  \bibinfo {author} {\bibfnamefont {H.}~\bibnamefont {Sundu}},\ }\href
  {\doibase 10.1103/PhysRevD.99.033002} {\bibfield  {journal} {\bibinfo
  {journal} {Phys. Rev. D}\ }\textbf {\bibinfo {volume} {99}},\ \bibinfo
  {pages} {033002} (\bibinfo {year} {2019}{\natexlab{a}})},\ \Eprint
  {http://arxiv.org/abs/1809.07791} {arXiv:1809.07791 [hep-ph]} \BibitemShut
  {NoStop}%
\bibitem [{\citenamefont {Agaev}\ \emph
  {et~al.}(2019{\natexlab{b}})\citenamefont {Agaev}, \citenamefont {Azizi},
  \citenamefont {Barsbay},\ and\ \citenamefont {Sundu}}]{Agaev:2019lwh}%
  \BibitemOpen
  \bibfield  {author} {\bibinfo {author} {\bibfnamefont {S.}~\bibnamefont
  {Agaev}}, \bibinfo {author} {\bibfnamefont {K.}~\bibnamefont {Azizi}},
  \bibinfo {author} {\bibfnamefont {B.}~\bibnamefont {Barsbay}}, \ and\
  \bibinfo {author} {\bibfnamefont {H.}~\bibnamefont {Sundu}},\ }\href@noop {}
  {\  (\bibinfo {year} {2019}{\natexlab{b}})},\ \Eprint
  {http://arxiv.org/abs/1912.07656} {arXiv:1912.07656 [hep-ph]} \BibitemShut
  {NoStop}%
\bibitem [{\citenamefont {Agaev}\ \emph
  {et~al.}(2020{\natexlab{a}})\citenamefont {Agaev}, \citenamefont {Azizi},\
  and\ \citenamefont {Sundu}}]{Agaev:2019kkz}%
  \BibitemOpen
  \bibfield  {author} {\bibinfo {author} {\bibfnamefont {S.}~\bibnamefont
  {Agaev}}, \bibinfo {author} {\bibfnamefont {K.}~\bibnamefont {Azizi}}, \ and\
  \bibinfo {author} {\bibfnamefont {H.}~\bibnamefont {Sundu}},\ }\href
  {\doibase 10.1016/j.nuclphysb.2019.114890} {\bibfield  {journal} {\bibinfo
  {journal} {Nucl. Phys. B}\ }\textbf {\bibinfo {volume} {951}},\ \bibinfo
  {pages} {114890} (\bibinfo {year} {2020}{\natexlab{a}})},\ \Eprint
  {http://arxiv.org/abs/1905.07591} {arXiv:1905.07591 [hep-ph]} \BibitemShut
  {NoStop}%
\bibitem [{\citenamefont {Agaev}\ \emph
  {et~al.}(2019{\natexlab{c}})\citenamefont {Agaev}, \citenamefont {Azizi},\
  and\ \citenamefont {Sundu}}]{Agaev:2019wkk}%
  \BibitemOpen
  \bibfield  {author} {\bibinfo {author} {\bibfnamefont {S.}~\bibnamefont
  {Agaev}}, \bibinfo {author} {\bibfnamefont {K.}~\bibnamefont {Azizi}}, \ and\
  \bibinfo {author} {\bibfnamefont {H.}~\bibnamefont {Sundu}},\ }\href
  {\doibase 10.1103/PhysRevD.100.094020} {\bibfield  {journal} {\bibinfo
  {journal} {Phys. Rev. D}\ }\textbf {\bibinfo {volume} {100}},\ \bibinfo
  {pages} {094020} (\bibinfo {year} {2019}{\natexlab{c}})},\ \Eprint
  {http://arxiv.org/abs/1907.04017} {arXiv:1907.04017 [hep-ph]} \BibitemShut
  {NoStop}%
\bibitem [{\citenamefont {Sundu}\ \emph {et~al.}(2019)\citenamefont {Sundu},
  \citenamefont {Agaev},\ and\ \citenamefont {Azizi}}]{Sundu:2019feu}%
  \BibitemOpen
  \bibfield  {author} {\bibinfo {author} {\bibfnamefont {H.}~\bibnamefont
  {Sundu}}, \bibinfo {author} {\bibfnamefont {S.}~\bibnamefont {Agaev}}, \ and\
  \bibinfo {author} {\bibfnamefont {K.}~\bibnamefont {Azizi}},\ }\href
  {\doibase 10.1140/epjc/s10052-019-7268-4} {\bibfield  {journal} {\bibinfo
  {journal} {Eur. Phys. J. C}\ }\textbf {\bibinfo {volume} {79}},\ \bibinfo
  {pages} {753} (\bibinfo {year} {2019})},\ \Eprint
  {http://arxiv.org/abs/1903.05931} {arXiv:1903.05931 [hep-ph]} \BibitemShut
  {NoStop}%
\bibitem [{\citenamefont {Agaev}\ \emph
  {et~al.}(2020{\natexlab{b}})\citenamefont {Agaev}, \citenamefont {Azizi},
  \citenamefont {Barsbay},\ and\ \citenamefont {Sundu}}]{Agaev:2020zag}%
  \BibitemOpen
  \bibfield  {author} {\bibinfo {author} {\bibfnamefont {S.}~\bibnamefont
  {Agaev}}, \bibinfo {author} {\bibfnamefont {K.}~\bibnamefont {Azizi}},
  \bibinfo {author} {\bibfnamefont {B.}~\bibnamefont {Barsbay}}, \ and\
  \bibinfo {author} {\bibfnamefont {H.}~\bibnamefont {Sundu}},\ }\href@noop {}
  {\  (\bibinfo {year} {2020}{\natexlab{b}})},\ \Eprint
  {http://arxiv.org/abs/2002.04553} {arXiv:2002.04553 [hep-ph]} \BibitemShut
  {NoStop}%
\bibitem [{\citenamefont {Wang}\ and\ \citenamefont
  {Chen}(2020)}]{Wang:2020jgb}%
  \BibitemOpen
  \bibfield  {author} {\bibinfo {author} {\bibfnamefont {Q.-N.}\ \bibnamefont
  {Wang}}\ and\ \bibinfo {author} {\bibfnamefont {W.}~\bibnamefont {Chen}},\
  }\href@noop {} {\  (\bibinfo {year} {2020})},\ \Eprint
  {http://arxiv.org/abs/2002.04243} {arXiv:2002.04243 [hep-ph]} \BibitemShut
  {NoStop}%
\bibitem [{\citenamefont {Tang}\ \emph {et~al.}(2019)\citenamefont {Tang},
  \citenamefont {Wan}, \citenamefont {Maltman},\ and\ \citenamefont
  {Qiao}}]{Tang:2019nwv}%
  \BibitemOpen
  \bibfield  {author} {\bibinfo {author} {\bibfnamefont {L.}~\bibnamefont
  {Tang}}, \bibinfo {author} {\bibfnamefont {B.-D.}\ \bibnamefont {Wan}},
  \bibinfo {author} {\bibfnamefont {K.}~\bibnamefont {Maltman}}, \ and\
  \bibinfo {author} {\bibfnamefont {C.-F.}\ \bibnamefont {Qiao}},\ }\href@noop
  {} {\  (\bibinfo {year} {2019})},\ \Eprint {http://arxiv.org/abs/1911.10951}
  {arXiv:1911.10951 [hep-ph]} \BibitemShut {NoStop}%
\bibitem [{\citenamefont {Shifman}\ \emph
  {et~al.}(1979{\natexlab{a}})\citenamefont {Shifman}, \citenamefont
  {Vainshtein},\ and\ \citenamefont {Zakharov}}]{Shifman:1978bx}%
  \BibitemOpen
  \bibfield  {author} {\bibinfo {author} {\bibfnamefont {M.~A.}\ \bibnamefont
  {Shifman}}, \bibinfo {author} {\bibfnamefont {A.}~\bibnamefont {Vainshtein}},
  \ and\ \bibinfo {author} {\bibfnamefont {V.~I.}\ \bibnamefont {Zakharov}},\
  }\href {\doibase 10.1016/0550-3213(79)90022-1} {\bibfield  {journal}
  {\bibinfo  {journal} {Nucl. Phys. B}\ }\textbf {\bibinfo {volume} {147}},\
  \bibinfo {pages} {385} (\bibinfo {year} {1979}{\natexlab{a}})}\BibitemShut
  {NoStop}%
\bibitem [{\citenamefont {Shifman}\ \emph
  {et~al.}(1979{\natexlab{b}})\citenamefont {Shifman}, \citenamefont
  {Vainshtein},\ and\ \citenamefont {Zakharov}}]{Shifman:1978by}%
  \BibitemOpen
  \bibfield  {author} {\bibinfo {author} {\bibfnamefont {M.~A.}\ \bibnamefont
  {Shifman}}, \bibinfo {author} {\bibfnamefont {A.}~\bibnamefont {Vainshtein}},
  \ and\ \bibinfo {author} {\bibfnamefont {V.~I.}\ \bibnamefont {Zakharov}},\
  }\href {\doibase 10.1016/0550-3213(79)90023-3} {\bibfield  {journal}
  {\bibinfo  {journal} {Nucl. Phys. B}\ }\textbf {\bibinfo {volume} {147}},\
  \bibinfo {pages} {448} (\bibinfo {year} {1979}{\natexlab{b}})}\BibitemShut
  {NoStop}%
\bibitem [{\citenamefont {Ackerstaff}\ \emph {et~al.}(1999)\citenamefont
  {Ackerstaff} \emph {et~al.}}]{Ackerstaff:1998yj}%
  \BibitemOpen
  \bibfield  {author} {\bibinfo {author} {\bibfnamefont {K.}~\bibnamefont
  {Ackerstaff}} \emph {et~al.} (\bibinfo {collaboration} {OPAL}),\ }\href
  {\doibase 10.1007/s100529901061} {\bibfield  {journal} {\bibinfo  {journal}
  {Eur. Phys. J. C}\ }\textbf {\bibinfo {volume} {7}},\ \bibinfo {pages} {571}
  (\bibinfo {year} {1999})},\ \Eprint {http://arxiv.org/abs/hep-ex/9808019}
  {arXiv:hep-ex/9808019} \BibitemShut {NoStop}%
\bibitem [{\citenamefont {Davier}\ \emph {et~al.}(2014)\citenamefont {Davier},
  \citenamefont {Höcker}, \citenamefont {Malaescu}, \citenamefont {Yuan},\
  and\ \citenamefont {Zhang}}]{Davier:2013sfa}%
  \BibitemOpen
  \bibfield  {author} {\bibinfo {author} {\bibfnamefont {M.}~\bibnamefont
  {Davier}}, \bibinfo {author} {\bibfnamefont {A.}~\bibnamefont {Höcker}},
  \bibinfo {author} {\bibfnamefont {B.}~\bibnamefont {Malaescu}}, \bibinfo
  {author} {\bibfnamefont {C.-Z.}\ \bibnamefont {Yuan}}, \ and\ \bibinfo
  {author} {\bibfnamefont {Z.}~\bibnamefont {Zhang}},\ }\href {\doibase
  10.1140/epjc/s10052-014-2803-9} {\bibfield  {journal} {\bibinfo  {journal}
  {Eur. Phys. J. C}\ }\textbf {\bibinfo {volume} {74}},\ \bibinfo {pages}
  {2803} (\bibinfo {year} {2014})},\ \Eprint {http://arxiv.org/abs/1312.1501}
  {arXiv:1312.1501 [hep-ex]} \BibitemShut {NoStop}%
\bibitem [{\citenamefont {Boito}\ \emph {et~al.}(2012)\citenamefont {Boito},
  \citenamefont {Golterman}, \citenamefont {Jamin}, \citenamefont {Mahdavi},
  \citenamefont {Maltman}, \citenamefont {Osborne},\ and\ \citenamefont
  {Peris}}]{Boito:2012cr}%
  \BibitemOpen
  \bibfield  {author} {\bibinfo {author} {\bibfnamefont {D.}~\bibnamefont
  {Boito}}, \bibinfo {author} {\bibfnamefont {M.}~\bibnamefont {Golterman}},
  \bibinfo {author} {\bibfnamefont {M.}~\bibnamefont {Jamin}}, \bibinfo
  {author} {\bibfnamefont {A.}~\bibnamefont {Mahdavi}}, \bibinfo {author}
  {\bibfnamefont {K.}~\bibnamefont {Maltman}}, \bibinfo {author} {\bibfnamefont
  {J.}~\bibnamefont {Osborne}}, \ and\ \bibinfo {author} {\bibfnamefont
  {S.}~\bibnamefont {Peris}},\ }\href {\doibase 10.1103/PhysRevD.85.093015}
  {\bibfield  {journal} {\bibinfo  {journal} {Phys. Rev. D}\ }\textbf {\bibinfo
  {volume} {85}},\ \bibinfo {pages} {093015} (\bibinfo {year} {2012})},\
  \Eprint {http://arxiv.org/abs/1203.3146} {arXiv:1203.3146 [hep-ph]}
  \BibitemShut {NoStop}%
\bibitem [{\citenamefont {Boito}\ \emph {et~al.}(2015)\citenamefont {Boito},
  \citenamefont {Golterman}, \citenamefont {Maltman}, \citenamefont {Osborne},\
  and\ \citenamefont {Peris}}]{Boito:2014sta}%
  \BibitemOpen
  \bibfield  {author} {\bibinfo {author} {\bibfnamefont {D.}~\bibnamefont
  {Boito}}, \bibinfo {author} {\bibfnamefont {M.}~\bibnamefont {Golterman}},
  \bibinfo {author} {\bibfnamefont {K.}~\bibnamefont {Maltman}}, \bibinfo
  {author} {\bibfnamefont {J.}~\bibnamefont {Osborne}}, \ and\ \bibinfo
  {author} {\bibfnamefont {S.}~\bibnamefont {Peris}},\ }\href {\doibase
  10.1103/PhysRevD.91.034003} {\bibfield  {journal} {\bibinfo  {journal} {Phys.
  Rev. D}\ }\textbf {\bibinfo {volume} {91}},\ \bibinfo {pages} {034003}
  (\bibinfo {year} {2015})},\ \Eprint {http://arxiv.org/abs/1410.3528}
  {arXiv:1410.3528 [hep-ph]} \BibitemShut {NoStop}%
\bibitem [{\citenamefont {Abazov}\ \emph {et~al.}(2016)\citenamefont {Abazov}
  \emph {et~al.}}]{D0:2016mwd}%
  \BibitemOpen
  \bibfield  {author} {\bibinfo {author} {\bibfnamefont {V.}~\bibnamefont
  {Abazov}} \emph {et~al.} (\bibinfo {collaboration} {D0}),\ }\href {\doibase
  10.1103/PhysRevLett.117.022003} {\bibfield  {journal} {\bibinfo  {journal}
  {Phys. Rev. Lett.}\ }\textbf {\bibinfo {volume} {117}},\ \bibinfo {pages}
  {022003} (\bibinfo {year} {2016})},\ \Eprint
  {http://arxiv.org/abs/1602.07588} {arXiv:1602.07588 [hep-ex]} \BibitemShut
  {NoStop}%
\bibitem [{\citenamefont {Abazov}\ \emph {et~al.}(2018)\citenamefont {Abazov}
  \emph {et~al.}}]{Abazov:2017poh}%
  \BibitemOpen
  \bibfield  {author} {\bibinfo {author} {\bibfnamefont {V.~M.}\ \bibnamefont
  {Abazov}} \emph {et~al.} (\bibinfo {collaboration} {D0}),\ }\href {\doibase
  10.1103/PhysRevD.97.092004} {\bibfield  {journal} {\bibinfo  {journal} {Phys.
  Rev. D}\ }\textbf {\bibinfo {volume} {97}},\ \bibinfo {pages} {092004}
  (\bibinfo {year} {2018})},\ \Eprint {http://arxiv.org/abs/1712.10176}
  {arXiv:1712.10176 [hep-ex]} \BibitemShut {NoStop}%
\bibitem [{\citenamefont {Aaij}\ \emph {et~al.}(2016)\citenamefont {Aaij} \emph
  {et~al.}}]{Aaij:2016iev}%
  \BibitemOpen
  \bibfield  {author} {\bibinfo {author} {\bibfnamefont {R.}~\bibnamefont
  {Aaij}} \emph {et~al.} (\bibinfo {collaboration} {LHCb}),\ }\href {\doibase
  10.1103/PhysRevLett.117.152003} {\bibfield  {journal} {\bibinfo  {journal}
  {Phys. Rev. Lett.}\ }\textbf {\bibinfo {volume} {117}},\ \bibinfo {pages}
  {152003} (\bibinfo {year} {2016})},\ \bibinfo {note} {[Addendum:
  Phys.Rev.Lett. 118, 109904 (2017)]},\ \Eprint
  {http://arxiv.org/abs/1608.00435} {arXiv:1608.00435 [hep-ex]} \BibitemShut
  {NoStop}%
\bibitem [{\citenamefont {Sirunyan}\ \emph {et~al.}(2018)\citenamefont
  {Sirunyan} \emph {et~al.}}]{Sirunyan:2017ofq}%
  \BibitemOpen
  \bibfield  {author} {\bibinfo {author} {\bibfnamefont {A.~M.}\ \bibnamefont
  {Sirunyan}} \emph {et~al.} (\bibinfo {collaboration} {CMS}),\ }\href
  {\doibase 10.1103/PhysRevLett.120.202005} {\bibfield  {journal} {\bibinfo
  {journal} {Phys. Rev. Lett.}\ }\textbf {\bibinfo {volume} {120}},\ \bibinfo
  {pages} {202005} (\bibinfo {year} {2018})},\ \Eprint
  {http://arxiv.org/abs/1712.06144} {arXiv:1712.06144 [hep-ex]} \BibitemShut
  {NoStop}%
\bibitem [{\citenamefont {Aaltonen}\ \emph {et~al.}(2018)\citenamefont
  {Aaltonen} \emph {et~al.}}]{Aaltonen:2017voc}%
  \BibitemOpen
  \bibfield  {author} {\bibinfo {author} {\bibfnamefont {T.}~\bibnamefont
  {Aaltonen}} \emph {et~al.} (\bibinfo {collaboration} {CDF}),\ }\href
  {\doibase 10.1103/PhysRevLett.120.202006} {\bibfield  {journal} {\bibinfo
  {journal} {Phys. Rev. Lett.}\ }\textbf {\bibinfo {volume} {120}},\ \bibinfo
  {pages} {202006} (\bibinfo {year} {2018})},\ \Eprint
  {http://arxiv.org/abs/1712.09620} {arXiv:1712.09620 [hep-ex]} \BibitemShut
  {NoStop}%
\bibitem [{\citenamefont {Aaboud}\ \emph {et~al.}(2018)\citenamefont {Aaboud}
  \emph {et~al.}}]{Aaboud:2018hgx}%
  \BibitemOpen
  \bibfield  {author} {\bibinfo {author} {\bibfnamefont {M.}~\bibnamefont
  {Aaboud}} \emph {et~al.} (\bibinfo {collaboration} {ATLAS}),\ }\href
  {\doibase 10.1103/PhysRevLett.120.202007} {\bibfield  {journal} {\bibinfo
  {journal} {Phys. Rev. Lett.}\ }\textbf {\bibinfo {volume} {120}},\ \bibinfo
  {pages} {202007} (\bibinfo {year} {2018})},\ \Eprint
  {http://arxiv.org/abs/1802.01840} {arXiv:1802.01840 [hep-ex]} \BibitemShut
  {NoStop}%
\bibitem [{\citenamefont {Cheung}\ \emph {et~al.}(2017)\citenamefont {Cheung},
  \citenamefont {Thomas}, \citenamefont {Dudek},\ and\ \citenamefont
  {Edwards}}]{Cheung:2017tnt}%
  \BibitemOpen
  \bibfield  {author} {\bibinfo {author} {\bibfnamefont {G.~K.~C.}\
  \bibnamefont {Cheung}}, \bibinfo {author} {\bibfnamefont {C.~E.}\
  \bibnamefont {Thomas}}, \bibinfo {author} {\bibfnamefont {J.~J.}\
  \bibnamefont {Dudek}}, \ and\ \bibinfo {author} {\bibfnamefont {R.~G.}\
  \bibnamefont {Edwards}} (\bibinfo {collaboration} {Hadron Spectrum}),\ }\href
  {\doibase 10.1007/JHEP11(2017)033} {\bibfield  {journal} {\bibinfo  {journal}
  {JHEP}\ }\textbf {\bibinfo {volume} {11}},\ \bibinfo {pages} {033} (\bibinfo
  {year} {2017})},\ \Eprint {http://arxiv.org/abs/1709.01417} {arXiv:1709.01417
  [hep-lat]} \BibitemShut {NoStop}%
\bibitem [{\citenamefont {Moore}\ and\ \citenamefont
  {Fleming}(2006)}]{Moore:2006ng}%
  \BibitemOpen
  \bibfield  {author} {\bibinfo {author} {\bibfnamefont {D.~C.}\ \bibnamefont
  {Moore}}\ and\ \bibinfo {author} {\bibfnamefont {G.~T.}\ \bibnamefont
  {Fleming}},\ }\href {\doibase 10.1103/PhysRevD.74.054504} {\bibfield
  {journal} {\bibinfo  {journal} {Phys. Rev.}\ }\textbf {\bibinfo {volume}
  {D74}},\ \bibinfo {pages} {054504} (\bibinfo {year} {2006})},\ \Eprint
  {http://arxiv.org/abs/hep-lat/0607004} {arXiv:hep-lat/0607004 [hep-lat]}
  \BibitemShut {NoStop}%
\bibitem [{\citenamefont {Michael}\ and\ \citenamefont
  {Teasdale}(1983)}]{Michael:1982gb}%
  \BibitemOpen
  \bibfield  {author} {\bibinfo {author} {\bibfnamefont {C.}~\bibnamefont
  {Michael}}\ and\ \bibinfo {author} {\bibfnamefont {I.}~\bibnamefont
  {Teasdale}},\ }\href {\doibase 10.1016/0550-3213(83)90674-0} {\bibfield
  {journal} {\bibinfo  {journal} {Nucl. Phys. B}\ }\textbf {\bibinfo {volume}
  {215}},\ \bibinfo {pages} {433} (\bibinfo {year} {1983})}\BibitemShut
  {NoStop}%
\bibitem [{\citenamefont {Luscher}\ and\ \citenamefont
  {Wolff}(1990)}]{Luscher:1990ck}%
  \BibitemOpen
  \bibfield  {author} {\bibinfo {author} {\bibfnamefont {M.}~\bibnamefont
  {Luscher}}\ and\ \bibinfo {author} {\bibfnamefont {U.}~\bibnamefont
  {Wolff}},\ }\href {\doibase 10.1016/0550-3213(90)90540-T} {\bibfield
  {journal} {\bibinfo  {journal} {Nucl. Phys. B}\ }\textbf {\bibinfo {volume}
  {339}},\ \bibinfo {pages} {222} (\bibinfo {year} {1990})}\BibitemShut
  {NoStop}%
\bibitem [{\citenamefont {Blossier}\ \emph {et~al.}(2009)\citenamefont
  {Blossier}, \citenamefont {Della~Morte}, \citenamefont {von Hippel},
  \citenamefont {Mendes},\ and\ \citenamefont {Sommer}}]{Blossier:2009kd}%
  \BibitemOpen
  \bibfield  {author} {\bibinfo {author} {\bibfnamefont {B.}~\bibnamefont
  {Blossier}}, \bibinfo {author} {\bibfnamefont {M.}~\bibnamefont
  {Della~Morte}}, \bibinfo {author} {\bibfnamefont {G.}~\bibnamefont {von
  Hippel}}, \bibinfo {author} {\bibfnamefont {T.}~\bibnamefont {Mendes}}, \
  and\ \bibinfo {author} {\bibfnamefont {R.}~\bibnamefont {Sommer}},\ }\href
  {\doibase 10.1088/1126-6708/2009/04/094} {\bibfield  {journal} {\bibinfo
  {journal} {JHEP}\ }\textbf {\bibinfo {volume} {04}},\ \bibinfo {pages} {094}
  (\bibinfo {year} {2009})},\ \Eprint {http://arxiv.org/abs/0902.1265}
  {arXiv:0902.1265 [hep-lat]} \BibitemShut {NoStop}%
\bibitem [{\citenamefont {Hörz}\ and\ \citenamefont
  {Hanlon}(2019)}]{Horz:2019rrn}%
  \BibitemOpen
  \bibfield  {author} {\bibinfo {author} {\bibfnamefont {B.}~\bibnamefont
  {Hörz}}\ and\ \bibinfo {author} {\bibfnamefont {A.}~\bibnamefont {Hanlon}},\
  }\href {\doibase 10.1103/PhysRevLett.123.142002} {\bibfield  {journal}
  {\bibinfo  {journal} {Phys. Rev. Lett.}\ }\textbf {\bibinfo {volume} {123}},\
  \bibinfo {pages} {142002} (\bibinfo {year} {2019})},\ \Eprint
  {http://arxiv.org/abs/1905.04277} {arXiv:1905.04277 [hep-lat]} \BibitemShut
  {NoStop}%
\bibitem [{\citenamefont {Billoire}\ \emph {et~al.}(1985)\citenamefont
  {Billoire}, \citenamefont {Marinari},\ and\ \citenamefont
  {Parisi}}]{Billoire:1985yn}%
  \BibitemOpen
  \bibfield  {author} {\bibinfo {author} {\bibfnamefont {A.}~\bibnamefont
  {Billoire}}, \bibinfo {author} {\bibfnamefont {E.}~\bibnamefont {Marinari}},
  \ and\ \bibinfo {author} {\bibfnamefont {G.}~\bibnamefont {Parisi}},\ }\href
  {\doibase 10.1016/0370-2693(85)91079-2} {\bibfield  {journal} {\bibinfo
  {journal} {Phys. Lett. B}\ }\textbf {\bibinfo {volume} {162}},\ \bibinfo
  {pages} {160} (\bibinfo {year} {1985})}\BibitemShut {NoStop}%
\bibitem [{\citenamefont {Gupta}\ \emph {et~al.}(1991)\citenamefont {Gupta},
  \citenamefont {Guralnik}, \citenamefont {Kilcup},\ and\ \citenamefont
  {Sharpe}}]{Gupta:1990mr}%
  \BibitemOpen
  \bibfield  {author} {\bibinfo {author} {\bibfnamefont {R.}~\bibnamefont
  {Gupta}}, \bibinfo {author} {\bibfnamefont {G.}~\bibnamefont {Guralnik}},
  \bibinfo {author} {\bibfnamefont {G.~W.}\ \bibnamefont {Kilcup}}, \ and\
  \bibinfo {author} {\bibfnamefont {S.~R.}\ \bibnamefont {Sharpe}},\ }\href
  {\doibase 10.1103/PhysRevD.43.2003} {\bibfield  {journal} {\bibinfo
  {journal} {Phys. Rev. D}\ }\textbf {\bibinfo {volume} {43}},\ \bibinfo
  {pages} {2003} (\bibinfo {year} {1991})}\BibitemShut {NoStop}%
\bibitem [{\citenamefont {Hudspith}(2015)}]{Hudspith:2014oja}%
  \BibitemOpen
  \bibfield  {author} {\bibinfo {author} {\bibfnamefont {R.}~\bibnamefont
  {Hudspith}} (\bibinfo {collaboration} {RBC, UKQCD}),\ }\href {\doibase
  10.1016/j.cpc.2014.10.017} {\bibfield  {journal} {\bibinfo  {journal}
  {Comput. Phys. Commun.}\ }\textbf {\bibinfo {volume} {187}},\ \bibinfo
  {pages} {115} (\bibinfo {year} {2015})},\ \Eprint
  {http://arxiv.org/abs/1405.5812} {arXiv:1405.5812 [hep-lat]} \BibitemShut
  {NoStop}%
\bibitem [{\citenamefont {Daniel}\ \emph {et~al.}(1992)\citenamefont {Daniel},
  \citenamefont {Gupta}, \citenamefont {Kilcup}, \citenamefont {Patel},\ and\
  \citenamefont {Sharpe}}]{Daniel:1992ek}%
  \BibitemOpen
  \bibfield  {author} {\bibinfo {author} {\bibfnamefont {D.}~\bibnamefont
  {Daniel}}, \bibinfo {author} {\bibfnamefont {R.}~\bibnamefont {Gupta}},
  \bibinfo {author} {\bibfnamefont {G.~W.}\ \bibnamefont {Kilcup}}, \bibinfo
  {author} {\bibfnamefont {A.}~\bibnamefont {Patel}}, \ and\ \bibinfo {author}
  {\bibfnamefont {S.~R.}\ \bibnamefont {Sharpe}},\ }\href {\doibase
  10.1103/PhysRevD.46.3130} {\bibfield  {journal} {\bibinfo  {journal} {Phys.
  Rev. D}\ }\textbf {\bibinfo {volume} {46}},\ \bibinfo {pages} {3130}
  (\bibinfo {year} {1992})},\ \Eprint {http://arxiv.org/abs/hep-lat/9204011}
  {arXiv:hep-lat/9204011} \BibitemShut {NoStop}%
\bibitem [{\citenamefont {Aoki}\ \emph {et~al.}(1996)\citenamefont {Aoki},
  \citenamefont {Fukugita}, \citenamefont {Hashimoto}, \citenamefont {Iwasaki},
  \citenamefont {Kanaya}, \citenamefont {Kuramashi}, \citenamefont {Mino},
  \citenamefont {Okawa}, \citenamefont {Ukawa},\ and\ \citenamefont
  {Yoshie}}]{Aoki:1995bb}%
  \BibitemOpen
  \bibfield  {author} {\bibinfo {author} {\bibfnamefont {S.}~\bibnamefont
  {Aoki}}, \bibinfo {author} {\bibfnamefont {M.}~\bibnamefont {Fukugita}},
  \bibinfo {author} {\bibfnamefont {S.}~\bibnamefont {Hashimoto}}, \bibinfo
  {author} {\bibfnamefont {Y.}~\bibnamefont {Iwasaki}}, \bibinfo {author}
  {\bibfnamefont {K.}~\bibnamefont {Kanaya}}, \bibinfo {author} {\bibfnamefont
  {Y.}~\bibnamefont {Kuramashi}}, \bibinfo {author} {\bibfnamefont
  {H.}~\bibnamefont {Mino}}, \bibinfo {author} {\bibfnamefont {M.}~\bibnamefont
  {Okawa}}, \bibinfo {author} {\bibfnamefont {A.}~\bibnamefont {Ukawa}}, \ and\
  \bibinfo {author} {\bibfnamefont {T.}~\bibnamefont {Yoshie}} (\bibinfo
  {collaboration} {JLQCD}),\ }\href {\doibase 10.1016/0920-5632(96)00072-2}
  {\bibfield  {journal} {\bibinfo  {journal} {Nucl. Phys. B Proc. Suppl.}\
  }\textbf {\bibinfo {volume} {47}},\ \bibinfo {pages} {354} (\bibinfo {year}
  {1996})},\ \Eprint {http://arxiv.org/abs/hep-lat/9510013}
  {arXiv:hep-lat/9510013} \BibitemShut {NoStop}%
\bibitem [{\citenamefont {Antonio}\ \emph {et~al.}(2007)\citenamefont {Antonio}
  \emph {et~al.}}]{Antonio:2006px}%
  \BibitemOpen
  \bibfield  {author} {\bibinfo {author} {\bibfnamefont {D.}~\bibnamefont
  {Antonio}} \emph {et~al.} (\bibinfo {collaboration} {RBC, UKQCD}),\ }\href
  {\doibase 10.1103/PhysRevD.75.114501} {\bibfield  {journal} {\bibinfo
  {journal} {Phys. Rev. D}\ }\textbf {\bibinfo {volume} {75}},\ \bibinfo
  {pages} {114501} (\bibinfo {year} {2007})},\ \Eprint
  {http://arxiv.org/abs/hep-lat/0612005} {arXiv:hep-lat/0612005} \BibitemShut
  {NoStop}%
\bibitem [{\citenamefont {Bedaque}(2004)}]{Bedaque:2004kc}%
  \BibitemOpen
  \bibfield  {author} {\bibinfo {author} {\bibfnamefont {P.~F.}\ \bibnamefont
  {Bedaque}},\ }\href {\doibase 10.1016/j.physletb.2004.04.045} {\bibfield
  {journal} {\bibinfo  {journal} {Phys. Lett. B}\ }\textbf {\bibinfo {volume}
  {593}},\ \bibinfo {pages} {82} (\bibinfo {year} {2004})},\ \Eprint
  {http://arxiv.org/abs/nucl-th/0402051} {arXiv:nucl-th/0402051} \BibitemShut
  {NoStop}%
\bibitem [{\citenamefont {Sachrajda}\ and\ \citenamefont
  {Villadoro}(2005)}]{Sachrajda:2004mi}%
  \BibitemOpen
  \bibfield  {author} {\bibinfo {author} {\bibfnamefont {C.}~\bibnamefont
  {Sachrajda}}\ and\ \bibinfo {author} {\bibfnamefont {G.}~\bibnamefont
  {Villadoro}},\ }\href {\doibase 10.1016/j.physletb.2005.01.033} {\bibfield
  {journal} {\bibinfo  {journal} {Phys. Lett. B}\ }\textbf {\bibinfo {volume}
  {609}},\ \bibinfo {pages} {73} (\bibinfo {year} {2005})},\ \Eprint
  {http://arxiv.org/abs/hep-lat/0411033} {arXiv:hep-lat/0411033} \BibitemShut
  {NoStop}%
\bibitem [{\citenamefont {Aoki}\ \emph {et~al.}(2011)\citenamefont {Aoki} \emph
  {et~al.}}]{Aoki:2010dy}%
  \BibitemOpen
  \bibfield  {author} {\bibinfo {author} {\bibfnamefont {Y.}~\bibnamefont
  {Aoki}} \emph {et~al.} (\bibinfo {collaboration} {RBC, UKQCD}),\ }\href
  {\doibase 10.1103/PhysRevD.83.074508} {\bibfield  {journal} {\bibinfo
  {journal} {Phys. Rev. D}\ }\textbf {\bibinfo {volume} {83}},\ \bibinfo
  {pages} {074508} (\bibinfo {year} {2011})},\ \Eprint
  {http://arxiv.org/abs/1011.0892} {arXiv:1011.0892 [hep-lat]} \BibitemShut
  {NoStop}%
\bibitem [{Note2()}]{Note2}%
  \BibitemOpen
  \bibinfo {note} {See App.~\ref {app:meson_amps} for more discussion of how to
  obtain physical amplitudes from these correlators}\BibitemShut {NoStop}%
\bibitem [{\citenamefont {Bhattacharya}\ \emph {et~al.}(1996)\citenamefont
  {Bhattacharya}, \citenamefont {Gupta}, \citenamefont {Kilcup},\ and\
  \citenamefont {Sharpe}}]{Bhattacharya:1995fz}%
  \BibitemOpen
  \bibfield  {author} {\bibinfo {author} {\bibfnamefont {T.}~\bibnamefont
  {Bhattacharya}}, \bibinfo {author} {\bibfnamefont {R.}~\bibnamefont {Gupta}},
  \bibinfo {author} {\bibfnamefont {G.}~\bibnamefont {Kilcup}}, \ and\ \bibinfo
  {author} {\bibfnamefont {S.~R.}\ \bibnamefont {Sharpe}},\ }\href {\doibase
  10.1103/PhysRevD.53.6486} {\bibfield  {journal} {\bibinfo  {journal} {Phys.
  Rev. D}\ }\textbf {\bibinfo {volume} {53}},\ \bibinfo {pages} {6486}
  (\bibinfo {year} {1996})},\ \Eprint {http://arxiv.org/abs/hep-lat/9512021}
  {arXiv:hep-lat/9512021} \BibitemShut {NoStop}%
\bibitem [{\citenamefont {Aoki}\ \emph {et~al.}(2010)\citenamefont {Aoki} \emph
  {et~al.}}]{Aoki:2009ix}%
  \BibitemOpen
  \bibfield  {author} {\bibinfo {author} {\bibfnamefont {S.}~\bibnamefont
  {Aoki}} \emph {et~al.} (\bibinfo {collaboration} {PACS-CS}),\ }\href
  {\doibase 10.1103/PhysRevD.81.074503} {\bibfield  {journal} {\bibinfo
  {journal} {Phys. Rev. D}\ }\textbf {\bibinfo {volume} {81}},\ \bibinfo
  {pages} {074503} (\bibinfo {year} {2010})},\ \Eprint
  {http://arxiv.org/abs/0911.2561} {arXiv:0911.2561 [hep-lat]} \BibitemShut
  {NoStop}%
\bibitem [{\citenamefont {Namekawa}\ \emph {et~al.}(2013)\citenamefont
  {Namekawa} \emph {et~al.}}]{Namekawa:2013vu}%
  \BibitemOpen
  \bibfield  {author} {\bibinfo {author} {\bibfnamefont {Y.}~\bibnamefont
  {Namekawa}} \emph {et~al.} (\bibinfo {collaboration} {PACS-CS}),\ }\href
  {\doibase 10.1103/PhysRevD.87.094512} {\bibfield  {journal} {\bibinfo
  {journal} {Phys. Rev. D}\ }\textbf {\bibinfo {volume} {87}},\ \bibinfo
  {pages} {094512} (\bibinfo {year} {2013})},\ \Eprint
  {http://arxiv.org/abs/1301.4743} {arXiv:1301.4743 [hep-lat]} \BibitemShut
  {NoStop}%
\bibitem [{ope()}]{openQCD}%
  \BibitemOpen
  \href@noop {} {}\bibinfo {howpublished}
  {\url{http://luscher.web.cern.ch/luscher/openQCD/.}}\BibitemShut {Stop}%
\bibitem [{Note3()}]{Note3}%
  \BibitemOpen
  \bibinfo {note} {We implemented this heavy-quark action using AVX/FMA2 vector
  intrinsics directly in openQCD. Typical charm-quark propagator inversions are
  comparable to those of our strange-quark propagators}\BibitemShut {NoStop}%
\bibitem [{\citenamefont {Luscher}(2004)}]{Luscher:2003qa}%
  \BibitemOpen
  \bibfield  {author} {\bibinfo {author} {\bibfnamefont {M.}~\bibnamefont
  {Luscher}},\ }\href {\doibase 10.1016/S0010-4655(03)00486-7} {\bibfield
  {journal} {\bibinfo  {journal} {Comput. Phys. Commun.}\ }\textbf {\bibinfo
  {volume} {156}},\ \bibinfo {pages} {209} (\bibinfo {year} {2004})},\ \Eprint
  {http://arxiv.org/abs/hep-lat/0310048} {arXiv:hep-lat/0310048} \BibitemShut
  {NoStop}%
\bibitem [{\citenamefont {Luscher}(2007{\natexlab{a}})}]{Luscher:2007es}%
  \BibitemOpen
  \bibfield  {author} {\bibinfo {author} {\bibfnamefont {M.}~\bibnamefont
  {Luscher}},\ }\href {\doibase 10.1088/1126-6708/2007/12/011} {\bibfield
  {journal} {\bibinfo  {journal} {JHEP}\ }\textbf {\bibinfo {volume} {12}},\
  \bibinfo {pages} {011} (\bibinfo {year} {2007}{\natexlab{a}})},\ \Eprint
  {http://arxiv.org/abs/0710.5417} {arXiv:0710.5417 [hep-lat]} \BibitemShut
  {NoStop}%
\bibitem [{\citenamefont {Luscher}(2007{\natexlab{b}})}]{Luscher:2007se}%
  \BibitemOpen
  \bibfield  {author} {\bibinfo {author} {\bibfnamefont {M.}~\bibnamefont
  {Luscher}},\ }\href {\doibase 10.1088/1126-6708/2007/07/081} {\bibfield
  {journal} {\bibinfo  {journal} {JHEP}\ }\textbf {\bibinfo {volume} {07}},\
  \bibinfo {pages} {081} (\bibinfo {year} {2007}{\natexlab{b}})},\ \Eprint
  {http://arxiv.org/abs/0706.2298} {arXiv:0706.2298 [hep-lat]} \BibitemShut
  {NoStop}%
\bibitem [{\citenamefont {Aoki}\ \emph {et~al.}(2003)\citenamefont {Aoki},
  \citenamefont {Kuramashi},\ and\ \citenamefont {Tominaga}}]{Aoki:2001ra}%
  \BibitemOpen
  \bibfield  {author} {\bibinfo {author} {\bibfnamefont {S.}~\bibnamefont
  {Aoki}}, \bibinfo {author} {\bibfnamefont {Y.}~\bibnamefont {Kuramashi}}, \
  and\ \bibinfo {author} {\bibfnamefont {S.-i.}\ \bibnamefont {Tominaga}},\
  }\href {\doibase 10.1143/PTP.109.383} {\bibfield  {journal} {\bibinfo
  {journal} {Prog. Theor. Phys.}\ }\textbf {\bibinfo {volume} {109}},\ \bibinfo
  {pages} {383} (\bibinfo {year} {2003})},\ \Eprint
  {http://arxiv.org/abs/hep-lat/0107009} {arXiv:hep-lat/0107009} \BibitemShut
  {NoStop}%
\bibitem [{\citenamefont {Aoki}\ \emph {et~al.}(2004)\citenamefont {Aoki},
  \citenamefont {Kayaba},\ and\ \citenamefont {Kuramashi}}]{Aoki:2003dg}%
  \BibitemOpen
  \bibfield  {author} {\bibinfo {author} {\bibfnamefont {S.}~\bibnamefont
  {Aoki}}, \bibinfo {author} {\bibfnamefont {Y.}~\bibnamefont {Kayaba}}, \ and\
  \bibinfo {author} {\bibfnamefont {Y.}~\bibnamefont {Kuramashi}},\ }\href
  {\doibase 10.1016/j.nuclphysb.2004.07.017} {\bibfield  {journal} {\bibinfo
  {journal} {Nucl. Phys. B}\ }\textbf {\bibinfo {volume} {697}},\ \bibinfo
  {pages} {271} (\bibinfo {year} {2004})},\ \Eprint
  {http://arxiv.org/abs/hep-lat/0309161} {arXiv:hep-lat/0309161} \BibitemShut
  {NoStop}%
\bibitem [{\citenamefont {Namekawa}\ \emph {et~al.}(2011)\citenamefont
  {Namekawa} \emph {et~al.}}]{Namekawa:2011wt}%
  \BibitemOpen
  \bibfield  {author} {\bibinfo {author} {\bibfnamefont {Y.}~\bibnamefont
  {Namekawa}} \emph {et~al.} (\bibinfo {collaboration} {PACS-CS}),\ }\href
  {\doibase 10.1103/PhysRevD.84.074505} {\bibfield  {journal} {\bibinfo
  {journal} {Phys. Rev. D}\ }\textbf {\bibinfo {volume} {84}},\ \bibinfo
  {pages} {074505} (\bibinfo {year} {2011})},\ \Eprint
  {http://arxiv.org/abs/1104.4600} {arXiv:1104.4600 [hep-lat]} \BibitemShut
  {NoStop}%
\bibitem [{\citenamefont {Mohler}\ \emph {et~al.}(2013)\citenamefont {Mohler},
  \citenamefont {Lang}, \citenamefont {Leskovec}, \citenamefont {Prelovsek},\
  and\ \citenamefont {Woloshyn}}]{Mohler:2013rwa}%
  \BibitemOpen
  \bibfield  {author} {\bibinfo {author} {\bibfnamefont {D.}~\bibnamefont
  {Mohler}}, \bibinfo {author} {\bibfnamefont {C.}~\bibnamefont {Lang}},
  \bibinfo {author} {\bibfnamefont {L.}~\bibnamefont {Leskovec}}, \bibinfo
  {author} {\bibfnamefont {S.}~\bibnamefont {Prelovsek}}, \ and\ \bibinfo
  {author} {\bibfnamefont {R.}~\bibnamefont {Woloshyn}},\ }\href {\doibase
  10.1103/PhysRevLett.111.222001} {\bibfield  {journal} {\bibinfo  {journal}
  {Phys. Rev. Lett.}\ }\textbf {\bibinfo {volume} {111}},\ \bibinfo {pages}
  {222001} (\bibinfo {year} {2013})},\ \Eprint {http://arxiv.org/abs/1308.3175}
  {arXiv:1308.3175 [hep-lat]} \BibitemShut {NoStop}%
\bibitem [{\citenamefont {Lepage}\ \emph {et~al.}(1992)\citenamefont {Lepage},
  \citenamefont {Magnea}, \citenamefont {Nakhleh}, \citenamefont {Magnea},\
  and\ \citenamefont {Hornbostel}}]{Lepage:1992tx}%
  \BibitemOpen
  \bibfield  {author} {\bibinfo {author} {\bibfnamefont {G.}~\bibnamefont
  {Lepage}}, \bibinfo {author} {\bibfnamefont {L.}~\bibnamefont {Magnea}},
  \bibinfo {author} {\bibfnamefont {C.}~\bibnamefont {Nakhleh}}, \bibinfo
  {author} {\bibfnamefont {U.}~\bibnamefont {Magnea}}, \ and\ \bibinfo {author}
  {\bibfnamefont {K.}~\bibnamefont {Hornbostel}},\ }\href {\doibase
  10.1103/PhysRevD.46.4052} {\bibfield  {journal} {\bibinfo  {journal} {Phys.
  Rev. D}\ }\textbf {\bibinfo {volume} {46}},\ \bibinfo {pages} {4052}
  (\bibinfo {year} {1992})},\ \Eprint {http://arxiv.org/abs/hep-lat/9205007}
  {arXiv:hep-lat/9205007} \BibitemShut {NoStop}%
\bibitem [{\citenamefont {Bhaduri}\ \emph {et~al.}(1981)\citenamefont
  {Bhaduri}, \citenamefont {Cohler},\ and\ \citenamefont
  {Nogami}}]{Bhaduri:1981pn}%
  \BibitemOpen
  \bibfield  {author} {\bibinfo {author} {\bibfnamefont {R.}~\bibnamefont
  {Bhaduri}}, \bibinfo {author} {\bibfnamefont {L.}~\bibnamefont {Cohler}}, \
  and\ \bibinfo {author} {\bibfnamefont {Y.}~\bibnamefont {Nogami}},\ }\href
  {\doibase 10.1007/BF02827441} {\bibfield  {journal} {\bibinfo  {journal}
  {Nuovo Cim. A}\ }\textbf {\bibinfo {volume} {65}},\ \bibinfo {pages} {376}
  (\bibinfo {year} {1981})}\BibitemShut {NoStop}%
\bibitem [{\citenamefont {Godfrey}\ and\ \citenamefont
  {Isgur}(1985)}]{Godfrey:1985xj}%
  \BibitemOpen
  \bibfield  {author} {\bibinfo {author} {\bibfnamefont {S.}~\bibnamefont
  {Godfrey}}\ and\ \bibinfo {author} {\bibfnamefont {N.}~\bibnamefont
  {Isgur}},\ }\href {\doibase 10.1103/PhysRevD.32.189} {\bibfield  {journal}
  {\bibinfo  {journal} {Phys. Rev. D}\ }\textbf {\bibinfo {volume} {32}},\
  \bibinfo {pages} {189} (\bibinfo {year} {1985})}\BibitemShut {NoStop}%
\bibitem [{\citenamefont {Capstick}\ and\ \citenamefont
  {Isgur}(1985)}]{Capstick:1986bm}%
  \BibitemOpen
  \bibfield  {author} {\bibinfo {author} {\bibfnamefont {S.}~\bibnamefont
  {Capstick}}\ and\ \bibinfo {author} {\bibfnamefont {N.}~\bibnamefont
  {Isgur}},\ }\href {\doibase 10.1103/PhysRevD.34.2809} {\bibfield  {journal}
  {\bibinfo  {journal} {AIP Conf. Proc.}\ }\textbf {\bibinfo {volume} {132}},\
  \bibinfo {pages} {267} (\bibinfo {year} {1985})}\BibitemShut {NoStop}%
\bibitem [{\citenamefont {Aoki}\ \emph {et~al.}(2020)\citenamefont {Aoki} \emph
  {et~al.}}]{Aoki:2019cca}%
  \BibitemOpen
  \bibfield  {author} {\bibinfo {author} {\bibfnamefont {S.}~\bibnamefont
  {Aoki}} \emph {et~al.} (\bibinfo {collaboration} {Flavour Lattice Averaging
  Group}),\ }\href {\doibase 10.1140/epjc/s10052-019-7354-7} {\bibfield
  {journal} {\bibinfo  {journal} {Eur. Phys. J. C}\ }\textbf {\bibinfo {volume}
  {80}},\ \bibinfo {pages} {113} (\bibinfo {year} {2020})},\ \Eprint
  {http://arxiv.org/abs/1902.08191} {arXiv:1902.08191 [hep-lat]} \BibitemShut
  {NoStop}%
\bibitem [{\citenamefont {Lewis}\ and\ \citenamefont
  {Woloshyn}(2009)}]{Lewis:2008fu}%
  \BibitemOpen
  \bibfield  {author} {\bibinfo {author} {\bibfnamefont {R.}~\bibnamefont
  {Lewis}}\ and\ \bibinfo {author} {\bibfnamefont {R.}~\bibnamefont
  {Woloshyn}},\ }\href {\doibase 10.1103/PhysRevD.79.014502} {\bibfield
  {journal} {\bibinfo  {journal} {Phys. Rev. D}\ }\textbf {\bibinfo {volume}
  {79}},\ \bibinfo {pages} {014502} (\bibinfo {year} {2009})},\ \Eprint
  {http://arxiv.org/abs/0806.4783} {arXiv:0806.4783 [hep-lat]} \BibitemShut
  {NoStop}%
\bibitem [{\citenamefont {Groote}\ and\ \citenamefont
  {Shigemitsu}(2000)}]{Groote:2000jd}%
  \BibitemOpen
  \bibfield  {author} {\bibinfo {author} {\bibfnamefont {S.}~\bibnamefont
  {Groote}}\ and\ \bibinfo {author} {\bibfnamefont {J.}~\bibnamefont
  {Shigemitsu}},\ }\href {\doibase 10.1103/PhysRevD.62.014508} {\bibfield
  {journal} {\bibinfo  {journal} {Phys. Rev. D}\ }\textbf {\bibinfo {volume}
  {62}},\ \bibinfo {pages} {014508} (\bibinfo {year} {2000})},\ \Eprint
  {http://arxiv.org/abs/hep-lat/0001021} {arXiv:hep-lat/0001021} \BibitemShut
  {NoStop}%
\bibitem [{\citenamefont {Nguyen}\ \emph {et~al.}(2011)\citenamefont {Nguyen},
  \citenamefont {Ishikawa}, \citenamefont {Ukawa},\ and\ \citenamefont
  {Ukita}}]{Nguyen:2011ek}%
  \BibitemOpen
  \bibfield  {author} {\bibinfo {author} {\bibfnamefont {O.~H.}\ \bibnamefont
  {Nguyen}}, \bibinfo {author} {\bibfnamefont {K.-I.}\ \bibnamefont
  {Ishikawa}}, \bibinfo {author} {\bibfnamefont {A.}~\bibnamefont {Ukawa}}, \
  and\ \bibinfo {author} {\bibfnamefont {N.}~\bibnamefont {Ukita}},\ }\href
  {\doibase 10.1007/JHEP04(2011)122} {\bibfield  {journal} {\bibinfo  {journal}
  {JHEP}\ }\textbf {\bibinfo {volume} {04}},\ \bibinfo {pages} {122} (\bibinfo
  {year} {2011})},\ \Eprint {http://arxiv.org/abs/1102.3652} {arXiv:1102.3652
  [hep-lat]} \BibitemShut {NoStop}%
\bibitem [{\citenamefont {Dowdall}\ \emph {et~al.}(2012)\citenamefont {Dowdall}
  \emph {et~al.}}]{Dowdall:2011wh}%
  \BibitemOpen
  \bibfield  {author} {\bibinfo {author} {\bibfnamefont {R.}~\bibnamefont
  {Dowdall}} \emph {et~al.} (\bibinfo {collaboration} {HPQCD}),\ }\href
  {\doibase 10.1103/PhysRevD.85.054509} {\bibfield  {journal} {\bibinfo
  {journal} {Phys. Rev. D}\ }\textbf {\bibinfo {volume} {85}},\ \bibinfo
  {pages} {054509} (\bibinfo {year} {2012})},\ \Eprint
  {http://arxiv.org/abs/1110.6887} {arXiv:1110.6887 [hep-lat]} \BibitemShut
  {NoStop}%
\bibitem [{\citenamefont {Manohar}(1997)}]{Manohar:1997qy}%
  \BibitemOpen
  \bibfield  {author} {\bibinfo {author} {\bibfnamefont {A.~V.}\ \bibnamefont
  {Manohar}},\ }\href {\doibase 10.1103/PhysRevD.56.230} {\bibfield  {journal}
  {\bibinfo  {journal} {Phys. Rev. D}\ }\textbf {\bibinfo {volume} {56}},\
  \bibinfo {pages} {230} (\bibinfo {year} {1997})},\ \Eprint
  {http://arxiv.org/abs/hep-ph/9701294} {arXiv:hep-ph/9701294} \BibitemShut
  {NoStop}%
\bibitem [{\citenamefont {Manohar}\ and\ \citenamefont
  {Wise}(1993)}]{Manohar:1992nd}%
  \BibitemOpen
  \bibfield  {author} {\bibinfo {author} {\bibfnamefont {A.~V.}\ \bibnamefont
  {Manohar}}\ and\ \bibinfo {author} {\bibfnamefont {M.~B.}\ \bibnamefont
  {Wise}},\ }\href {\doibase 10.1016/0550-3213(93)90614-U} {\bibfield
  {journal} {\bibinfo  {journal} {Nucl. Phys. B}\ }\textbf {\bibinfo {volume}
  {399}},\ \bibinfo {pages} {17} (\bibinfo {year} {1993})},\ \Eprint
  {http://arxiv.org/abs/hep-ph/9212236} {arXiv:hep-ph/9212236} \BibitemShut
  {NoStop}%
\bibitem [{\citenamefont {Czarnecki}\ \emph {et~al.}(2018)\citenamefont
  {Czarnecki}, \citenamefont {Leng},\ and\ \citenamefont
  {Voloshin}}]{Czarnecki:2017vco}%
  \BibitemOpen
  \bibfield  {author} {\bibinfo {author} {\bibfnamefont {A.}~\bibnamefont
  {Czarnecki}}, \bibinfo {author} {\bibfnamefont {B.}~\bibnamefont {Leng}}, \
  and\ \bibinfo {author} {\bibfnamefont {M.}~\bibnamefont {Voloshin}},\ }\href
  {\doibase 10.1016/j.physletb.2018.01.034} {\bibfield  {journal} {\bibinfo
  {journal} {Phys. Lett. B}\ }\textbf {\bibinfo {volume} {778}},\ \bibinfo
  {pages} {233} (\bibinfo {year} {2018})},\ \Eprint
  {http://arxiv.org/abs/1708.04594} {arXiv:1708.04594 [hep-ph]} \BibitemShut
  {NoStop}%
\end{thebibliography}%

\end{document}